\documentclass[hyper,11pt,paper]{article}
\usepackage{jheppub}
\usepackage{graphicx,amsmath,amssymb,multirow,array,bm,mathrsfs}
\usepackage{epsf,amsfonts,bbold}
\usepackage[numbers,sort&compress]{natbib}
\usepackage{hyperref}
\usepackage{slashed}
\usepackage{braket}

\newcommand{\beq}{\begin{equation}}
\newcommand{\eeq}{\end{equation}}
\newcommand{\s}{{\mathcal{S}}}

\newcommand{\bea}{\begin{eqnarray}}
\newcommand{\eea}{\end{eqnarray}}
\newcommand{\bdm}{\begin{displaymath}}
\newcommand{\edm}{\end{displaymath}}
\newcommand{\bal}{\begin{align}}
\newcommand{\eal}{\end{align}}
\newcommand{\nn}{\nonumber}

\newcommand{\NO}{\nonumber}
\newcommand{\p}{\pi}

\newcommand{\one}{\textrm{I}}
\newcommand{\two}{\textrm{II}}
\newcommand{\K}{\textrm{K}}
\newcommand{\RKKY}{\textrm{RKKY}}

\def\pa{\partial}

\def\le{\left}
\def\ri{\right}

\def\dif{\mathrm{d}}
\def\a{\alpha}
\def\b{\beta}
\def\g{\gamma}
\def\d{\delta}
\def\D{\Delta}
\def\l{\lambda}

\def\F{\Phi}
\def\f{\phi}
\def\e{\epsilon}

\def\m{\mu}
\def\n{\nu}

\def\r{\rho}
\def\s{\sigma}
\def\t{\theta}

\def\k{\chi}

\def\aplus{\mathcal{A}_t^+}
\def\aminus{\mathcal{A}_t^-}
\def\ca{\mathcal{A}}
\def\cq{\mathcal{Q}}
\def\co{\mathcal{O}}
\def\ch{\mathcal{H}}
\def\G{\Gamma}
\def\car{\mathcal{R}}
\def\cj{{\cal J}}

\def\){\right)}
\def\({\left( }
\def\]{\right] }
\def\[{\left[ }

\def\NO{\nonumber}

\newcommand{\be}{\begin{equation}}
\newcommand{\ee}{\end{equation}}

\def\bea{\begin{eqnarray}}
\def\eea{\end{eqnarray}}

\def\bal#1\eal{\begin{align}#1\end{align}}

\def\bald{\begin{aligned}}
\def\eald{\end{aligned}}

\def\beqx{\begin{displaymath}}
\def\eeqx{\end{displaymath}}

\newcommand{\bmat}{\left(\begin{array}}
\newcommand{\emat}{\end{array}\right)}

\def\a{\alpha}
\def\b{\beta}

\def\d{\delta}
\def\e{\epsilon}
\def\f{\phi}
\def\g{\gamma}

\def\k{\kappa}
\def\l{\lambda}
\def\m{\mu}
\def\n{\nu}

\def\p{\pi}

\def\r{\rho}
\def\s{\sigma}
\def\t{\tau}

\def\D{\Delta}
\def\F{\Phi}
\def\G{\Gamma}

\def\ca{{\cal A}}

\def\ch{{\cal H}}

\def\cj{{\cal J}}

\def\co{{\cal O}}

\def\cq{{\cal Q}}
\def\car{{\cal R}}

\def\bo{{\raise-.3ex\hbox{\large$\Box$}}}               
\def\pa{\partial}                                       
\def\face{{\raise.2ex\hbox{$\displaystyle \bigodot$}\mskip-2.2mu \llap {$\ddot
        \smile$}}}                                   
\def\>{\rangle}                                      
\def\<{\langle}                                      

\def\Hat#1{\widehat{#1}}                             
\def\leftrightarrowfill{$\mathsurround=0pt \mathord\leftarrow \mkern-6mu
        \cleaders\hbox{$\mkern-2mu \mathord- \mkern-2mu$}\hfill
        \mkern-6mu \mathord\rightarrow$}        
\def\dvec#1{\vbox{\ialign{##\crcr
        \leftrightarrowfill\crcr\noalign{\kern-1pt\nointerlineskip}
        $\hfil\displaystyle{#1}\hfil$\crcr}}}           

\def\-{\hphantom{-}}

\title{A Holographic Two-Impurity Kondo Model}

\author[a]{Andy O'Bannon,}
\author[b]{Ioannis Papadimitriou,}
\author[c]{Jonas Probst,}

\affiliation[a]{STAG Research Centre, Physics and Astronomy, University of Southampton, \\ \hspace{.3cm} Highfield, Southampton SO17 1BJ, United Kingdom}
\affiliation[b]{SISSA and INFN - Sezione di Trieste, Via Bonomea 265, I 34136 Trieste, Italy}
\affiliation[c]{Rudolf Peierls Centre for Theoretical Physics, University of Oxford, \\ \hspace{.3cm} 1 Keble Road, Oxford OX1 3NP, United Kingdom}

\emailAdd{a.obannon@soton.ac.uk}
\emailAdd{ioannis.papadimitriou@sissa.it}
\emailAdd{jonas.probst@physics.ox.ac.uk}

\abstract{We propose a model of a strongly-interacting two-impurity Kondo system based on the Anti-de Sitter/Conformal Field Theory (AdS/CFT) correspondence, also known as holography. In a Landau Fermi Liquid, the single-impurity Kondo effect is the screening of an impurity spin at low temperature $T$. The two-impurity Kondo model then describes the competition between the Kondo interaction and the Heisenberg interaction between two impurity spins, also called the Ruderman-Kittel-Kasuya-Yosida (RKKY) interaction. For spin-$1/2$ impurities, that competition leads to a quantum phase transition from a Kondo-screened phase to a phase in which the two impurity spins screen one another. Our holographic model is based on a $(1+1)$-dimensional CFT description of the two-impurity Kondo model, reliable for two impurities with negligible separation in space. We consider only impurity spins in a totally anti-symmetric representation of an $SU(N)$ spin symmetry. We employ a large-$N$ limit, in which both Kondo and RKKY couplings are double-trace, and both Kondo and inter-impurity screening appear as condensation of single-trace operators at the impurities' location. We perform the holographic renormalization of our model, which allows us to identify the Kondo and RKKY couplings as boundary conditions on fields in AdS. We numerically compute the phase diagram of our model in the plane of RKKY coupling versus $T$, finding evidence for a quantum phase transition from a trivial phase, with neither Kondo nor inter-impurity screening, to a non-trivial phase, with both Kondo and anti-ferromagnetic inter-impurity screening. More generally we show, just using $SU(N)$ representation theory, that ferromagnetic correlations must be absent at leading order in the large-$N$ limit. Our holographic model may be useful for studying many open problems involving strongly-interacting quantum impurities, including for example the Kondo lattice, relevant for describing the heavy fermion compounds.}

\preprint{OUTP-15-28P, SISSA 49/2015/FISI}

\begin{document}
\maketitle

\section{Introduction and Summary}
\label{intro}

The heavy fermion compounds are rare-earth or actinide based alloys, many of which exhibit a quantum phase transition, that is, a phase transition when the temperature $T$ is zero, as a function of pressure, magnetic field, or chemical doping~\cite{2006cond.mat.12006C,2008NatPh...4..186G,2009arXiv0912.0040S,2010uqpt.book..193S,2010Sci...329.1161S,2015arXiv150905769C}. Typically, on one side of the quantum phase transition is an anti-ferromagnetic (AFM) metallic phase, while on the other side is a Landau Fermi Liquid (LFL) phase with fermionic quasi-particle excitations hundreds or thousands of times heavier than those of normal metals, hence the name ``heavy fermions.'' Both phases have a Fermi surface, although the Fermi surface volume is larger in the LFL phase by an amount proportional to the concentration of the rare-earth or actinide atoms.

Heavy fermion quantum phase transitions are typically continuous, occurring at a quantum critical point.  When heated up, the quantum critical degrees of freedom give rise to a ``strange metal,'' whose characteristic feature is an electrical resistivity $\rho \propto T$~\cite{2006cond.mat.12006C}. In contrast, the AFM and LFL phases have $\rho \propto T^2$. Although the strange metal has a Fermi surface, similar to the AFM and LFL phases, the strange metal degrees of freedom are not merely weakly-interacting quasi-particle excitations of the Fermi surface. Indeed, the strange metal degrees of freedom appear to be strongly-interacting~\cite{2014arXiv1409.4673K}. As a result, formulating a microscopic theory of strange metals remains a major challenge.

That challenge is especially urgent because strange metals often exhibit transitions to superconductivity, which is necessarily unconventional, \textit{i.e.}\ non-BCS. Moreover, among heavy fermion superconductors, the strange metals have the highest transition temperatures, reaching as high as $\approx 20^{\circ}$K~\cite{2006cond.mat.12006C}. Remarkably, in the hole-doped cuprate superconductors, the highest transition temperatures also appear in strange metal states~\cite{2014arXiv1409.4673K}. Clearly, any microscopic description of these forms of unconventional superconductivity must be based on a theory of strange metals.

Doniach proposed a theoretical description of the heavy fermion compounds at lattice scales, called the ``Kondo lattice''~\cite{1977PhyBC..91..231D}. The rare-earth or actinide atoms' $f$-orbital electrons act as localized magnetic moments, that is, spins fixed in a regular, periodic array, while the other atoms provide conduction electrons that form a LFL. The Kondo lattice Hamiltonian thus includes a LFL kinetic term plus two types of interaction terms. The first type are Kondo interaction terms: each spin couples to the spin current of the LFL. The second type are Heisenberg interaction terms between spins.

In fact, in the heavy fermion compounds the latter is induced by the former: via the Kondo coupling, each spin produces Friedel oscillations in the LFL that subsequently induce Heisenberg couplings between spins. The strength of these induced Heisenberg interactions decays with separation as a sinusoid (with period fixed by the Fermi momentum, $k_F$) in a power-law envelope~\cite{2006cond.mat.12006C}. These induced Heisenberg interactions are called Ruderman-Kittel-Kasuya-Yosida (RKKY) interactions. In a (standard) abuse of terminology, we will always refer to the Heisenberg interactions as ``RKKY,'' even if they are not induced by the Kondo interactions. Indeed, following the most general formulation of the Kondo lattice, we will treat the Kondo and Heisenberg/RKKY couplings as independent parameters.

The Kondo lattice ``problem'' is to determine the eigenstates of the Kondo lattice Hamiltonian for any strengths of the Kondo and RKKY couplings and for any $T$, and from them to derive observables such as $\rho$. A complete solution, employing no approximations, remains a major challenge. Existing solutions exploit simplifying limits. Indeed, the simplest limits just ignore one or the other interaction. For example, suppose the Kondo interaction energy is negligible relative to that of the RKKY interaction. In other words, suppose we just ignore the Kondo interaction terms in the Hamiltonian, which thus reduces to the sum of the LFL kinetic term and a Heisenberg Hamiltonian. AFM RKKY couplings then trivially lead to an AFM metal, as observed on one side of many heavy fermion quantum phase transitions.

Conversely, suppose the RKKY interaction terms can be neglected, so that the Hamiltonian describes a LFL with Kondo couplings to a lattice of spins. In fact, consider an even more extreme simplification: a Hamiltonian with only two terms, a LFL kinetic term and a Kondo coupling to a \textit{single} spin. Such a Hamiltonian has been realized experimentally in quantum dots~\cite{Goldhaber1998,Cronenwett24071998,vanderWiel22092000} and in metals doped with a dilute concentration of magnetic impurities~\cite{PTP.32.37,0034-4885-37-2-001,GrŸner1978591}, and is thus called the ``single-impurity Kondo Hamiltonian.''

The corresponding ``single-impurity Kondo problem'' has been solved completely via a combination of complementary techniques: numerical Renormalization Group (RG) techniques~\cite{Wilson:1974mb,PhysRevB.21.1003,PhysRevB.21.1044}, integrability~\cite{PhysRevLett.45.379,Wiegmann:1980,RevModPhys.55.331,doi:10.1080/00018738300101581,0022-3719-19-17-017,1994cond.mat..8101A,ZinnJustin1998, PhysRevB.58.3814}, large-$N$ limits~\cite{PhysRevB.35.5072,RevModPhys.59.845,1997PhRvL..79.4665P,1998PhRvB..58.3794P,2006cond.mat.12006C,2015arXiv150905769C}, Conformal Field Theory (CFT) techniques~\cite{1998PhRvB..58.3794P,Affleck:1990zd,Affleck:1990by,Affleck:1990iv,Affleck:1991tk,PhysRevB.48.7297,Affleck:1995ge}, and more. For reviews of many of these, see for example refs.~\cite{Hewson:1993,doi:10.1080/000187398243500}. The solution is most succinctly described as an RG flow from the ultra-violet (UV) to the infra-red (IR), with $T$ playing the role of RG scale. What happens as $T$ decreases depends on the sign of the Kondo coupling, that is, whether the Kondo coupling is ferromagnetic (FM) or AFM. The former renormalizes to zero in the IR: the low-$T$ limit is a LFL and a free spin. An AFM Kondo coupling, however, increases as $T$ decreases, eventually diverging at a characteristic, dynamically-generated scale, the ``Kondo temperature,'' $T_{\K}$.

In that case, when $T<T_{\K}$ the LFL fermions form a screening cloud around the impurity, the ``Kondo cloud''~\cite{2009arXiv0911.2209A}. When $T=0$, the Kondo cloud has characteristic size $\propto1/T_{\K}$ and net spin $1/2$, which locks with the impurity spin into the anti-symmetric singlet of the spin $SU(2)$ symmetry. This ``Kondo singlet'' is the ground state of the single-impurity AFM Kondo Hamiltonian~\cite{Hewson:1993,doi:10.1080/000187398243500}. Remarkably, the solutions of the AFM Kondo problem reveal that the excitations about this highly-entangled many-body ground state are simply those of a LFL, albeit with different properties from the LFL at high $T$. In particular, at low $T$ the LFL fermion spectral function exhibits a ``Kondo resonance'' at the Fermi level, and the LFL fermions acquire an $s$-wave phase shift of $\pi/2$, the maximum allowed by unitarity~\cite{Hewson:1993,doi:10.1080/000187398243500}. In practice, the name ``Kondo effect'' is used to refer to any, or all, of these phenomena (the screening of the impurity spin, the Kondo resonance, the phase shift, etc.).

Returning to the Kondo lattice, but still ignoring RKKY interactions, as the impurity concentration increases the single-impurity approximation will eventually break down because the Kondo clouds will significantly overlap~\cite{2009arXiv0911.2209A}. In that case, mean-field limits can restore control. Examples of mean-field limits include promoting the $SU(2)$ spin symmetry to $SU(N)$ and then sending $N \to \infty$, \textit{i.e.}\ the large-$N$ limit~\cite{PhysRevB.35.5072,2006cond.mat.12006C,2015arXiv150905769C}, or sending the number of spatial dimensions to infinity, which is the basis for (extended) dynamical mean field theory~\cite{RevModPhys.68.13,2006cond.mat.12006C,2009arXiv0912.0040S,2010uqpt.book..193S}. Generically in such mean-field limits, Kondo resonances do appear, and in fact hybridize with the conduction band, injecting new fermions into the spectrum. These new fermions contribute to the Fermi surface, which thus becomes ``large,'' and are also heavy, intuitively because they arise from spins fixed in place.

These simplifying limits can thus describe the two phases on either side of the heavy fermion quantum critical point. What about the quantum critical point itself, which arises from a competition between RKKY and Kondo interactions in a regime where neither is negligible? What about the resulting strange metal?

Here the simplest starting point is the two-impurity Kondo Hamiltonian, which includes the LFL kinetic term, two spins with equal Kondo couplings to the LFL, and an RKKY interaction between the spins~\cite{PhysRevLett.47.737,PhysRevLett.58.843,PhysRevB.35.5072,PhysRevB.35.4901,PhysRevLett.61.125,PhysRevB.39.3415,PhysRevB.40.324,PhysRevB.40.4780,Jones1990,Millis1990,Affleck:1991yq,PhysRevLett.72.916,1995PhRvB..52.9528A,INGERSENT1994402,PhysRevLett.74.2583,PhysRevB.51.8287,PhysRevLett.74.2808,PhysRevLett.76.275,Jones:2007}. The two-impurity Kondo problem has not been solved completely, however a combination of various methods, including large-$N$~\cite{PhysRevB.35.5072,PhysRevB.39.3415,Millis1990} and CFT techniques~\cite{PhysRevB.40.324,Affleck:1991yq,1995PhRvB..52.9528A,PhysRevLett.74.2583,PhysRevB.51.8287,PhysRevLett.74.2808}, have provided substantial progress.

Crucially, in the two-impurity Kondo model the LFL decomposes into two ``channels'' of fermions, of even and odd parity with respect to the mid-plane between the impurity spins. For AFM Kondo coupling, what happens at low $T$ depends on whether the RKKY coupling is FM or AFM. For example, with $SU(2)$ spin symmetry, an infinitely large FM RKKY coupling forces the spins to lock into a triplet, effectively forming a single spin-$1$ impurity. Upon lowering $T$, that effective impurity is screened in a multi-stage Kondo effect. The odd channel has larger Kondo coupling and so screens half of the impurity spin first, leaving behind a single spin-$1/2$ impurity, which is then screened at lower $T$ by the even channel. At sufficiently low $T$ a LFL emerges, where each channel acquires a $\pi/2$ phase shift. For infinitely large AFM RKKY coupling, the spins lock into an anti-symmetric singlet, and so effectively decouple from the LFL. The low-$T$ LFL is then the same as at high $T$, and in particular no phase shift occurs. However, at $T=0$ particle-hole symmetry allows phase shifts of only $\pi/2$ or zero~\cite{Millis1990}. Changing the RKKY coupling from FM to AFM thus necessarily leads to a quantum phase transition where the phase shift jumps discontinuously from $\pi/2$ to zero~\cite{Millis1990}. That quantum phase transition turns out to be second order~\cite{PhysRevLett.58.843,PhysRevLett.61.125,PhysRevB.40.324,Jones1990,Jones:2007}. Crucially, however, no change of symmetry occurs at the quantum critical point: the $T=0$ ground state is always an $SU(2)$ singlet.

The two-impurity Kondo problem teaches us two lessons relevant for the Kondo lattice problem. The first lesson is that the Kondo effect does not occur at each impurity, rather, a Kondo effect occurs in each symmetry channel, even or odd. For AFM Kondo and infinite FM RKKY couplings, the ground state is not two Kondo clouds, one around each impurity, but a more complicated state in which the even channel's Kondo cloud screens the impurity left over by the odd channel's Kondo cloud. The second lesson is that a competition between RKKY and Kondo couplings can produce a non-trivial critical point. However, the two-impurity Kondo critical point is qualitatively distinct from that of the Kondo lattice: the latter involves a genuine change of symmetry, from the AFM metal phase to the paramagnetic LFL phase.

To date, for the Kondo lattice problem in the regime where Kondo and RKKY interactions are comparable, the most successful simplifying limits are mean-field~\cite{2006cond.mat.12006C,2008NatPh...4..186G,2009arXiv0912.0040S,2010uqpt.book..193S,2010Sci...329.1161S,2015arXiv150905769C}. Indeed, a patchwork of various mean-field limits can reproduce many features of the heavy fermion phase diagram, at least qualitatively. However to our knowledge no single mean-field limit, or collection of mean-field limits, describes all features, qualitatively or quantitatively.

We thus turn to an alternative approach: the Anti-de Sitter/CFT (AdS/CFT) Correspondence, also known as gauge-gravity duality or holography~\cite{Maldacena:1997re,Gubser:1998bc,Witten:1998qj}. AdS/CFT equates a weakly-coupled theory of gravity in $(d+1)$-dimensional AdS spacetime, $AdS_{d+1}$, with a strongly-coupled $d$-dimensional CFT ``living'' on the boundary of $AdS_{d+1}$. Typically the strongly-coupled CFT is a non-Abelian Yang-Mills (YM)~\cite{Aharony:1999ti} or Chern-Simons theory~\cite{Aharony:2008ug} in the 't Hooft large-$N$ limit.

Various holographic single-impurity Kondo models exist: see for example refs.~\cite{Kachru:2009xf,Sachdev:2010um,Kachru:2010dk,Sachdev:2010uj,Mueck:2010ja,Faraggi:2011bb,Jensen:2011su,Karaiskos:2011kf,Harrison:2011fs,Benincasa:2011zu,Faraggi:2011ge,Benincasa:2012wu,Matsueda:2012gc,Itsios:2012ev,Erdmenger:2013dpa}. In all cases, the $SU(2)$ spin symmetry is replaced by $SU(N)$, which is then gauged, that is, $SU(N)$ gauge fields are introduced, and often additional fields, such as supersymmetric partners for the $SU(N)$ gauge fields. A magnetic impurity is then described by an $SU(N)$ Wilson line~\cite{Maldacena:1998im,Rey:1998ik,Camino:2001at,Yamaguchi:2006tq,Gomis:2006sb,Gomis:2006im}. (For a different approach, using a delta-function source to describe a point-like impurity, see ref.~\cite{Blake:2014lva}.)

These changes to the original Kondo problem have dramatic consequences. The $SU(N)$ gauge fields bring with them a new coupling constant, the 't Hooft coupling. A simple holographic description, in terms of a classical Einstein-Hilbert action, requires two limits, the 't Hooft large-$N$ limit, followed by large 't Hooft coupling. In other words, holography not only employs large $N$, but also replaces the weakly-coupled quasi-particles of a LFL with a strongly-interacting gauge theory. In that sense, holographic Kondo models are always strongly-interacting mean-field limits.

The differences between holography and other mean-field limits raise many questions. How do the phase diagrams of holographic Kondo lattices compare to those of the heavy fermion compounds? In holographic Kondo lattices, does a competition between Kondo and RKKY couplings universally give rise to quantum criticality and strange metal phases? If so, then what can holography teach us about the strange metal degrees of freedom? After all, holography has distinct advantages over other methods, for example in studying entanglement entropy~\cite{Nishioka:2009un}, quantum quenches~\cite{Chesler:2013lia}, and more, which could provide unique insights into the nature of the strange metal degrees of freedom.

Various attempts to build holographic lattices of impurities appear in refs.~\cite{Kachru:2009xf,Sachdev:2010um,Kachru:2010dk,Sachdev:2010uj,Jensen:2011su,Horowitz:2012ky}. However, these include neither the Kondo nor RKKY couplings, so although some may have $\rho \propto T$~\cite{Jensen:2011su}, whether they really describe the strange metal state that arises from heavy fermion quantum criticality is unclear. Indeed, some have properties distinctly different from the heavy fermion strange metal phase, such as non-zero extensive entropy at $T=0$.

In fact, among holographic single-impurity Kondo models, only the model of ref.~\cite{Erdmenger:2013dpa} includes a Kondo coupling at all, and also describes many essential single-impurity Kondo phenomena, such as the appearance of $T_{\K}$ and a phase shift. Our goal in this paper is to extend the holographic single-impurity model of ref.~\cite{Erdmenger:2013dpa} to a holographic two-impurity Kondo model, including an RKKY coupling, and to study whether the competition between the Kondo and RKKY couplings may produce a quantum phase transition.

The holographic single-impurity Kondo model of ref.~\cite{Erdmenger:2013dpa} uses the holographic version of the large-$N$ limit, described above, and two other ingredients: the CFT description of the Kondo problem and the Abrikosov pseudo-fermion representation of the impurity spins.

The CFT approach to the single-impurity Kondo problem~\cite{Affleck:1990zd,Affleck:1990by,Affleck:1990iv,Affleck:1991tk,PhysRevB.48.7297,Affleck:1995ge} begins with a partial wave decomposition of the LFL fermions, retaining only the $s$-wave around the impurity. That limit produces a one-dimensional problem: left- and right-moving fermions (in- and out-going $s$-waves) on a half line (the radial distance to the impurity), interacting with the impurity at the origin. Mapping the right-movers to the negative half of the real line and re-labeling them as left-movers produces the simplest description: left-movers alone on the entire real line, interacting with the impurity at the origin. The advantage of the CFT approach is an infinite accidental symmetry: the left-movers form a chiral CFT, which has an infinite number of symmetry generators, including in particular Virasoro and Kac-Moody generators. In the CFT description, the Kondo effect reduces to a change in boundary conditions at the origin, the simplest example being a $\pi/2$ phase shift.

The CFT approach extends to the two-impurity Kondo problem as well, in the limit that the separation between the impurities is negligible~\cite{PhysRevB.40.324,Affleck:1991yq,1995PhRvB..52.9528A,PhysRevLett.74.2583,PhysRevB.51.8287,PhysRevLett.74.2808}. Now, however, the two channels of fermions, even and odd, lead in the CFT description to two channels of left-moving fermions coupled to a single effective impurity at the origin.

The Abrikosov pseudo-fermion representation is most appropriate for an impurity spin in a totally anti-symmetric representation of $SU(N)$. In that case, we can write the spin operator as a bi-linear product of auxiliary fermions, the pseudo-fermions. That introduces an additional, redundant ``auxiliary'' symmetry, namely a $U(1)$ that shifts the phase of the pseudo-fermion but leaves the spin operator invariant. In the large-$N$ limit, after introducing pseudo-fermions, the Kondo coupling becomes double-trace with respect to $SU(N)$, and the Kondo effect appears as symmetry breaking at the impurity: below a critical temperature, on the order of $T_{\K}$, a charged scalar operator condenses at the impurity~\cite{0022-3719-19-17-017,PhysRevB.35.5072,2003PhRvL..90u6403S,2004PhRvB..69c5111S}. That scalar is built from a LFL fermion and a pseudo-fermion, and is a singlet of $SU(N)$ but bi-fundamental under the electromagnetic $U(1)$ (which shifts the LFL fermion's phase) and auxiliary $U(1)$ symmetries. Its condensation represents Kondo screening, and breaks $U(1) \times U(1)$ to the diagonal. Of course, this phase transition is an artifact of the large-$N$ limit. The actual Kondo effect is a smooth crossover.

With two impurity spins we must introduce two distinct species, or ``flavors,'' of pseudo-fermions. The auxiliary $U(1)$ symmetry is enhanced to $U(2)$, under which the pseudo-fermions transform in the fundamental representation. The RKKY coupling is quadratic in pseudo-fermions, and double-trace with respect to $SU(N)$. Crucially, two totally anti-symmetric impurities cannot lock into a singlet of $SU(N)$ unless $N$ is even and the representation's Young tableau has exactly $N/2$ boxes. In that special case, at large $N$ the competition between Kondo and RKKY couplings leads to a first-order quantum phase transition~\cite{PhysRevB.39.3415,Millis1990}.

On the gravity side of our holographic two-impurity Kondo model, the classical action includes four terms (not counting boundary terms). First is a $(2+1)$-dimensional Einstein-Hilbert action with negative cosmological constant,~\textit{i.e.}\ gravity in $AdS_3$. Roughly speaking, this term is dual to a large-$N$, strongly-coupled $(1+1)$-dimensional CFT. Second is the action of a Chern-Simons gauge field, dual to Kac-Moody currents~\cite{Kraus:2006wn}, as in the CFT description of the Kondo model. In other words, we replace the free left-moving fermions in the CFT description of the Kondo effect with a strongly-coupled CFT with Kac-Moody algebra. Third is a $U(2)$ YM gauge field localized at a co-dimension one brane, meaning an $AdS_2$ subspace of $AdS_3$, dual to the auxiliary $U(2)$ charges of the Abrikosov pseudo-fermions localized at the impurity. Fourth is a complex scalar field also localized to $AdS_2$, bi-fundamental under the Chern-Simons and $U(2)$ YM gauge groups, and dual to the scalar operator that condenses in the large-$N$ Kondo effect. We treat the Chern-Simons gauge field, $U(2)$ YM fields, and bi-fundamental scalar field in the probe limit: in the classical action at large $N$, the Einstein-Hilbert term scales as $N^2$, while the three other terms each scale as $N$. To leading order in $N$, we can thus neglect the matter fields' contribution to Einstein's equation, and solve the matter fields' equations of motion in the fixed background geometry. We study the dual CFT only in $(1+1)$-dimensional Minkowski space with non-zero temperature $T$, so the bulk geometry will always be the Poincar\'e patch of the BTZ black hole~\cite{Aharony:1999ti}.

We have two main results. The first is the holographic renormalization~\cite{deHaro:2000xn,Bianchi:2001kw,Papadimitriou:2005ii,Papadimitriou:2010as} of our model. The main challenge there is the well-known fact that a YM gauge field in $AdS_2$ diverges asymptotically near the $AdS_2$ boundary, in stark contrast to gauge fields in higher-dimensional AdS spaces. That divergence can alter the asymptotics of any other fields coupled to the YM field, and indeed alters the asymptotics of our complex scalar field. Recalling that the near-boundary region of $AdS_2$ corresponds to the UV of the dual field theory~\cite{Aharony:1999ti}, holography thus suggests that the charge dual to the $AdS_2$ YM field behaves much like an irrelevant operator. In particular, changing the auxiliary $U(2)$ charge can change the dimension of the scalar operator at the UV fixed point. Our holographic renormalization will indeed be very similar to that for fields dual to irrelevant operators~\cite{vanRees:2011fr,vanRees:2011ir}. The holographic renormalization will provide the complete set of covariant boundary counterterms, which allows us to compute renormalized correlators, including the renormalized thermodynamic free energy. Moreover, the holographic renormalization allows us to identify the double-trace Kondo and RKKY couplings of our model as boundary conditions on the bi-fundamental scalar and $U(2)$ YM field, respectively. This is the first explicit identification of an RKKY coupling in holography, which is a necessary first step towards building a holographic Kondo lattice.

Our second main result is the phase diagram of our model in the plane of RKKY coupling versus $T/T_{\K}$, which we obtain by solving the equations of motion of our model numerically. For any FM RKKY coupling, and for AFM RKKY coupling below a critical value, we find only a trivial phase, in which neither Kondo nor inter-impurity screening occurs, dual to trivial solutions for the complex scalar and $U(2)$ YM gauge fields. Above a critical value of AFM RKKY coupling and below a critical $T/T_{\K}$, a phase transition occurs: the complex scalar condenses, signaling Kondo screening, and simultaneously the off-diagonal components of the $U(2)$ YM gauge field condense, signaling AFM correlations of order $N^2$ between the spins. Indeed, the coexistence of Kondo and inter-impurity screening is generic in two-impurity Kondo models when the two impurity spins do not lock into a spin singlet, and in fact the coexistence of Kondo and inter-impurity screening is widely believed to occur in the Kondo lattice~\cite{2009arXiv0912.0040S,2010uqpt.book..193S,Jones:2007}. In fact, we will present an argument, to our knowledge novel, that at leading order at large $N$ and with totally anti-symmetric impurity spins, only AFM correlations of order $N^2$ are visible, while FM correlations are absent. That argument is based only on $SU(N)$ representation theory, and thus may have implications for many other large-$N$ descriptions of magnetism, in holography and beyond.  Our numerical evidence suggests that the transition is first order near the critical value of the AFM RKKY coupling, but upon increasing the AFM RKKY coupling becomes second order. Our numerical evidence also suggests that the first order-transition near the critical AFM RKKY coupling persists to $T/T_{\K}=0$, similar to the large-$N$ quantum phase transition with two impurities in an anti-symmetric representation with exactly $N/2$ boxes~\cite{PhysRevB.39.3415,Millis1990}. In other words, our numerical evidence suggests that a first-order quantum phase transition occurs in our model, as we increase the RKKY coupling through a critical AFM value, from a trivial phase, with neither Kondo screening nor inter-impurity correlations, to a non-trivial phase, with both Kondo screening and AFM inter-impurity correlations of order $N^2$.

This paper is organized as follows. In section~\ref{review} we review the details of the original two-impurity Kondo model that we will need for our holographic model. In section~\ref{model} we present our holographic model. In section~\ref{holorg} we describe our ansatz for solutions and perform the holographic renormalization of our model. In section~\ref{phase} we present our results for the phase diagram. We conclude in section~\ref{outlook} with suggestions for future research, and especially for building holographic Kondo lattices. We collect in an appendix some technical results about the normalizability of massless and massive gauge fields in $AdS_{d+1}$ with $d\geq 1$, where the case of massless gauge fields in $d=1$ ($AdS_2$) will be useful to us throughout the paper.

\section{Review: the Single- and Two-Impurity Kondo Models}
\label{review}

In this section we review the details of the single- and two-impurity Kondo models that we will need for our holographic model. In particular, we review the CFT and large-$N$ approaches to both the single- and two-impurity Kondo models.

\subsection{The Single-Impurity Kondo Model}
\label{sikm}

The single-impurity Kondo Hamiltonian density, $\hat{H}_{\K}$, describes the interaction of a LFL with a single localized quantum impurity spin~\cite{Hewson:1993,doi:10.1080/000187398243500}:
\begin{align}
\hat{H}_{\K}=c_\a^\dagger\frac{-\nabla^2}{2m}c_\a+\hat{\l}_{\K} \, \d(\vec{x}) \, S^A \, \hat{\mathcal{J}}^A,
\label{KondoH}
\end{align}
where $c_\a^\dagger$ and $c_\a$ are creation and annihilation operators for LFL fermions of spin $\a$, that is, the $c_\a$ are in the fundamental representation of the $SU(2)$ spin symmetry, $m$ is the fermion mass, $\hat{\l}_{\K}$ is the Kondo coupling constant, $S^A$ is the spin of the impurity localized at the origin, also in the fundamental representation of $SU(2)$, and $\hat{\mathcal{J}}^A=c_\a^\dagger \,T^A_{\a\b} \, c_\b$ is the LFL spin current at the impurity's location, with $T^A_{\a\b}$ the generators of $SU(2)$ (so $A=1,2,3$).

The beta function for $\hat{\l}_{\K}$ is negative to one-loop order in perturbation theory in $\hat{\l}_{\K}$~\cite{doi:10.1080/000187398243500,Hewson:1993}. Consequently, a FM Kondo coupling, $\hat{\l}_{\K}<0$, is marginally irrelevant in the IR. On the other hand, an AFM Kondo coupling, $\hat{\l}_{\K}>0$, is asymptotically free in the UV, but in the IR appears to diverge at a dynamically-generated scale, the Kondo temperature, $T_{\K}$. In the high-temperature regime, $T\gg T_{\K}$, perturbation theory in $\hat{\l}_{\K}$ is thus reliable for calculating observables, including (at one loop) the characteristic $-\ln(T/T_{\K})$ contribution to the resistivity, $\rho$~\cite{PTP.32.37}. At low temperatures, the renormalization of AFM $\hat{\l}_{\K}$ to large values is the main obstacle to solving the Kondo problem, \textit{i.e.}\ to determining the eigenstates of $\hat{H}_{\K}$ and the resulting thermodynamic and transport properties.

The single-impurity Kondo model has been realized experimentally in quantum dots~\cite{Goldhaber1998,Cronenwett24071998,vanderWiel22092000} and in metals doped with a dilute concentration of magnetic impurities~\cite{PTP.32.37,0034-4885-37-2-001,GrŸner1978591}. In many of these cases, multiple conduction bands, or ``channels'' (or in particle physics language, ``flavors''), couple to the same impurity, and in many cases the impurity has a spin degeneracy greater than two~\cite{PhysRevB.35.5072,RevModPhys.59.845,2006cond.mat.12006C,2015arXiv150905769C,Hewson:1993,doi:10.1080/000187398243500}. To describe these cases, the Kondo model has been generalized to the case of $k$ fermion channels, each in the fundamental representation of an $SU(N)$ spin symmetry, with an impurity $S^A$ in a general representation $\r_{\textrm{UV}}$ of $SU(N)$, with dimension $\mathrm{dim}(\r_{\textrm{UV}})$. The symmetry group that leaves $\hat{H}_{\K}$ invariant is then $SU(N) \times SU(k) \times U(1)$, with channel symmetry $SU(k)$ and electromagnetic symmetry $U(1)$, which acts by shifting the phase of $c_\a$. The single-impurity Kondo problem has been solved for general $N$, $k$ and $\r_{\textrm{UV}}$ using a number of complementary techniques, including numerical RG~\cite{Wilson:1974mb,PhysRevB.21.1003,PhysRevB.21.1044}, integrability~\cite{PhysRevLett.45.379,Wiegmann:1980,RevModPhys.55.331,doi:10.1080/00018738300101581,0022-3719-19-17-017,1994cond.mat..8101A,ZinnJustin1998, PhysRevB.58.3814}, large-$N$ techniques~\cite{PhysRevB.35.5072,RevModPhys.59.845,1997PhRvL..79.4665P,1998PhRvB..58.3794P,2006cond.mat.12006C,2015arXiv150905769C}, CFT techniques~\cite{Affleck:1990zd,Affleck:1990by,Affleck:1990iv,Affleck:1991tk,PhysRevB.48.7297,Affleck:1995ge}, and more. For reviews of many of these, see for example refs.~\cite{doi:10.1080/000187398243500,Hewson:1993}.

Without going into details, we can reach a basic understanding of the solution to the Kondo problem as follows. Let us assume that, starting from the LFL in the UV, the RG flow takes us all the way to $\hat{\l}_{\K}\rightarrow+\infty$ in the IR. In that case, the ground state must minimize the Kondo interaction $S^A\hat{\mathcal{J}}^A$. Concretely, among all the eigenstates of $S^A+\hat{\mathcal{J}}^A$ that can be formed by the impurity $\r_{\textrm{UV}}$ and the LFL fermions, subject to Pauli exclusion and $SU(k)$ channel symmetry, the ground state will have the minimal eigenvalue of $S^A\hat{\mathcal{J}}^A$~\cite{1997PhRvL..79.4665P, 1998PhRvB..58.3794P}. Let $\r_{\textrm{IR}}$ denote the corresponding $SU(N)$ representation of any impurity remaining in the IR, with dimension $\mathrm{dim}(\r_{\textrm{IR}})$. The $\hat{\l}_{\K}\rightarrow\infty$ fixed point must fall into one of the following three classes, depending on how $\mathrm{dim}(\r_{\textrm{IR}})$ compares to $\mathrm{dim}(\r_{\textrm{UV}})$:
\begin{enumerate}
	\item \emph{Critical Screening}: If $\r_{\textrm{IR}}$ is a singlet of $SU(N)$, $\mathrm{dim}(\r_{\textrm{IR}})=0$, then the impurity has been screened completely. This occurs for instance in the original single-channel $SU(2)$ Kondo model: the ground state is the Kondo singlet, that is, the Kondo cloud has net spin $1/2$, which locks with the impurity spin into the anti-symmetric singlet of $SU(2)$.
	\item \emph{Underscreening}: If $0<\mathrm{dim}(\r_{\textrm{IR}})\leq\mathrm{dim}(\r_{\textrm{UV}})$, then the impurity is either partially screened or unscreened, and whatever net impurity spin remaining in the IR interacts with the LFL via a marginally irrelevant FM Kondo coupling~\cite{1997PhRvL..79.4665P, PhysRevB.73.224445}.
	\item \emph{Overscreening}: If $\mathrm{dim}(\r_{\textrm{IR}})>\mathrm{dim}(\r_{\textrm{UV}})$, then the na\"ive strong coupling fixed point $\hat{\l}_{\K}\rightarrow\infty$ cannot be the actual IR fixed point, since that would lead to a greater number of impurity degrees of freedom (greater impurity entropy) in the IR than in the UV, which is impossible for a physical RG flow~\cite{Affleck:1991tk,1998PhRvB..58.3794P}. In fact, the overscreened impurity interacts with neighboring LFL fermions via a marginally relevant AFM Kondo coupling, rendering the na\"ive IR fixed point unstable. The true IR fixed point is not at $\hat{\lambda}_{\K} \to \infty$, but at a non-trivial, intermediate value of $\hat{\l}_{\K}$~\cite{PhysRevB.73.224445,Nozieres:1980}, and gives rise to non-Fermi liquid behavior.
\end{enumerate}
With critical or underscreening, the excitations about the ground state arrange themselves again into a LFL. However, the IR LFL is distinct from the UV LFL. In the IR, the LFL fermions are subject to special boundary conditions: their wave function must vanish at the location of the impurity. Intuitively, the reason is that, due to Pauli exclusion, a LFL fermion can penetrate that location only by destroying the screened impurity in representation $\r_{\textrm{IR}}$, whose binding energy is $\propto \hat{\l}_{\K}\rightarrow\infty$~\cite{Affleck:1995ge}. The vanishing of the wave function is equivalent to an $s$-wave $\pi/2$ phase shift in the IR relative to the UV.

Our holographic model is mainly based on the CFT and large-$N$ approaches to the Kondo problem. We shall therefore quickly review the features of these approaches that will be essential to our holographic model.

\subsubsection{CFT Techniques}

The single-impurity Kondo model is spherically symmetric about the impurity: if we perform a partial-wave decomposition of the $c_\a$, then only the $s$-wave couples to the impurity. The CFT approach~\cite{Affleck:1990zd,Affleck:1990by,Affleck:1990iv,Affleck:1991tk,PhysRevB.48.7297,Affleck:1995ge} begins by discarding all higher partial waves (in real space), followed by linearizing the dispersion relation about $k_F$ (in momentum space). The result is a $(1+1)$-dimensional model on the positive real axis, representing the radial distance to the impurity, with left- and right-moving fermions (in-coming and out-going $s$-waves) interacting with the impurity at the origin. Linearizing the dispersion relation about $k_F$ trivially leads to a relativistic model, with the Fermi velocity $v_F\equiv k_F/m$ playing the role of the speed of light. After extending the positive real axis to negative values, reflecting the right-movers about the origin, and re-labeling them as left-movers, we obtain the simplest description of the single-impurity Kondo model: left-movers alone, moving on the entire real line, interacting with the impurity at the origin. The resulting $(1+1)$-dimensional Kondo Hamiltonian density is (suppressing $SU(k)$ channel indices)
\begin{align}
	H_{\K}=\frac{v_F}{2\pi}\psi^\dagger_\a i\pa_x\psi_\a+v_F\l_{\K} \d(x)S^A\psi_\a^\dagger T^A_{\a\b}\psi_\b, \label{KondoCFT}
\end{align}
where $\psi_\a^\dagger$ creates a left-moving fermion with spin $\a$, $\l_{\K}\propto\hat{\l}_{\K}$ is the classically marginal $(1+1)$-dimensional Kondo coupling, and $T^A_{\a\b}$ are now the generators of $SU(N)$ ($A = 1, \ldots, N^2-1$), in the fundamental representation. We will henceforth choose units with $v_F \equiv 1$.

The free left-moving fermions form a chiral CFT, invariant under a single Virasoro algebra. Moreover, the $SU(N) \times SU(k) \times U(1)$ symmetry of the original Hamiltonian density, $\hat{H}_{\K}$ in eq.~\eqref{KondoH}, has now been enhanced to an $SU(N)_k \times SU(k)_N \times U(1)$ Kac-Moody symmetry. This infinite (accidental) symmetry is the main advantage of the CFT approach.

In the CFT approach, with AFM Kondo coupling, $\l_{\K}>0$, the UV fixed point (high $T$) is simply free left-moving fermions and a decoupled impurity. The Kac-Moody symmetry determines the spectrum of eigenstates completely~\cite{Affleck:1990zd,Affleck:1990by,Affleck:1990iv,Affleck:1995ge}. The Kondo problem then reduces to determining the IR fixed point CFT (low $T$). The CFT solution of the Kondo problem is based on two proposals. The first is that the IR CFT must have the same Kac-Moody symmetry as the UV fixed point, which will thus determine the spectrum of eigenstates in the IR completely~\cite{Affleck:1990zd}. The second is that the IR eigenstates are obtained from those in the UV by ``fusion'' with the impurity representation, $\r_{\textrm{UV}}$~\cite{Affleck:1990by}. The CFT results for the spectrum agree with other methods, including in particular integrability~\cite{Affleck:1990zd,Affleck:1990by,Affleck:1990iv,Affleck:1995ge}. However, the CFT approach also provides novel information. For example, the spectrum of irrelevant deformations about the IR fixed point determines the low-$T$ scaling exponents of the entropy, magnetic susceptibility, and electrical resistivity~\cite{Affleck:1990zd,Affleck:1990by,Affleck:1990iv,PhysRevB.48.7297,Affleck:1995ge}.

For our holographic Kondo model, the key ingredient we need from the CFT approach is the existence of a $(1+1)$-dimensional chiral CFT description of the Kondo Hamiltonian, invariant under an $SU(N)_k \times SU(k)_N \times U(1)$ Kac-Moody symmetry.

\subsubsection{Large-$N$ Techniques}

Our holographic Kondo model will employ a large-$N$ limit~\cite{PhysRevB.35.5072,RevModPhys.59.845,1997PhRvL..79.4665P,1998PhRvB..58.3794P,2006cond.mat.12006C,2015arXiv150905769C}, and in particular will employ the large-$N$ description of the Kondo effect as symmetry breaking at the impurity's location~\cite{0022-3719-19-17-017,PhysRevB.35.5072,2003PhRvL..90u6403S,2004PhRvB..69c5111S}. That description begins by representing the impurity spin $S^A$ in terms of Abrikosov pseudo-fermions, in the fundamental representation of $SU(N)$:
\begin{align}
	S^A=\chi_\a^\dagger T^A_{\a\b}\chi_\b,
\end{align}
where $\chi_\a^\dagger$ and $\chi_\a$ are creation and annihilation operators for an Abrikosov pseudo-fermion. The $\chi_\a$ obey fermionic anti-commutation relations, which ensures that $S^A$ indeed obeys the $SU(N)$ algebra. The Hilbert space on which $S^A$ acts is built by acting on the vacuum with the $\chi_\a^\dagger$. Because the $\chi^{\dagger}_\a$ anti-commute, the states in the Hilbert space form totally anti-symmetric tensor products of the fundamental representation of $SU(N)$, with the rank of a tensor given by the total number $q$ of Abrikosov pseudo-fermions in a particular state (so $q$ ranges from zero to $N$). To obtain an irreducible representation, we must fix the rank $q$ of the anti-symmetric tensor, by imposing a constraint on the states in the Hilbert space:
\begin{align}
	\chi_\a^\dagger\chi_\a=q.
	\label{qconstraint}
\end{align}
This constraint can also be understood from a different perspective: writing an $SU(N)$ spin as $S^A=\chi^\dagger_\a T^A_{\a\b}\chi_\b$ introduces an additional $U(1)$ symmetry, which acts by shifting the phase of $\chi_\a$, but leaves $S^A$ invariant, and hence is redundant or ``auxiliary.'' The auxiliary $U(1)$ artificially enlarges the Hilbert space, so to project onto the subspace of physical states we must ``gauge-fix" the charge of the auxiliary $U(1)$, leading to the constraint in eq.~\eqref{qconstraint}.

Totally symmetric representations of $SU(N)$ can be obtained by representing $S^A$ via Schwinger bosons, rather than Abrikosov pseudo-fermions~\cite{1997PhRvL..79.4665P, 2008arXiv0809.4836A, 2010JMP....51i3504M}. General representations of $SU(N)$ can be realized in several different ways: by mixing Abrikosov pseudo-fermions and Schwinger bosons~\cite{PhysRevB.73.224445}, by replacing the fundamental representation generators $T^A_{\a\b}$ by generators of another representation~\cite{ZinnJustin1998, PhysRevB.58.3814}, or by introducing multiple flavors of Abrikosov pseudo-fermion, subject to a more complicated constraint~\cite{Gomis:2006sb}. However, in what follows, we will exclusively consider impurity spins in totally anti-symmetric, rank $q$ tensor representations of $SU(N)$, and we will always represent $S^A$ using Abrikosov pseudo-fermions.

Representing $S^A$ in terms of Abrikosov pseudo-fermions allows a convenient re-writing of the Kondo interaction, as follows. Using the completeness relation satisfied by the fundamental-representation $SU(N)$ generators,
\begin{align}
	T^A_{\a\b}T^A_{\g\d}=\frac{1}{2}\le(\d_{\a\d}\d_{\b\g}-\frac{1}{N}\d_{\a\b}\d_{\g\d}\ri), \label{completeness}
\end{align}
and $\chi_\a$'s anti-commutation relations, we can re-write the Kondo interaction in eq.~\eqref{KondoCFT} as, after dropping an unimportant constant $\propto q$,
\begin{align}
	\l_{\K}S^A\psi^\dagger_\g T^A_{\g\d}\psi_\d=\l_{\K}\le(\chi^\dagger_\a T^A_{\a\b}\chi_\b\ri)\le(\psi^\dagger_\g T^A_{\g\d}\psi_\d\ri)=-\frac{1}{2}\l_{\K}\le(\mathcal{O}^\dagger\mathcal{O}+\frac{q}{N}\le(\psi^\dagger_\a\psi_\a\ri)\ri), \label{KondoAbrikosov}
\end{align}
where we have defined the scalar operator $\mathcal{O} \equiv \psi^{\dagger}_\a \chi_\a$, which is a function of time $t$ only, because $\chi_{\alpha}$ cannot propagate away from the impurity's location, $x=0$. In other words, $\mathcal{O}$ is a $(0+1)$-dimensional operator. Clearly, $\mathcal{O}$ is a singlet of the spin $SU(N)_k$ symmetry, has the same channel $SU(k)_N$ and electromagnetic $U(1)$ representation as $\psi^{\dagger}_\a$, and has the same auxiliary $U(1)$ charge as $\chi_\a$. Classically $\psi_\a$ has dimension $1/2$ and $\chi_\a$ has dimension zero, so $\mathcal{O}$ has dimension $1/2$. The Kondo interaction eq.~\eqref{KondoAbrikosov} is then classically marginal, \textit{i.e.}\ $\l_{\K}$ is classically dimensionless.

So far our discussion has actually been valid for any value of $N$, but let us now consider the large-$N$ limit: we take $N \to \infty$, keeping both $N \l_{\K}$ and $q/N$ fixed and of order one. In that case, in the Kondo coupling of eq.~\eqref{KondoAbrikosov} the $(q/N) \psi^{\dagger}_\a \psi_\a$ term is sub-leading in $N$ relative to the $\mathcal{O}^\dagger\mathcal{O}$ term. We thus find that the Kondo interaction, when written in terms of Abrikosov pseudo-fermions and in the large-$N$ limit, is a classically-marginal ``double-trace'' interaction, of the form $- \l_{\K} \mathcal{O}^\dagger\mathcal{O}$. We put ``double-trace'' in quotation marks because $\mathcal{O}$ is not the trace of a matrix in the adjoint of $SU(N)$, but a contraction of a field in the anti-fundamental representation of $SU(N)$, $\psi^\dagger_\a$, with a field in the fundamental representation, $\chi_\a$. In what follows we will drop the quotation marks. The double-trace form of the Kondo interaction will be extremely useful for our holographic model: a double-trace interaction will be realized holographically by a simple linear boundary condition on the complex scalar field dual to $\mathcal{O}$, as we will discuss in section~\ref{holorg}.

 The solution of the large-$N$ saddle point equations reveals a second-order mean-field phase transition at a critical temperature, on the order of but distinct from $T_{\K}$, below which $\mathcal{O}$ acquires a non-zero expectation value, $\langle \mathcal{O} \rangle \neq 0$~\cite{0022-3719-19-17-017,PhysRevB.35.5072,2003PhRvL..90u6403S,2004PhRvB..69c5111S}. The condensation of $\mathcal{O}$ spontaneously breaks the channel symmetry down to $SU(k-1)_N$, and breaks the $U(1) \times U(1)$ electromagnetic and auxiliary symmetry down to the diagonal. Intuitively, the condensation of $\mathcal{O}$ represents the formation of a Kondo cloud around $x=0$, screening the impurity spin.

Of course, spontaneous symmetry breaking in $(0+1)$ dimensions is impossible for finite $N$: the phase transition is an artifact of the large-$N$ limit. Corrections in $1/N$ will change the phase transition to a smooth cross-over~\cite{0022-3719-19-17-017}, as observed in experimental realizations of the single-impurity Kondo effect~\cite{Goldhaber1998,Cronenwett24071998,vanderWiel22092000,PTP.32.37,0034-4885-37-2-001,GrŸner1978591}. In this sense the large-$N$ limit of the single-impurity Kondo model is singular. Moreover, in the large-$N$ limit, for $T$ above the critical temperature, where $\langle \mathcal{O}\rangle=0$, all physics reduces to that of the UV chiral CFT, free left-moving fermions. In particular, the characteristic $-\ln(T/T_{\K})$ contribution to the resistivity at high $T$ is absent at large $N$. Nevertheless, the large-$N$ limit captures much of the essential single-impurity Kondo physics at  low $T$, including low-$T$ scaling exponents and the phase shift~\cite{PhysRevB.35.5072,RevModPhys.59.845,1997PhRvL..79.4665P,1998PhRvB..58.3794P,2006cond.mat.12006C,2015arXiv150905769C}.

To summarize our review of the single-impurity Kondo model: at low $T$ and large $N$, the single-impurity Kondo effect can be described as a $(1+1)$-dimensional chiral CFT, free left-moving fermions, deformed by a marginally-relevant, double-trace coupling to an impurity spin, leading to spontaneous symmetry breaking at the impurity's location.

\subsection{The Two-Impurity Kondo Model}
\label{tikm}

The two-impurity Kondo model~\cite{PhysRevLett.47.737,PhysRevLett.58.843,PhysRevB.35.5072,PhysRevB.35.4901,PhysRevLett.61.125,PhysRevB.39.3415,PhysRevB.40.324,PhysRevB.40.4780,Jones1990,Millis1990,Affleck:1991yq,PhysRevLett.72.916,1995PhRvB..52.9528A,INGERSENT1994402,PhysRevLett.74.2583,PhysRevB.51.8287,PhysRevLett.74.2808,PhysRevLett.76.275,Jones:2007} is the simplest model that features the competition between the Kondo and RKKY interactions, and is a natural first step towards building a Kondo lattice. The two-impurity Kondo model consists of two localized impurity spins $S^A_{\one}$ and $S^A_{\two}$, both in the same representation $\r_{\textrm{UV}}$ of $SU(N)$, separated by a distance $\ell$, and interacting with a LFL via two AFM Kondo couplings of equal strengths.

The most general formulation of the two-impurity Kondo model also includes a Heisenberg interaction between the impurity spins, of the form $S_{\one}^A S_{\two}^A$. However, even if such an interaction is absent in the UV, the Kondo interactions produce Friedel oscillations in the LFL that induce a Heisenberg interaction between $S^A_{\one}$ and $S^A_{\two}$~\cite{PhysRevLett.47.737,2006cond.mat.12006C}. In the high-$T$ regime, where perturbation theory in $\l_{\K}$ is reliable, the leading induced Heisenberg interaction is order $\lambda_{\K}^2$~\cite{PhysRevLett.47.737,2006cond.mat.12006C}. Strictly speaking, the term ``RKKY interaction'' refers only to that second-order induced Heisenberg interaction. Due to the Friedel oscillations, that RKKY coupling depends on $\ell$ as a sinusoid in $2k_F \ell$ decaying inside a power-law envelope, which in three spatial dimensions is $(k_F \ell)^{-3}$~\cite{PhysRevLett.47.737,2006cond.mat.12006C}. The sign of the RKKY interaction, that is, whether the RKKY interaction is FM or AFM, thus depends on $\ell$. Heuristically, $k_F\ell$ dictates how many layers of screening fermions lie between the two impurity spins. If the two spins are very close, then both spins are effectively screened by the same fermions, which therefore mediate a FM RKKY interaction.

In some cases the RKKY coupling vanishes, and hence the only contribution to the Heisenberg coupling constant is what we add ``by hand.'' A vanishing RKKY coupling obviously occurs for any non-zero $\ell$ where the sinusoid in $2k_F \ell$ vanishes. Crucially for our holographic model, in the large-$N$ limit the RKKY interaction is sub-leading in $N$ relative to the Kondo coupling~\cite{PhysRevB.39.3415}, so if we work only to leading order in the large-$N$ limit, then the RKKY coupling effectively vanishes. We will thus treat the Heisenberg coupling constant as a free parameter. However, in a (standard) abuse of terminology, we will always refer to the Heisenberg coupling constant as the ``RKKY coupling,'' even though it is not necessarily induced by the Kondo interactions.

The RKKY coupling constant (induced or otherwise), $\l_{\RKKY}$, has classical dimension $1$, and so is classically relevant. The two-impurity Kondo model thus has two intrinsic scales, $T_{\K}$ and $\l_{\RKKY}$, although only the latter is explicit in the Hamiltonian density.

Despite its apparent simplicity, the two-impurity Kondo problem has been solved only for certain values of $N$, $k$ and $\r_{\textrm{UV}}$, using many of the techniques developed for the single-impurity Kondo problem.

For example, at low energy, or equivalently large distances, where $\ell$ is negligible, a CFT description of the two-impurity Kondo model becomes reliable~\cite{Affleck:1991yq,1995PhRvB..52.9528A}. The CFT approach to the two-impurity Kondo model begins with spatial averages over the momentum directions of the LFL fermion wave function, leading to a $(1+1)$-dimensional description, analogous to the $s$-wave reduction in the single-impurity Kondo model~\cite{Affleck:1991yq,1995PhRvB..52.9528A}. However, now two modes per channel participate in the interactions, namely modes with even and odd parity about the mid-plane between the two impurities. In other words, the reduction to $(1+1)$ dimensions leads to an effective doubling of the number of channels, from $k$ to $K\equiv2k$. Additionally, the even and odd modes have momentum-dependent Kondo couplings of different strengths. However, at low energies, where $\ell$ is negligible, these Kondo couplings can be evaluated at $k_F$, in which case the differences between them are irrelevant in the RG sense~\cite{Affleck:1991yq,1995PhRvB..52.9528A}. Ultimately, the CFT description thus involves $K$ channels of $(1+1)$-dimensional left-moving fermions interacting with identical Kondo couplings to two identical impurity spins at the origin.

The original two-impurity Kondo model has $N=2$, $k=1$, and $\r_{\textrm{UV}}$ the fundamental representation of $SU(2)$, and has been studied using a combination of numerical RG techniques~\cite{PhysRevLett.58.843, PhysRevLett.61.125,PhysRevB.40.324,Jones1990,PhysRevLett.76.275,Jones:2007} and CFT techniques~\cite{PhysRevB.40.324,Affleck:1991yq,1995PhRvB..52.9528A,PhysRevLett.74.2583,PhysRevB.51.8287,PhysRevLett.74.2808}. The results conform to intuition. In the FM RKKY limit, $\l_{\RKKY}/T_{\K} \to -\infty$, the two impurities lock into the triplet of $SU(2)$, in order to minimize the RKKY interaction. Upon lowering $T$, this effective spin-$1$ impurity is completely screened in a two-stage Kondo effect. In the first step, the fermions from the more strongly coupled odd channel screen half of the spin-$1$ impurity, and in the second step the remaining spin-$1/2$ impurity is screened by the even channel. The IR fixed point is a LFL with a $\pi/2$ phase shift. In the AFM RKKY limit, $\l_{\RKKY}/T_{\K} \to + \infty$, the two impurities lock into the anti-symmetric singlet of $SU(2)$, and effectively disappear from the spectrum. Consequently, no impurity remains that could be screened by the LFL, so the IR fixed point is a LFL with no phase shift.

In fact, at $T=0$ particle-hole symmetry allows only two values of the phase shift, $\pi/2$ and zero, so the FM and AFM RKKY limits must be separated by a quantum phase transition where the phase shift jumps discontinuously from one value to the other~\cite{Millis1990,1995PhRvB..52.9528A}. Numerical RG and CFT techniques show that the transition occurs at a non-zero AFM value $\l_{\RKKY}/T_{\K}\approx 2.2$ and is second order, and hence gives rise to a quantum critical point~\cite{PhysRevLett.61.125,PhysRevB.40.324,Jones1990,PhysRevLett.76.275,Jones:2007}. However, no change of symmetry occurs at the critical point: the ground state on both sides of the transition is a singlet of $SU(2)$.

Surprisingly, numerical RG techniques reveal that the spin-spin correlator, $\langle S^A_{\one} S^A_{\two}\rangle$, monotonically decreases as $\l_{\RKKY}/T_{\K}$ increases, and is continuous, even through the phase transition~\cite{PhysRevLett.61.125,PhysRevB.40.324,Jones1990,Jones:2007}. Indeed, $\langle S^A_{\one} S^A_{\two}\rangle$ decreases smoothly and monotonically as $\l_{\RKKY}/T_{\K}$ increases from the FM limit, $\l_{\RKKY}/T_{\K} \to -\infty$, where $\langle S^A_{\one} S^A_{\two}\rangle=1/4$, the triplet value, to the AFM RKKY limit $\l_{\RKKY}/T_{\K} \to + \infty$, where $\langle S^A_{\one} S^A_{\two}\rangle=-3/4$, the singlet value.

Although we lack a complete solution of the two-impurity Kondo problem for general $N$, $k$ and $\r_\mathrm{UV}$, the results for the original two-impurity Kondo problem suggests the following intuition for the general case. In the limit of infinitely strong FM RKKY coupling, $\l_{\RKKY}/T_{\K}\rightarrow-\infty$, or AFM RKKY coupling, $\l_{\RKKY}/T_{\K}\rightarrow+\infty$, the ground state of the two-impurity system should be an eigenstate of $S_{\one}^AS_{\two}^A$ with maximum or minimum eigenvalue, respectively. We shall denote the corresponding $SU(N)$ representations by $\r_{\textrm{FM}}$ and $\r_{\textrm{AFM}}$, respectively. For general values of $\l_{\RKKY}/T_{\K}$, the ground state will be a superposition of the eigenstates of $S_{\one}^AS_{\two}^A$ that appear in the tensor product $\r_{\textrm{UV}} \otimes \r_{\textrm{UV}}$. In the AFM or FM RKKY limits, $\l_{\RKKY}/T_{\K} \to \pm \infty$, the system effectively reduces to a $K$-channel $SU(N)$ Kondo model with a single impurity in a representation $\r_{\textrm{AFM}}$ or $\r_{\textrm{FM}}$, respectively. In the AFM case, in some special cases $\r_{\textrm{UV}}$ is such that the two impurity spins can lock into a singlet. In those cases, no Kondo screening will occur, and the IR fixed point will be a LFL with no phase shift. On the other hand, in the more general case that $\r_{\textrm{AFM}}$ is non-trivial, then the residual impurity spin will be Kondo screened to the extent possible by the $K$ channels. The coexistence of inter-impurity and Kondo screening is thus generic in the AFM limit. The AFM IR fixed point will then be either a non-LFL (overscreening) or a phase-shifted LFL (under- or critical- screening), depending on the values of $N$, $K$ and $\r_\mathrm{AFM}$. By contrast, in the FM case, $\r_{\textrm{FM}}$ is always non-trivial. In that case, Kondo screening will occur, and again, the IR fixed point will be either a non-LFL or a phase-shifted LFL. In the special case that the IR fixed point in the AFM limit does not have a phase shift while that in the FM limit does, then the two must be separated by a quantum phase transition~\cite{Millis1990}, while in the more general case that both limits have phase shifts, the evolution from one limit to the other may or may not be continuous. 

Large-$N$ results for the two-impurity Kondo problem, in the case of $K=2$ channels, appear in refs.~\cite{PhysRevB.35.5072,PhysRevB.39.3415,Millis1990}. As mentioned above, at leading order in large $N$ the genuine RKKY interaction, induced by the Kondo interaction, is absent~\cite{PhysRevB.39.3415}, so to obtain a Heisenberg interaction between the impurity spins, we must add an RKKY coupling by hand, scaled appropriately with $N$ to contribute at the same order as the Kondo coupling. The authors of refs.~\cite{PhysRevB.39.3415,Millis1990} carefully chose a totally anti-symmetric $\r_{\textrm{UV}}$ whose Young tableau had exactly $q=N/2$ boxes, to ensure that the two spins can lock into a singlet. Indeed, their large-$N$ saddle-point solution reveals a first-order quantum phase transition between an AFM phase with no phase shift and a FM phase with a $\pi/2$ phase shift, indicating Kondo screening.

Our holographic model will also contain two totally anti-symmetric $SU(N)$ spin impurities. However, our holographic model will be too crude to allow us to identify the exact number of boxes $q$ in the corresponding Young tableau. We will only know that $q \propto N$. As a result, in our holographic model, in the AFM limit, typically the ground state will not be a singlet of $SU(N)$, and hence Kondo screening and a phase shift will occur. Indeed, as we have seen, the coexistence of Kondo and inter-impurity screening is in fact generic in two-impurity Kondo models when the two impurity spins do not lock into a spin singlet. The coexistence of Kondo and inter-impurity screening is also widely believed to occur in the Kondo lattice~\cite{2009arXiv0912.0040S,2010uqpt.book..193S,Jones:2007}.

As in the single-impurity case, with totally anti-symmetric impurity spins the Abrikosov pseudo-fermion representation allows us to write the Kondo couplings of the two impurity spins as double-trace couplings with respect to $SU(N)$. The pseudo-fermions also allow us to write the RKKY coupling as a double-trace coupling of $SU(N)$, as follows. We introduce two species of pseudo-fermion, one for each spin:
\begin{align}
	S^A_i=\chi_{i\a}^\dagger T^A_{\a\b}\chi_{i\b}, \qquad i = \one, \two.
\end{align}
We can then define $\mathcal{O}_{\one} \equiv \psi^{\dagger}_\a \chi_{\one \a}$, which in the large-$N$ limit produces double-trace Kondo couplings of the form $-\l_{\K}^{\one} \mathcal{O}_{\one}^{\dagger} \mathcal{O}_{\one}$, and similarly for $\mathcal{O}_{\two}$ and $\l_{\K}^{\two}$. In our holographic model, we will always take $\l_{\K}^{\one} = \l_{\K}^{\two} = \l_{\K}$, following the CFT approach to the two-impurity Kondo model, in which the difference $\l_{\K}^{\one} - \l_{\K}^{\two}$ is irrelevant in the RG sense~\cite{Affleck:1991yq,1995PhRvB..52.9528A}.

Generically, when we introduce pseudo-fermions, we introduce an auxiliary $U(1)$ at each impurity site: each $U(1)$ acts by shifting the phase of the pseudo-fermions at that site. However, if the impurities are coincident, and the RKKY coupling vanishes, then the auxiliary $U(1) \times U(1)$ is enhanced to $U(2)$, under which $\chi_{\one\a}$ and $\chi_{\two\a}$ combine into a doublet. The two scalars $\mathcal{O}_{\one}$ and $\mathcal{O}_{\two}$ thus also combine into a doublet of that $U(2)$, which we denote $\mathcal{O} \equiv (\mathcal{O}_{\one},\mathcal{O}_{\two})^{\textrm{T}}$. We use $U(2)$ generators
\begin{align}
\tau^b=\frac{1}{2}(\mathbb{1},\s^1,\s^2,\s^3), \qquad b = 0, \ldots, 3,
\end{align}
with $\s^1$, $\s^2$, and $\s^3$ the Pauli matrices. The components of the auxiliary $U(2)$ Noether charges are then
\begin{align}
	R^b\equiv \chi_{i\a}^\dagger\tau_{ij}^b\chi_{j\a},
	\label{AbrikosovCharge}
\end{align}
which obey the $(0+1)$-dimensional conservation equation, $\partial_t R^b = 0$. The constraint on the auxiliary charge in the single-impurity case, eq.~\eqref{qconstraint}, is generalized in the two-impurity case to constraints on the elements of $R^b$ in the Cartan of the auxiliary $U(2)$: if $S_{\one}^A$ and $S_{\two}^A$ have $q_{\one}$ and $q_{\two}$ boxes in their Young tableaux, respectively, then we must impose
\begin{align}
	R^0=\frac{1}{2}(q_{\one}+q_{\two}), \qquad R^3=\frac{1}{2}(q_{\one}-q_{\two}).
	\label{cartanconstraint}
\end{align}
Using the completeness relation in eq.~\eqref{completeness} and $\chi_{i\a}$'s anti-commutation relations, the RKKY interaction can be recast as a double-trace interaction with respect to $SU(N)$,
\begin{align}
	\l_{\RKKY} S^A_{\one} S^A_{\two} = -\frac{1}{2}\l_{\RKKY}\le((R^1)^2+(R^2)^2-\frac{1}{2}(q_{\one}+q_{\two})+\frac{q_{\one}q_{\two}}{N}\ri).
	\label{RKKYdt}
\end{align}
Upon dropping the insignificant constants $(q_{\one}+q_{\two})/2$ and $q_{\one}q_{\two}/N$, we thus find that the RKKY interaction, when written in terms of pseudo-fermions, is a classically-relevant double-trace interaction, of the form $-\l_{\RKKY}((R^1)^2+(R^2)^2)$. Clearly the RKKY interaction explicitly breaks the auxiliary $U(2)$ symmetry down to the subgroup that commutes with $(R^1)^2+(R^2)^2$, namely, down to the Cartan of $U(2)$. Furthermore, if the two impurities are identical, then $q_{\one} = q_{\two} = q$, so from eq.~\eqref{cartanconstraint} we have $R^0 = q$ and $R^3 = 0$.  The double-trace form of the RKKY interaction will be extremely useful for our holographic model: a double-trace interaction will be realized holographically by a boundary condition on the $U(2)$ YM gauge field dual to $R^b$, as we will discuss in section~\ref{holorg}.

Let us summarize our review of the two-impurity Kondo model. At large distances compared to $\ell$, and at large $N$, the two-impurity Kondo model reduces to a $(1+1)$-dimensional chiral CFT consisting of $K=2k$ channels of left-moving fermions, with one marginally-relevant, double-trace Kondo coupling for each impurity spin, and a relevant, double-trace RKKY coupling between the impurity spins, which breaks the auxiliary $U(2)$ symmetry down to the $U(1) \times U(1)$ Cartan subgroup. As in the single-impurity case, we expect the Kondo effect to appear as condensation of $\mathcal{O}_{\one}$ and $\mathcal{O}_{\two}$ below some critical temperature. Via large-$N$ factorization, $\langle S^A_{\one}S^A_{\two} \rangle \propto - \langle (R^1)^2+(R^2)^2\rangle \propto - \langle R^1 \rangle^2-\langle R^2\rangle^2$, so we expect non-zero spin-spin correlations, $\langle S^A_{\one}S^A_{\two}\rangle \neq 0$, to appear as condensation of $R^1$ and/or $R^2$. Both effects will indeed appear in our holographic model.

\section{A Holographic Two-Impurity Kondo Model}
\label{model}

In this section we present the field content and classical action of our holographic two-impurity Kondo model, and then derive the classical equations of motion, which we will study in the subsequent sections.

We begin with the CFT description of the two-impurity Kondo model reviewed in the previous section, with $K$ channels of $(1+1)$-dimensional left-moving fermions and two coincident impurity spins expressed in terms of Abrikosov pseudo-fermions. The chiral fermions form a chiral CFT with $SU(N)_K \times SU(K)_N \times U(1)$ Kac-Moody algebra. To reach a holographic description, our first step is to introduce additional degrees of freedom in the adjoint representation of $SU(N)_K$, including in particular $SU(N)_K$ gauge fields. This introduces an additional coupling, besides the Kondo and RKKY couplings, namely the 't Hooft coupling. We then take the 't Hooft large-$N$ limit, and the additional limit of large 't Hooft coupling. We choose the adjoint degrees of freedom such that, in these limits, we obtain a CFT holographically dual to Einstein-Hilbert gravity in $AdS_3$. For a specific example of such a construction, see ref.~\cite{Erdmenger:2013dpa}.

Of course, we cannot blithely gauge the $SU(N)_K$ symmetry, because the $SU(N)_K$ currents have chiral anomalies, due to the left-moving fermions. If we obtained our Kondo model from a string theory construction, then the net gauge anomalies would vanish. However, following ref.~\cite{Erdmenger:2013dpa}, instead of such a ``top-down'' model, we will work with a ``bottom-up'' model, built from the minimal ingredients that must be present in any holographic two-impurity Kondo model, but with enough structure to describe the essential phenomena. In that case, to suppress the gauge anomalies, we take the probe limit: when $N \to \infty$, we keep $K$ fixed, so that $K \ll N$, and compute all expectation values only to order $N$. At leading order in that limit, the anomalies do not appear~\cite{Buchbinder:2007ar,Erdmenger:2013dpa}, and effectively $SU(N)_K \to SU(N)$.

Each $SU(N)$-invariant, single-trace, low-dimension (\textit{i.e.}\ dimension of order $N^0$) operator is holographically dual to a field in the gravity description. The stress-energy tensor of the $(1+1)$-dimensional CFT is dual to the metric in $AdS_3$. The $SU(N)$ currents are not gauge-invariant, and hence do not appear explicitly in the gravity description. The $SU(K)_N \times U(1)$ Kac-Moody currents are dual to a $SU(K)_N \times U(1)$ Chern-Simons gauge field~\cite{Kraus:2006wn}, which in form notation we call $A$. The auxiliary $U(2)$ charges $R^b$ at the impurities' location, $x=0$, are dual to a $U(2)$ YM gauge field, which in form notation we call $a = a^b \tau^b$, localized to $x=0$, that is, localized to an $AdS_2$ subspace of $AdS_3$. The complex scalar $\mathcal{O}$ at the impurities' location is bi-fundamental under $SU(K)_N \times U(1)$ and the auxiliary $U(2)$, and is dual to a complex scalar field, $\Phi$, localized to the $AdS_2$ subspace and bi-fundamental under the $SU(K)_N \times U(1)$ Chern-Simons and $U(2)$ YM gauge fields.

An $\mathcal{O}$ of different spin or charge could be an essential ingredient for holographic duals of other quantum impurity models. For example, suppose our impurity spins were in a totally symmetric representation of $SU(N)$, and that we represented them using Schwinger bosons~\cite{1997PhRvL..79.4665P, 2008arXiv0809.4836A, 2010JMP....51i3504M}. That case would involve a fermionic, rather than bosonic, $\mathcal{O}$. An $\mathcal{O}$ with different charges could be essential for a holographic dual of an Anderson impurity model, which describes the formation of a localized magnetic moment, and which gives rise, at low energies, to the single-impurity Kondo model~\cite{2006cond.mat.12006C}. In the Anderson model, the impurity is a bi-linear product of two physical, rather than auxiliary, $f$ electrons, which are charged under the $U(1)$ of electromagnetism, in contrast to the pseudo-fermions, which are neutral under that $U(1)$. A holographic dual of the Anderson model would thus require a complex scalar similar to our $\mathcal{O}$, but built from a LFL fermion and the $f$ electron, and hence neutral under the $U(1)$ of electromagnetism. The dual complex scalar field $\Phi$ would then be neutral under the $U(1)$ factor of the $SU(K)_N \times U(1)$ Chern-Simons gauge group. Our choices of spin and charge for $\mathcal{O}$ indicate unambiguously that our holographic model is dual to a Kondo model with impurity spins in totally anti-symmetric representations of $SU(N)$.

As mentioned above, we will work in a probe limit. In our holographic model, that means the Einstein-Hilbert action will scale as $N^2$, but the matter action will scale as $N$. The matter fields' contribution to the Einstein equation is then suppressed by a factor of $N$, and so can be neglected in the large-$N$ limit. As a result, we only need to solve the matter fields' equations of motion in a fixed background metric, obtained by solving the vacuum Einstein equation with negative cosmological constant. To describe a $(1+1)$-dimensional CFT on the real line with non-zero $T$, we must use the $AdS_3$-Schwarzschild, or BTZ, black brane metric,
\begin{align}
\label{ads3metric}
	ds^2_{\textrm{BTZ}}= \frac{1}{z^2}\le(\frac{1}{h(z)}\dif z^2 -h(z)\dif t^2+\dif x^2\ri), \qquad h(z) = 1-\frac{z^2}{z_H^2},
\end{align}
where $z$ is the holographic radial coordinate, with the boundary at $z=0$ and the horizon at $z=z_H$, while $t$ and $x$ are the CFT time and space directions. We have chosen units in which the $AdS_3$ radius of curvature is unity. The Hawking temperature of the black brane, and hence the temperature of the dual CFT, is $T=1/(2\pi z_H)$. The impurity is located at $x=0$, which at $T=0$ is an $AdS_2$ subspace of $AdS_3$. More generally, for any $T$ the induced metric of the $x=0$ subspace is
\begin{align}
	g_{mn}\dif x^m x^n=\frac{1}{z^2}\le(\frac{1}{h(z)}\dif z^2 -h(z)\dif t^2\ri),\label{metricAdS2}
\end{align}
where $m,n=z,t$. The determinant of this induced metric is $g \equiv \textrm{det}\left(g_{mn}\right) = -1/z^4$.

For simplicity, we henceforth take $K=1$, unless stated otherwise. In that case, $SU(K)_N \times U(1)$ reduces to $U(1)$, so our Chern-Simons gauge field $A$ is Abelian, with field strength $F=dA$. We will discuss the generalization to $K>1$ later in this section.

For the classical action of our holographic two-impurity Kondo model, following ref.~\cite{Erdmenger:2013dpa} we choose the simplest two-derivative action quadratic in the fields. (Indeed, our action will be identical to that of ref.~\cite{Erdmenger:2013dpa}, but with a $U(2)$ YM gauge field in $AdS_2$, instead of a $U(1)$ gauge field.) The action of our model, $S$, splits into two terms, one for the Chern-Simons gauge field, $S_{\textrm{CS}}$, and one for the fields $a$ and $\Phi$ in the $AdS_2$ subspace, $S_{AdS_2}$,
\begin{subequations}
\label{action}
\beq
S=S_{\textrm{CS}}+S_{AdS_2},
\eeq
\beq
S_{\textrm{CS}} = -\frac{N}{4\pi}\int\limits_{AdS_3} A\wedge\dif A,
\eeq
\beq	
S_{AdS_2} = -N\int\limits_{AdS_2}\dif^2x\: \sqrt{-g}\le[\frac{1}{2}\mathrm{tr}\le(f^{mn}f_{mn}\ri) + \ \le(D^m\F\ri)^\dagger \le(D_m\F\ri) +M^2 \Phi^{\dagger} \Phi \ri], \label{actionAdS2}
\eeq
\end{subequations}
where $f_{mn}$ is the field strength of the $AdS_2$ YM field, while $D_m$ is the $U(2)$ gauge-covariant derivative, which acts on $f_{mn}$ and $\Phi$ as
\beq
D_m f^{np} = \nabla_m f^{np} - i [a_m,f^{np}], \qquad D_m\F =\le(\pa_m+iA_m-ia_m \ri) \F,
\eeq
and $M^2$ is $\Phi$'s mass-squared. The symmetries completely determine the form of the action at the two-derivative, quadratic level, except for the value of the scalar's mass-squared, $M^2$, which we will fix in section~\ref{holorg}. Although simple, we will see in section~\ref{phase} that the action in eq.~\eqref{action} is sufficient to capture the basic physics of the large-$N$ two-impurity Kondo model, and can thus serve as a foundation for further model-building, for example by adding terms higher-order in derivatives or in the fields.

If we define the $U(2)$ gauge current
\beq
J^b_m\equiv-i\le(\F^\dagger \tau^b(D_m\F)-(D_m\F)^\dagger \tau^b\F\ri), \label{jbig}
\eeq
then the equations of motion that follow from the action in eq.~\eqref{action} are, for $A$, $a$, and $\Phi$, respectively,
\begin{subequations}
\label{eoms}
\beq
\e^{n\m\n}F_{\m\n} = - 8 \pi \d\le(x\ri)\sqrt{-g} \, g^{nm}J^0_m, \qquad F_{zt}=0,
\label{CSeom}
\eeq
\beq
\le(D_mf^{mn}\ri)^b = - g^{nm} J^b_m,
\label{2YM}
\eeq
\beq
(D_mD^m -M^2)\F = 0,
\label{phieom}
\eeq
\end{subequations}
where $\mu,\nu=z,t,x$, and we choose $(z,t,x)$ to be a right-handed coordinate chart, $\e^{ztx}=1$.

We will work in radial gauge for both gauge fields, $A_z = 0$ and $a_z^b=0$, which we achieve via the gauge transformations
\beq
A_\m \longrightarrow \Gamma\le(A_\m+i\pa_\m\ri)\Gamma^{-1}, \qquad a_m\longrightarrow \gamma\le(a_m+i\pa_m\ri)\gamma^{-1}\nn,
\eeq
with gauge transformation parameters $\Gamma \in U(1)$ and $\g \in U(2)$ given by
\beq
\Gamma^{-1}=\exp\le(i\int^z\dif z' A_z\ri), \qquad \g^{-1}=\mathcal{P}\exp\le(i\int^z\dif z' a_z\ri),\nn
\eeq
where the lower endpoints of the integrations over the dummy variable $z'$ are arbitrary but fixed, and $\mathcal{P}$ denotes path-ordering. The residual gauge invariance then consists of $z$-independent gauge transformations, which in the following we will fix by imposing boundary conditions at $z=z_H$.

We are interested in time-independent solutions, in which case the equation of motion for the Chern-Simons gauge field $A$, eq.~\eqref{CSeom}, simplifies to 
\begin{subequations}
\beq
\pa_x A_t=4 \pi \d(x)\sqrt{-g} \, g^{zz}J^0_z,\label{CS1}
\eeq
\beq
\pa_zA_x=4 \pi \d(x)\sqrt{-g} \, g^{tt}J^0_t,\label{CS2}
\eeq
\beq
\pa_z A_t=0\label{CS3},
\eeq
\end{subequations}
while the equation of motion for the $U(2)$ YM field $a$, eq.~\eqref{2YM}, simplifies to $J_z^0=0$, plus a constraint (first order in derivatives)
\beq
\e^{bcd}g^{tt}a_t^c\pa_za^d_t=J^b_z, \qquad b,c,d = 1,2,3,
\label{YM2}
\eeq
and a dynamical equation (second order in derivatives)
\beq
\frac{1}{\sqrt{-g}}\pa_z\le(\sqrt{-g}g^{zz}g^{tt}\pa_za_t^b\ri)=-g^{tt}J^b_t.\label{YM3}
\eeq
Eqs.~\eqref{CS1} and \eqref{CS3} together with $J_z^0=0$ imply that $A_t$ is a constant. Regularity requires $A_t=0$ to vanish at $z=z_H$, hence $A_t=0$ everywhere. The only remaining non-trivial component of the Chern-Simons gauge field is then $A_x$, which is a function only of $z$, and which does not appear in the equations of motion for $a$ and $\Phi$. In particular, the equation of motion for $\Phi$, eq.~\eqref{phieom}, simplifies to
\beq
\frac{1}{\sqrt{-g}}\pa_z\le(\sqrt{-g}g^{zz}\pa_z\F\ri)-(M^2+g^{tt} a_t^b a_t^c \tau^b \tau^c) \F = 0.
\label{scalar}
\eeq
We can thus solve for $a$ and $\Phi$, and then plug those solutions into eq.~\eqref{CS2} to find $A_x$. However, we will not present explicit solutions for $A_x$ in the following. We will only need to know that non-trivial solutions for $A_x$ exist.

Let us now consider $K>1$, so that the Chern-Simons gauge field becomes non-Abelian, with gauge group $SU(K)_N \times U(1)$. In that case, the Chern-Simons gauge field $A$ does not decouple from the $AdS_2$ fields $a$ and $\Phi$ so easily. Specifically, $A$ decouples if and only if the solutions for $a$ and $\Phi$ preserve the full $SU(K)_N \times U(1)$ symmetry. Since $\Phi$ is in the fundamental of $SU(K)_N \times U(1)$, that requires all the components of $\Phi$ to be proportional to one another. Recalling that $\Phi$ is dual to $\mathcal{O}\equiv (\mathcal{O}_{\one},\mathcal{O}_{\two})^{\textrm{T}}$, and that the Kondo coupling to the first impurity spin is $\propto \mathcal{O}_{\one}^{\dagger} \mathcal{O}_{\one}$ and similarly for the second Kondo coupling, that implies that the two Kondo couplings respect the $SU(K)_N \times U(1)$ channel symmetry: all of the $K$ channels couple to each impurity with the same strength. In that case, a straightforward exercise shows that the equations of motion for time-independent $a$ and $\Phi$ are identical to those in eqs.~\eqref{YM3} and~\eqref{scalar}. In sum, as long as we preserve the channel symmetry, we can work with $K=1$ without loss of generality, as we have done.

On the other hand, if we break the channel symmetry, so that in the holographic model $A$ appears in the equations of motion of the $AdS_2$ fields, then we expect a multi-stage Kondo effect, as we reviewed for the original two-impurity Kondo model in subsection~\ref{tikm}. However, in the original model, the difference in Kondo couplings between channels is irrelevant in the IR~\cite{Affleck:1991yq,1995PhRvB..52.9528A}. Recalling that in holography $z$ plays the role of energy scale, with $z=0$ corresponding to the UV, in our holographic model we then expect $A$ to decouple from the $AdS_2$ fields \textit{dynamically}, deep in the bulk (at large $z$). We leave for future research the question of whether that actually occurs in our holographic model.

In the remainder of the paper, we focus on the equations of motion for static solutions of $a$ and $\Phi$, eqs.~\eqref{YM3} and~\eqref{scalar}. To solve these we need to determine the boundary conditions on $a$ and $\Phi$ at $z=0$, and the value of $M^2$. That requires holographic renormalization of our model, which we perform in the next section.

\section{Holographic Renormalization and Boundary Conditions}
\label{holorg}


In this section we perform the holographic renormalization~\cite{deHaro:2000xn,Bianchi:2001kw,Papadimitriou:2005ii,Papadimitriou:2010as} of the holographic two-impurity Kondo model introduced in section~\ref{model}. The essence of holographic renormalization is formulating a well-posed variational problem for the bulk fields, which among other things requires identifying the boundary conditions on the fields at the AdS boundary allowed by normalizability and regularity. Holographic renormalization will allow us to identify the Kondo and RKKY couplings in our model and to compute renormalized correlators, including the renormalized thermodynamic free energy, both of which we will use to study the phase diagram of our model in section~\ref{phase}.

In our case, holographic renormalization is non-trivial because our model includes a $U(2)$ YM gauge field $a_m^b$ in an $AdS_2$ subspace of $AdS_3$, with the induced metric in eq.~\eqref{metricAdS2}.  As is well-known (see for example~\cite{Castro:2008ms,Erdmenger:2013dpa,Fujita:2014mqa}), a solution of the YM equations in $AdS_2$ typically \textit{diverges} near the $AdS_2$ boundary, in contrast to YM gauge fields in higher-dimensional AdS spacetimes. Indeed, solving our eqs.~\eqref{YM2} and~\eqref{YM3} for $a_t^b$ and expanding about the $AdS_2$ boundary, $z=0$, we find $a_t^b = Q^b/z + \ldots$, where $\ldots$ denotes terms sub-leading in $z$ as $z \to 0$, compared to the term shown. The constants $Q^b$ are the fluxes of the YM gauge field components $a^b$ at the $AdS_2$ boundary: with $\star$ the Hodge star of $AdS_2$,
\beq
\label{asympflux}
\lim_{z \to 0}\star f^b=\lim_{z \to 0}\sqrt{-g} \, g^{zz} \, g^{tt}f^b_{zt}=Q^b.
\eeq
The fluxes $Q^b$ determine the expectation values of the conserved $U(2)$ charges $R^b$. If our model was top-down, then in principle we could derive an exact relation between $Q^b$ and $R^b$. However, in our bottom-up model, we can only assume that the relation between the two is monotonic, and that if $Q^b=0$ then $R^b=0$. As discussed below eq.~\eqref{RKKYdt}, two identical impurity spins, each in a totally anti-symmetric representation with $q$ boxes, must obey the constraints $R^0=q$ and $R^3=0$. To describe two identical impurity spins in our holographic model, we will therefore consider various values of $Q^0$, but will always take $Q^3=0$. Crucially, our choice $K=1$ then guarantees, based on $SU(N)$ representation theory arguments and Pauli exclusion alone, that overscreening cannot occur in our model~\cite{1998PhRvB..58.3794P}.

Although the leading solution $Q^a/z$ diverges as $z \to 0$, we show in the appendix that it is nevertheless normalizable, according to the criteria of refs.~\cite{Klebanov:1999tb,Marolf:2006nd}. More generally, in the appendix we determine the allowed boundary conditions for a massless or massive gauge field in $AdS_{d+1}$ with $d \geq 1$. For the $AdS_2$ case, $d=1$, we find that Dirichlet, Neumann, and ``mixed'' (also called Robin) boundary conditions are all allowed.

However, the divergence of $Q^a/z$ can affect the asymptotic behavior of other fields coupled to the YM gauge field, including in particular charged matter, such as our $\Phi$. In $\Phi$'s equation of motion eq.~\eqref{scalar}, the coupling to the YM gauge field asymptotically approaches a constant, $\lim_{z \to 0} g^{tt} a_t^b a_t^c = -Q^b Q^c$, which is the same order in $z$ as the mass-squared, $M^2$. The YM gauge field thus effectively shifts $\Phi$'s mass-squared matrix from $M^2$ times the $U(2)$ identity matrix to $M^2 - Q^b Q^c \tau^b \tau^c$. The powers of $z$ that appear in $\Phi$'s asymptotic expansion will thus be determined not by $M^2$, but by $M^2 - Q^b Q^c \tau^b \tau^c$. Those powers determine the dimension of $\Phi$'s dual operator $\mathcal{O}$ at the UV fixed point. As a result, fixing $M^2$ and changing the $Q^b$ will change $\mathcal{O}$'s UV dimension, and thus change the UV fixed point. This does not happen in the original two-impurity Kondo model, where $\mathcal{O}$'s UV dimension is always the free-field value, $1/2$, regardless of the choice of $R^b$ (or equivalently of $\r_{\textrm{UV}}$). In other words, this is a special feature of the holographic model, which by process of elimination must be due to the additional, strongly-interacting degrees of freedom we introduced. The same effect appeared in the holographic single-impurity Kondo model of ref.~\cite{Erdmenger:2013dpa} and the holographic Bose-Hubbard model of ref.~\cite{Fujita:2014mqa}.

However, a well-defined theory in asymptotically AdS spacetime requires a well-defined boundary value problem, with boundary conditions imposed on a conformally-equivalent class of asymptotic solutions. In other words, in order to obtain a sensible variational problem, we must fix the asymptotics of all fields. We will therefore take an unusual step: when $Q^b$ changes, we will change $M^2$, that is, we will change the Lagrangian of our theory in AdS, in order to maintain $\Phi$'s asymptotics. Specifically, we will demand that $\mathcal{O}$ always has dimension $1/2$ in the UV, so that the Kondo couplings in our model are always classically marginal in the UV. We will thus be comparing UV fixed points with various values of the $Q^b$ (though always with $Q^3=0$, as explained above), but otherwise identical. Something similar was done in the holographic single-impurity Kondo model of ref.~\cite{Erdmenger:2013dpa}, although how we maintain $\Phi$'s asymptotics as the $Q^a$ change will be very different from the single-impurity case, due to crucial differences in boundary conditions, as we will discuss in the rest of this section.

We fix $\Phi$'s asymptotics as follows. We diagonalize $\Phi$'s mass matrix,
\beq
M^2-Q^bQ^c\tau^b\tau^c = \mathcal{S} \begin{pmatrix} M^2_- & \\ & M^2_+ \end{pmatrix} \mathcal{S}^\dagger,
\eeq
using the unitary matrix
\begin{align}
	\mathcal{S}=\frac{1}{\sqrt{2}}\begin{pmatrix}\mathrm{sgn}(Q^0)\frac{Q^1-iQ^2}{\sqrt{(Q^1)^2+(Q^2)^2}}&-\mathrm{sgn}(Q^0)\frac{Q^1-iQ^2}{\sqrt{(Q^1)^2+(Q^2)^2}}\\1&1\end{pmatrix},
\end{align}
and eigenvalues
\begin{align}
	M^2_\mp=M^2-\frac{1}{4}\le(\le|Q^0\ri|\pm\sqrt{(Q^1)^2+(Q^2)^2}\ri)^2.\label{masses}
\end{align}
The modes with these values of mass-squared are the components $\phi_-$ and $\phi_+$ of $\mathcal{S}^{-1} \Phi \equiv (\phi_-,\phi_+)^{\textrm{T}}$. As a result, the powers of $z$ in $\phi_-$'s asymptotic expansion are determined by $M_-$, and similarly for $\phi_+$ and $M_+$.

Our Kondo interactions are of the form $\mathcal{O}_{\one}^{\dagger} \mathcal{O}_{\one}$ and $\mathcal{O}_{\two}^{\dagger} \mathcal{O}_{\two}$, where $\mathcal{O}_{\one}$ and $\mathcal{O}_{\two}$ are dual to $\Phi_{\one}$ and $\Phi_{\two}$, the components of $\Phi = (\Phi_{\one},\Phi_{\two})^{\textrm{T}}$. To obtain classically marginal Kondo couplings at the UV fixed point, we want both $\mathcal{O}_{\one}$ and $\mathcal{O}_{\two}$ to have dimension $1/2$, as mentioned above. That requires $\Phi_{\one}$ and $\Phi_{\two}$ to have asymptotic powers of $z$ identical to those of a scalar field that saturates the Breitenlohner-Freedman (BF) bound, which in $AdS_2$ means leading asymptotic terms $\sqrt{z}$ and $\sqrt{z} \ln(z)$.

The components $\Phi_{\one}$ and $\Phi_{\two}$ are linear combinations of $\phi_-$ and $\phi_+$, so each of $\Phi_{\one}$ and $\Phi_{\two}$ has an asymptotic expansion with powers of $z$ determined by both $M_-$ and $M_+$. We guarantee that $\Phi_{\one}$ and $\Phi_{\two}$ each has the asymptotics of a scalar at the BF bound as follows. First, we set $M_-^2$ to the $AdS_2$ BF bound, $M^2_-=-1/4$, which via eq.~\eqref{masses} fixes $M^2$ in terms of $Q^0$, $Q^1$, and $Q^2$. Second, we choose an ansatz in which $\phi_+$ vanishes identically. In that way, $\Phi_{\one}$ and $\Phi_{\two}$ each has the asymptotics of a scalar at the BF bound, determined entirely by $\phi_-$, without ``contamination'' from $\phi_+$. (We do not set $M_+^2 = -1/4$, because then $M^2_-$ would violate the BF bound, producing an instability.)

Setting $\phi_+= 0$ is consistent if and only if $\phi_+$ is not sourced by other fields, \textit{i.e.}\ the couplings to other fields vanish from $\phi_+$'s equation of motion, namely eq.~\eqref{scalar} multiplied by $\mathcal{S}^{-1}$. That leads to three constraints. The first constraint is that $a_t^3=0$, which is indeed a solution of $a_t^3$'s equation of motion, eq.~\eqref{YM3} with $b=3$, when $\phi_+=0$. Moreover, $a_t^3=0$ implies $Q^3=0$, which is required to describe two identical impurities, as explained above. The second constraint is that $a_t^1 = \frac{Q^1}{Q^2} a_t^2$. However, $a_t^2$'s equation of motion, eq.~\eqref{YM3} with $b=2$, implies that $\frac{Q^1}{Q^2} a_t^2$ satisfies $a_t^1$'s equation of motion, eq.~\eqref{YM3} with $b=1$, if and only if $Q^1 = Q^2$. We therefore take $Q^1 = Q^2$ and $a_t^1 = a_t^2$. The third constraint is $\textrm{Re} \,\phi_- \propto \textrm{Im} \,\phi_-$, which comes from the $U(2)$ constraint in eq.~\eqref{YM2}. Eqs.~\eqref{YM3} and~\eqref{scalar} then imply $\textrm{Re} \,\phi_- = \pm \textrm{Im} \,\phi_-$. We will choose $\textrm{Re} \,\phi_- = \textrm{Im} \,\phi_-$, and define
\beq
\phi \equiv \textrm{Re} \,\phi_- = \textrm{Im} \,\phi_-.
\eeq
In summary, in addition to our choices of section~\ref{model} (radial gauge and static fields), our ansatz includes
\beq
\mathcal{S}^{-1}\F\equiv \phi \begin{pmatrix}1+i\\0 \end{pmatrix}, \qquad a_t^3=0, \qquad a_t^1 = a_t^2.\label{ansatz}
\eeq
Clearly a non-trivial solution for $\phi$ picks a direction in $U(2)$, breaking $U(2)$ down to a single $U(1)$ generated by $\t^0-\t^3$. We will discuss the symmetry breaking pattern in our model in more detail in section~\ref{phase}.

For the ansatz in eq.~\eqref{ansatz}, a straightforward exercise shows that
\beq
\phi = |\Phi_{\one}| = |\Phi_{\two}|,
\eeq
and therefore our two Kondo couplings will be equal, $\l_{\one} = \l_{\two} = \l_{\K}$, and the strengths of the Kondo screening clouds will be equal, $\langle |\mathcal{O}_{\one}|\rangle = \langle |\mathcal{O}_{\two}|\rangle$, as desired.

Additionally, with the ansatz in eq.~\eqref{ansatz}, the equations of motion simplify dramatically. For convenience, we will define a rescaled $a_t^1$,
\beq
\label{caladef}
\mathcal{A}_t \equiv \sqrt{2}\,\mathrm{sgn}(Q^0Q^1) \, a_t^1,
\eeq
which is holographically dual to
\beq
\label{calRdef}
\mathcal{R} \equiv \sqrt{2}\,\mathrm{sgn}(Q^0Q^1)\,R^1,
\eeq
and which has an asymptotic expansion
\beq
\label{calaexpansion}
\mathcal{A}_t = \frac{\mathcal{Q}}{z} + \mu+\ldots, \qquad \mathcal{Q} \equiv \sqrt{2}\,\mathrm{sgn}(Q^0Q^1) Q^1, \qquad \mu \equiv \sqrt{2}\,\mathrm{sgn}(Q^0Q^1) \mu^1,
\eeq
where $\ldots$ represents terms sub-leading in $z$ as $z \to 0$, compared to the terms shown. We then define
\beq
\label{gaugedefs}
\mathcal{A}_t^{\pm}\equiv\frac{1}{2}(a_t^0\pm\mathcal{A}_t),
\eeq
which are holographically dual to
\beq
\mathcal{R}^{\pm} \equiv \frac{1}{2} \left(R^0 \pm \mathcal{R}\right),
\eeq
and which have the asymptotic expansions
\beq
\label{calapmasymp}
\mathcal{A}_t^{\pm} =  \frac{\mathcal{Q}^{\pm}}{z} + \mu^{\pm} + \ldots, \qquad \mathcal{Q}^{\pm} = \frac{1}{2} (Q^0 \pm \mathcal{Q}),
\eeq
where again $\ldots$ represents terms sub-leading in $z$ as $z \to 0$, compared to the terms shown. Inserting $\mathcal{Q}^+$ into eq.~\eqref{masses} we find that our choice $M^2_-=-1/4$ implies $M^2 = -1/4 + (\mathcal{Q}^+)^2$. The equations of motion, eqs.~\eqref{YM3} and~\eqref{scalar}, then reduce to
\begin{subequations}
\label{EOMS}
\beq
\pa_z\le(\sqrt{-g}\,g^{zz}g^{tt}\partial_z\mathcal{A}_t^-\ri)=0,\label{EOMatilde}
\eeq
\beq
\pa_z\le(\sqrt{-g}\,g^{zz}g^{tt}\partial_z \mathcal{A}^+_t\ri)=\sqrt{-g}\,g^{tt} 2\mathcal{A}^+_t\f^2,\label{EOMa}
\eeq
\beq
\pa_z\le(\sqrt{-g}\,g^{zz}\pa_z\f\ri)-\sqrt{-g}\le(M^2+g^{tt}(\mathcal{A}_t^+)^2\ri)\f=0. \label{EOMphi}
\eeq
\end{subequations}
Clearly $\mathcal{A}_t^-$ decouples from $\mathcal{A}_t^+$ and $\phi$, and in fact $\mathcal{A}_t^-$'s equation of motion, eq.~\eqref{EOMatilde}, is trivial to solve: $\mathcal{A}_t^- = \mathcal{Q}^-/z + \mu^-$. On the other hand, $\mathcal{A}_t^+$ and $\phi$ remain coupled, and we have been able to solve their equations of motion, eqs.~\eqref{EOMa} and~\eqref{EOMphi} respectively, only numerically, as we will discuss in section~\ref{phase}.

Eqs.~\eqref{EOMa} and~\eqref{EOMphi} are actually identical in form to the equations of motion in the holographic single-impurity Kondo model of ref.~\cite{Erdmenger:2013dpa}, but where ref.~\cite{Erdmenger:2013dpa} had $a_t^0$ and $\phi$ we have $\mathcal{A}_t^+$ and $\phi$. However, in the following subsection we will see that the boundary conditions on $\mathcal{A}_t^+$ and $\phi$ in our holographic two-impurity Kondo model are very different from those in ref.~\cite{Erdmenger:2013dpa}, and furthermore the boundary conditions will effectively couple $\mathcal{A}_t^-$ to $\mathcal{A}_t^+$ and $\phi$.

\subsection{Asymptotic Expansions and Boundary Counterterms}
\label{asymptotic}

In the most general terms, holographic renormalization~\cite{Bianchi:2001kw,deHaro:2000xn} can be understood as the process of rendering the variational problem on certain non-compact manifolds well-posed~\cite{Papadimitriou:2005ii,Papadimitriou:2010as}. This process has two essential ingredients. First, the variational problem must be defined within the space of general asymptotic solutions of the equations of motion. Second, boundary terms must be added to the action to ensure that the variational problem within the class of asymptotic solutions is well-posed.

In this subsection we obtain the general static asymptotic solutions of the equations of motion, eqs.~\eqref{EOMS}, and determine the boundary terms that render the Dirichlet variational problem for the action in eq.~\eqref{actionAdS2} well-posed. These boundary terms are commonly referred to as ``counterterms,'' because in holography we identify the bulk on-shell action with the generating functional of connected correlation functions, and the boundary terms cancel the near-boundary divergences of the on-shell action, which map to UV divergences of the generating functional. Having determined the boundary counterterms, we can also introduce additional, finite, boundary terms to change the boundary conditions from Dirichlet to Neumann or to ``mixed'' (also called Robin). We will address finite boundary terms explicitly in the following subsection, where we will relate mixed boundary conditions to the double-trace Kondo and RKKY couplings.   

The ``holographic dictionary'' is the map between integration constants parameterizing the asymptotic solutions in the bulk and observables in the dual field theory.  To express the holographic dictionary in the simplest possible terms, we must partially gauge-fix the bulk diffemorphisms, and any other gauge symmetries, by imposing what is commonly called Fefferman-Graham gauge~\cite{Bianchi:2001kw,deHaro:2000xn,Fefferman}. Such a gauge amounts to picking a radial coordinate $r$ such that $g_{rt}=0$, while $g_{rr}$ can be any function of $r$, provided that function remains the same for all solutions of the equations of motion. In particular, since $h(z)$ in eq.~\eqref{metricAdS2} depends on the BTZ black hole temperature, $z$ is not a Fefferman-Graham coordinate. In the rest of this section, we will therefore switch to a canonical radial coordinate, $r$, such that the induced metric on the $AdS_2$ defect takes the form      
\beq\label{FG-gauge}
g_{mn}dx^mdx^n=d r^2+\g(r)d t^2,
\eeq
where the $AdS_2$ boundary is now at $r \to \infty$, in which limit $\gamma(r)$ diverges as $-e^{2r}$. Explicitly, the radial coordinate $r$ is related to the coordinate $z$ in eq.~\eqref{metricAdS2} as
\beq
\label{rzdef}
r=\ln\le(\frac{1+\sqrt{1-z^2/z_H^2}}{2z}\ri),
\eeq
which implies, in the near-boundary region, $z \to 0$,
\beq
\label{rasymptotics}
r=-\ln(z)-\frac{z^2}{4z_H^2}+\co(z^4/z_H^4).
\eeq
We use the coordinate $r$ only in this section. In all other sections, we use the coordinate $z$. For the gauge fields $\ca_m^{\pm}$ we will choose radial gauge in Fefferman-Graham coordinates, $\ca^\pm_r=0$. Via eq.~\eqref{rzdef}, $\mathcal{A}_r^{\pm}\propto \mathcal{A}_z^{\pm}$, so our gauge choice is equivalent to that of section~\ref{model}, namely radial gauge in the $z$ coordinate, $\mathcal{A}_z^{\pm}=0$.

In the Fefferman-Graham gauge of eq.~\eqref{FG-gauge} the equations of motion  \eqref{EOMatilde}, \eqref{EOMa} and \eqref{EOMphi} become, with a dot denoting $\partial_r$ (for example $\dot \ca^{\pm}_t \equiv \partial_r \ca^{\pm}_t$),
\begin{subequations}
\label{reduced-eoms}
\beq
\ddot \ca^-_t+\frac12\g^{-1}\dot\g\dot \ca^-_t=0,
\eeq
\beq
\ddot \ca^+_t+\frac12\g^{-1}\dot\g\dot \ca^+_t-2\f^2\ca^+_t=0,
\eeq
\beq
\ddot\f-\frac12\g^{-1}\dot\g\dot\f-\g^{-1}(\ca^+_t)^2\f-M^2\f=0.
\eeq
\end{subequations}
The general asymptotic solutions of these equations are, for our choice $M^2 = -1/4 + (\cq^+)^2$,
\begin{subequations}
\label{UV-asymptotics-nl}
\begin{alignat}{2}
\ca_t^-&=e^r\cq^-+\m^-+\ldots,\\
\ca^+_t&=e^r\cq^+-2\cq^+\left(\frac13\a^2r^3+(\a^2-\a\b)r^2+(2\a^2-2\a\b+\b^2)r\right)+\m^++\ldots,\\
\f&=e^{-r/2}\left(-\a \, r+\b\right)+\ldots,
\end{alignat}
\end{subequations}
where $\a$, $\b$, $\cq^-$, and $\m^\pm$ are integration constants, and $\ldots$ represents terms sub-leading in $r$ as $r \to \infty$, compared to the terms shown. As mentioned above, although $\cq^+$ is an integration constant of the $\ca_t^+$ equation of motion, $\cq^+$ also determines the asymptotic behavior of the scalar field $\f$. The asymptotic expansions in eq.~\eqref{UV-asymptotics-nl} are for $M^2 = -1/4 + (\cq^+)^2$ only, which we chose so that $\f$ would have the same asymptotic expansion as a free scalar field in $AdS_2$ whose mass saturates the BF bound, and is thus dual to an operator of dimension $1/2$.

Our choice of $M^2$ is actually dramatically different from that in the holographic single-impurity Kondo model of ref.~\cite{Erdmenger:2013dpa}. In the model of ref.~\cite{Erdmenger:2013dpa}, $M^2 = -1/4+(Q^0)^2$, where $Q^0$ was held fixed in $a_t^0$'s variational principle. In other words, $Q^0$ was an \textit{input}. In contrast, in our model $M^2 = -1/4 + (\mathcal{Q}^+)^2$, where $\mathcal{Q}^+$ will be an \textit{output}: in principle, we should again fix $Q^0$, which fixes the impurities' representation, and then solve the equations of motion and extract $\mathcal{Q}^+$ from the solution for $\mathcal{A}^+_t$. In other words, $M^2$, and hence $\phi$'s equation of motion, will depend on the solutions of the equations of motion. Though that may sound unusual, allowing $M^2$ to depend on the solutions maintains the form of $\phi$'s asymptotic expansion, as required for a well-posed variational problem---which is fundamental to define holography in the classical gravity limit. For a quantum gravity theory in asymptotically locally AdS space, we must perform a path integral over all fields, subject to fixed asymptotic boundary conditions. In the classical limit, we must thus fix the asymptotics of all fields to obtain a well-posed variational problem. In our model, guaranteeing a well-posed variational problem for $\f$ requires allowing $M^2$ to depend on $\mathcal{Q}^+$.

We now proceed to determine the boundary counterterms for our model. These can be determined systematically a number of ways, but we will adopt a Hamiltonian approach where the radial coordinate $r$ plays the role of Hamiltonian time~\cite{deBoer:1999xf,Martelli:2002sp,Papadimitriou:2004ap}. In this approach, we obtain the counterterms by solving asymptotically the radial Hamilton-Jacobi (HJ) equation, with the leading forms of the asymptotic expansions in eq.~\eqref{UV-asymptotics-nl} as boundary conditions. 

We will write the $AdS_2$ action in eq.~\eqref{actionAdS2} in the Fefferman-Graham gauge of eq.~\eqref{FG-gauge}, dropping time derivatives, and using the ansatz in eq.~\eqref{ansatz}. We will also introduce a cut-off by integrating not to $r \to \infty$ but up to some large but finite $r$, producing a regulated action,
\beq\label{pp-action}
S_{\textrm{reg}}\equiv-2N\int^r d r'd t \sqrt{-\g}\le(\frac{1}{2} \g^{-1}\left((\dot \ca^+_t)^2+(\dot \ca^-_t)^2\right) +\dot\f^2+\g^{-1}(\ca_t^+)^2\f^2+M^2\f^2\ri),
\eeq
which is the starting point of the radial Hamiltonian analysis for systematically deriving the boundary counterterms directly in covariant form. We will also define a corresponding Lagrangian via $S_{\textrm{reg}}(r) \equiv \int^r dr' \, \mathcal{L}_{\textrm{reg}}$.

For the radial Hamiltonian analysis we will need the canonical momenta,
\beq\label{momenta}
\pi^t_{\ca_t^\pm}\equiv\frac{\pa\mathcal{L}_{\textrm{reg}}}{\pa\dot{\ca}^\pm_t}=-2N\sqrt{-\g}\g^{-1}\dot\ca_t^\pm,\qquad
\pi_\f\equiv\frac{\pa\mathcal{L}_{\textrm{reg}}}{\pa\dot{\f}}=-2N\sqrt{-\g}\;2\dot\f,
\eeq
which can also be expressed as derivatives of the {\em on-shell} $S_{\mathrm{reg}}$ with respect to the induced fields at the cut-off surface,
\beq\label{HJ-momenta}
\pi^t_{\ca_t^\pm}=\frac{\d S_\mathrm{reg}}{\d \ca_t^\pm},\qquad
\pi_\f=\frac{\d S_\mathrm{reg}}{\d\f}.
\eeq

Legendre-transforming $\mathcal{L}_{\textrm{reg}}$ using the canonical momenta in eq.~\eqref{momenta}, we obtain the Hamiltonian, $\ch$, which is a functional of $\ca_t^\pm$, $\f$, $\p_{\ca_t^\pm}^t$, and $\p_\f$. Crucially, $S_{\textrm{reg}}$ depends on $r$ through the induced metric $\g$ on the cut-off surface, which we treat as a non-dynamical background field. The radial HJ equation is thus
\begin{align}\label{HJ-eq}
	\ch+\frac{\pa S_\mathrm{reg}}{\pa r}=0.
\end{align}
Our goal is to solve eq.~\eqref{HJ-eq} for $S_{\textrm{reg}}$. More specifically, we are interested in determining only the divergent part of $S_\mathrm{reg}$, so in eq.~\eqref{HJ-eq} we can replace $\frac{\partial S_{\mathrm{reg}}}{\partial r}$ with $2\g\frac{\partial S_{\mathrm{reg}}}{\partial \g}$ using the fact that $\g$ asymptotically approaches $-e^{2r}$. With that replacement, and writing the canonical momenta in terms of $S_{\mathrm{reg}}$ as in \eqref{HJ-momenta}, eq.~\eqref{HJ-eq} becomes
\begin{align}
\label{HJ-simple}
&-\frac{1}{2N}\int dt\frac{1}{\sqrt{-\g}}\left(\frac12\g\left(\frac{\d S_\mathrm{reg}}{\d \ca^+_t}\right)^2+\frac12\g\left(\frac{\d S_\mathrm{reg}}{\d \ca_t^-}\right)^2+\frac14\left(\frac{\d S_\mathrm{reg}}{\d \f}\right)^2\right)\NO\\
&\hskip2.5cm+2N\int dt \sqrt{-\g}\left(\rule{0cm}{0.4cm}\g^{-1}\f^2(\ca^+_t)^2+M^2\f^2\right)+2\g\frac{\d  S_\mathrm{reg}}{\d \g}=0.
\end{align}
We can simplify the HJ equation further: we switch from $\mathcal{A}_t^+$ and $\phi$ to
\beq
u\equiv\frac{1}{2}\g^{-1}(\ca_t^+)^2,\qquad v\equiv\f^2,
\eeq
and then, because $\ca_t^+$ and $\f$ are coupled to one another in the bulk while $\ca_t^-$ decouples, we introduce a covariant ansatz for $S_{\mathrm{reg}}$ in which we separate $u$ and $v$ from $\ca_t^-$,
\beq
\label{HJ-ansatz}
S_\mathrm{reg}=2N\int\dif t\sqrt{-\g}\le(F(u,v)-u-\frac{1}{2}\g^{-1}(\ca_t^-)^2\ri).
\eeq
From the HJ equation, eq.~\eqref{HJ-eq}, we then find a first order partial differential equation for $F(u,v)$,
\begin{equation}
\label{master}
u\left(\partial_u F \right)^2+v\left(\partial_v F\right)^2-F-2\left(\frac{M^2}{2}+u\right)v=0,
\end{equation}
which is independent of $\ca_t^-$, indicating that our ansatz is self-consistent.

Crucially, a complete integral of the HJ equation eq.~\eqref{HJ-eq}, rather than the most general solution, suffices to describe all possible solutions of the second order equations of motion, eq.~\eqref{reduced-eoms}. A complete integral, in this case, corresponds to a solution of eq.~\eqref{HJ-eq} that contains one integration constant for each of the induced fields $\ca^\pm_t$ and $\f$, plus one integration constant related to the radial coordinate, or equivalently to the metric $\g$. However, all these integration constants enter only in the finite part of the on-shell action~\cite{Papadimitriou:2010as}. To determine the divergent part of the on-shell action, therefore, we can simply look for solutions of eq.~\eqref{HJ-eq} with fixed values of the integration constants that parameterize a complete integral. In fact, ignoring the integration constant associated with the induced metric $\g$, namely the BTZ black hole temperature $T$, allowed us to go from eq.~\eqref{HJ-eq} to eq.~\eqref{HJ-simple}. Similarly, ignoring the integration constant associated with $\ca_t^-$ allowed us to work within the ansatz in eq.~\eqref{HJ-ansatz}. However, a complete integral of eq.~\eqref{HJ-simple} still contains two integration constants, and so can be used to obtain a subclass of exact solutions of the second order equations of motion, eq.~\eqref{reduced-eoms}, but at $T=0$. Again, given that we are only interested in determining the near-boundary (or UV) divergences, we can consider special solutions of eq.~\eqref{HJ-simple} only, corresponding to the correct asymptotics.    

In particular, as mentioned above, in order for $\f$'s asymptotic behavior to be well-defined, $\cq^+$ must be related to $M^2$ as $M^2 = -1/4 + (\cq^+)^2$, which implies that $u$ satisfies a (second class) constraint asymptotically as $r \to \infty$,
\beq\label{constraint}
2u+M^2+\frac{1}{4}= \co\le(r^3e^{-r}\ri).
\eeq
The solution $F(u,v)$ of eq.~\eqref{master} takes the form of a Taylor-expansion in this constraint~\cite{Chemissany:2014xsa},
\begin{align}
	F(u,v)=\sum\limits_{k\geq0}f_k(v)\le(u+\frac{M^2+1/4}{2}\ri)^k, \label{HJTaylor}
\end{align}
so that our task now is to solve for the coefficients $f_k(v)$. As in the class of holographic field theories with Lifshitz scaling symmetry discussed in ref.~\cite{Chemissany:2014xsa}, relaxing the constraint in eq.~\eqref{constraint} corresponds to an irrelevant deformation of the theory, in the sense that such a deviation from the constraint will modify the asymptotic behavior of the bulk fields, in our case of $\f$. In such cases, while the $k=0$ term in the Taylor expansion of eq.~\eqref{HJTaylor} contributes to the near-boundary divergences of the on-shell action, we also need terms of higher order in $k$ to renormalize higher-point functions of the operator dual to $\ca_t^+$. Specifically, to renormalize a correlation function with $k$ insertions of the operator dual to $\ca_t^+$, we must keep all terms up to and including order $k$ in the Taylor expansion of eq.~\eqref{HJTaylor}.

In other words, deviating from $M^2 = -1/4 + (\cq^+)^2$, for example by changing $\cq^+$ while keeping $M^2$ fixed, changes $\f$'s asymptotics, and hence looks like an irrelevant deformation of the dual UV CFT. Put differently, $\mathcal{A}^+_t$ is dual to an operator that is ``effectively'' irrelevant. Of course, $\mathcal{A}_t^+$ is not actually dual to an irrelevant operator: $\mathcal{A}^+_t$ is dual to a conserved charge in $(0+1)$ dimensions, whose dimension is protected by symmetry to be zero. Indeed, in the original two-impurity Kondo model, the dimensions of operators do not change upon changing $\mathcal{Q}^+$, \textit{i.e.}\ the auxiliary charge does not act as an irrelevant operator. Holography is telling us that the additional strongly-coupled degrees of freedom we added to obtain a classical gravity dual leads to a conserved charge that mimics, in many ways, an irrelevant operator. As a result, the holographic renormalization of our model is similar to, though strictly speaking distinct from, that of a field dual to an irrelevant operator.

In any case, in this paper we will only consider the free energy, which is related to the renormalized on-shell action, and also one-point functions, and therefore we will only need to determine $f_0(v)$ and $f_1(v)$. A calculation of two-point functions in the holographic single-impurity Kondo model of ref.~\cite{Erdmenger:2013dpa}, which requires $f_2(v)$, will appear in ref.~\cite{usfuture}.     

Inserting the Taylor expansion for $F(u,v)$ in eq.~\eqref{HJTaylor} into eq.~\eqref{master} leads to a set of ordinary differential equations for the $f_k(v)$, which could in principle be solved to obtain $f_0(v)$ and $f_1(v)$. However, given that we have already determined the full asymptotic expansions in eq.~\eqref{UV-asymptotics-nl}, we can determine $f_0(v)$ and $f_1(v)$ by integrating the first order ``flow equations,'' obtained by equating the two alternative expressions for the canonical momenta, eq.~\eqref{momenta} and eq.~\eqref{HJ-momenta}. For our ansatz in eq.~\eqref{HJ-ansatz} the flow equations give us
\beq
\label{flow-eqs0}
\dot\f=-\f\,\partial_vF, \qquad \dot\ca_t^+=(1-\partial_uF)\ca_t^+.
\eeq
Plugging the Taylor expansion of $F(u,v)$ from eq.~\eqref{HJTaylor} into eq.~\eqref{flow-eqs0}, and retaining only the leading term, gives
\beq
\label{flow-eqs}
\dot\f\approx -f_0'(v)\f, \qquad \dot\ca_t^+ \approx (1-f_1(v))\ca_t^+.
\eeq
Plugging the asymptotic expansions in eq.~\eqref{UV-asymptotics-nl} into eq.~\eqref{flow-eqs} and integrating gives
\begin{subequations}
\label{fs}
\begin{alignat}{1}
f_0(v)&=v\left(\frac12+\frac{1}{\ln v}+\co\(\frac{\ln(-\ln v)}{(\ln v)^2}\)\right),\\
f_1(v)&=\frac23 v\ln v\left(1-\frac{2\ln(-\ln v)}{\ln v}+\co\(\frac{1}{\ln v}\)\right).
\end{alignat}
\end{subequations}

In principle, the results in eq.~\eqref{fs} are sufficient to capture the near-boundary divergences of both the regularized action and the regularized one-point functions, in covariant form, and hence are sufficient to determine the boundary counterterms that we need. However, the $f_0(k)$ and $f_1(k)$ in eq.~\eqref{fs} involve factors of $\ln \f$, that is, certain divergences involve expressions \textit{non-analytic} in $\f$. Given that for Dirichlet boundary conditions $\f$ is the covariant source of the dual scalar operator $\mathcal{O}_{\one}=\mathcal{O}_{\two}$, such non-analyticity would imply a certain kind of non-locality in the dual field theory. We thus face a choice: preserve either covariance or locality, but not both simultaneously. In other words, our system has an anomaly. In such cases, following the usual course of action in field theory, we demand locality at the expense of partially breaking covariance, by allowing explicit cut-off dependence in the counterterms~\cite{Bianchi:2001de,Papadimitriou:2004rz}. The resulting cut-off dependence indeed gives rise to a conformal anomaly in the dual field theory~\cite{Henningson:1998gx}. In the present case, demanding locality leads to a boundary counterterm action, $S_{\mathrm{ct}}$, that depends explicitly on the cut-off, $r$,      
\begin{align} \label{counterterms}
S_{\mathrm{ct}}=&-2N\int\dif t\sqrt{-\g}\le\{\f^2\le(\frac{1}{2}-\frac1r\ri)-\frac{1}{2}\g^{-1}(\ca^+_t)^2-\frac{1}{2}\g^{-1}(\ca^-_t)^2\right.\NO\\
&\left.\hskip5cm-\frac{2}{3}\f^2\le(r-\frac{\b}{\a}\ri)\le(\frac{1}{2} \gamma^{-1} \left(\ca^+\right)^2+\frac{M^2+1/4}{2}\ri)\ri\}.\end{align}

Explicit cut-off dependence in counterterms is actually a standard result for scalars saturating the BF bound~\cite{Bianchi:2001de,Papadimitriou:2004rz}, such as our $\f$. However, explicit cut-off dependence also appears in the counterterms in the second line of eq.~\eqref{counterterms}, although those counterterms do not contribute to the near-boundary divergences of the action, and so do not modify the conformal anomaly of the dual theory. Those counterterms also explicitly depend on the ratio $\b/\a$ of $\f$'s asymptotic coefficients. We will shortly show, via the boundary conditions we are going to impose on $\f$, that $\b/\a$ is proportional to the Kondo coupling in our model, and so this counterterm is well-defined. More generally, in the presence of fields dual to irrelevant operators, counterterms typically depend explicitly on the renormalized canonical momenta, and are well-defined for boundary conditions dual to multi-trace couplings, as first observed in refs.~\cite{vanRees:2011fr,vanRees:2011ir}.   

Given the regulated action $S_{\mathrm{reg}}$ in eq.~\eqref{pp-action} and the counterterm action $S_{\mathrm{ct}}$ in eq~\eqref{counterterms}, we can now define the renormalized action,
\beq
S_{\mathrm{ren}} \equiv \lim_{r \to \infty} \left(S_{\mathrm{reg}}+S_{\mathrm{ct}}\right).
\eeq

\subsection{One-Point Functions for Dirichlet Boundary Conditions}
\label{dirichletbc}

Our next task is to determine the finite counterterms required to enforce mixed boundary conditions on the bulk fields, and hence implement the Kondo and RKKY couplings in the dual field theory, as mentioned above. As a necessary first step, we must determine the form of renormalized one-point functions for Dirichlet boundary conditions. We thus define ``subtracted'' canonical momenta as
\beq\label{HJ-momenta-ren}
\left({\pi}^t_{\ca_t^\pm}\right)_{\mathrm{sub}}\equiv\frac{\d }{\d \ca_t^\pm}\left[S_{\textrm{reg}}+ S_{\textrm{ct}}\right], \qquad
\left(\pi_\f\right)_{\mathrm{sub}}\equiv\frac{\d }{\d\f}\left[S_{\textrm{reg}}+ S_{\textrm{ct}}\right],
\eeq
where the derivatives in eq.~\eqref{HJ-momenta-ren} must be evaluated at the radial cut-off. To remove the cut-off we need to extract the leading asymptotic behavior of the fields from eqs.~\eqref{FG-gauge} and~\eqref{UV-asymptotics-nl}, and define the renormalized one-point functions $\Hat\pi^t_{\ca_t^{\pm}}$ and $\Hat\pi_\f$ as the finite limits 
\begin{subequations}
\label{ren-lim}
\begin{alignat}{2}
\Hat\pi^t_{\ca_t^-}&\equiv\lim_{r\to\infty}\left[\frac{e^{2r}}{\sqrt{-\g}}\left(\pi^t_{\ca_t^-}\right)_{\mathrm{sub}}\right]=-2N\m^-,\\
\Hat\pi^t_{\ca_t^+}&\equiv\lim_{r\to\infty}\left[\frac{e^{2r}}{\sqrt{-\g}}\left(\pi^t_{\ca_t^+}\right)_{\mathrm{sub}}\right]=-2N\m^+-4N\cq^+\le(2\a^2-2\a\b+\b^2-\frac{1}{3}\frac{\b^3}{\a}\ri),\\
\Hat\pi_\f &\equiv\lim_{r\to\infty}\left[-\frac{e^{r/2}r}{\sqrt{-\g}}\left(\pi_\f\right)_{\mathrm{sub}}\right]=-4N\b.
\end{alignat}
\end{subequations}
From the definitions of $\ca_t^{\pm}$ in terms of $a_t^0$ and $\ca_t$ in eq.~\eqref{gaugedefs}, we then find
\begin{subequations}
\begin{alignat}{2}
\Hat\pi^t_{a^0}&=\frac12(\Hat\pi^t_{\ca_t^+}+\Hat\pi^t_{\ca_t^-})=-N\m^0-2N\cq^+\le(2\a^2-2\a\b+\b^2-\frac{1}{3}\frac{\b^3}{\a}\ri),\\
\label{calamomdef}
\Hat\pi^t_{\ca_t}&=\frac12(\Hat\pi^t_{\ca_t^+}-\Hat\pi^t_{\ca_t^-})=-N\m-2N\cq^+\le(2\a^2-2\a\b+\b^2-\frac{1}{3}\frac{\b^3}{\a}\ri),
\end{alignat}
\end{subequations}
with $\mu$ defined from $\ca_t$'s asymptotics in eq.~\eqref{calaexpansion}, and $\m^0=\m^++\m^-$.

\subsection{Double-Trace Kondo and RKKY Couplings from Boundary Conditions}
\label{dtbc}

As discussed in section~\ref{review}, introducing Abrikosov pseudo-fermions allows us to write the Kondo and RKKY couplings as double-trace with respect to the $SU(N)$ spin group. Such a description is particularly useful in the large-$N$ limit, since large-$N$ factorization means that the effect of multi-trace deformations takes a simple and generic form, independent of the details of the CFT or the particular deformation~\cite{Papadimitriou:2007sj}, as we now review.

Consider a gauge theory (not necessarily with classical gravity dual) in $d$ spacetime dimensions, with Lagrangian $\mathcal{L}$. In the $N \to \infty$ limit, we normalize $\mathcal{L}$ to be $\co(N^0)$. In the gauge theory, let $\Hat\co$ be a local, gauge-invariant, single-trace scalar operator of conformal dimension $\D$, normalized such that $\langle \Hat\co\rangle$ is $\co(N^0)$ as $N\to\infty$. Moreover, let $W[J]$ be the generating functional of connected correlation functions of $\Hat\co$, with source $J$, and let $\Gamma[\s]$ be the quantum effective action, with $\s=\d W/\d J$ the one-point function of $\Hat\co$, all normalized to be $\co(N^0)$ as $N\to\infty$. Now suppose we deform $\mathcal{L}$ by a generic multi-trace deformation, that is, by a polynomial $f(\Hat\co)$ in $\Hat\co$,
\beq
\mathcal{L}\longrightarrow \mathcal{L}+f(\Hat\co),
\eeq 
where the degree of $f(\Hat\co)$ is $\in[2,d/\D]$. At leading order in the large-$N$ limit, the effect of such a multi-trace deformation takes a simple, universal form, namely the multi-trace deformation amounts to the following transformation~\cite{Papadimitriou:2007sj}:
\begin{subequations}
\label{multi-trace}
\begin{alignat}{2}
J & \quad \longrightarrow \quad J_f\equiv J-f'(\s),\\
\s &\quad \longrightarrow \quad \s_f\equiv\s,\\
\G[\s] &\quad \longrightarrow \quad \G_f[\s_f]\equiv\G[\s]+\int d^dxf(\s),\\
W[J]& \quad \longrightarrow \quad W_f[J_f]\equiv W[J]+\int d^dx\left.(f(\s)-\s f'(\s))\right|_{\s=\d W/\d J}.
\end{alignat}
\end{subequations}
Once the deformation polynomial $f(\Hat\co)$ is specified, eq.~\eqref{multi-trace} allows us to determine straightforwardly the finite boundary terms that must be added to the renormalized action to implement the multi-trace deformation holographically. Those finite boundary terms then determine the boundary conditions that must be imposed on the bulk fields. Our task is to determine the finite boundary terms for our double-trace Kondo and RKKY couplings.

As discussed in subsection~\ref{sikm}, Abrikosov pseudo-fermions allow us to write a Kondo interaction as a double-trace deformation of the Hamiltonian of the form in eq.~\eqref{KondoAbrikosov}. For two impurities with equal Kondo couplings of strength $\l_{\K}$, the Kondo double-trace deformation of the Hamiltonian takes the form
\begin{align}
\label{kondodt1}
\ch_\mathrm{K}=-\frac{1}{2}N\l_{\K}\le(\frac{\mathcal{O}_{\rm I}^\dagger\mathcal{O}_{\rm I}}{N^2}+\frac{\mathcal{O}_{\rm II}^\dagger\mathcal{O}_{\rm II}}{N^2}\ri),
\end{align}
where now we have normalized the scalar operators $\co_{\rm I}$ and $\co_{\rm II}$ to be $\co(N)$ as $N\to\infty$. We also kept $N\l_{\K}$ fixed as $N \to \infty$, so that $\ch_{\rm K}$ is $\co(N^0)$ in the large-$N$ limit, in contrast to the Kondo Hamiltonian in eq.~\eqref{KondoCFT}, which is $\co(N)$. Our ansatz in eq.~\eqref{ansatz} only allows us to describe the subspace of the Hilbert space where $\co_{\rm I}=\co_{\rm II}$. Within that subspace, the Kondo interaction in eq.~\eqref{kondodt1} becomes
\beq
\label{kondodt2}
\ch_\mathrm{K}=-N\l_{\K}\frac{|\co_{\one}|^2}{N^2},
\eeq 
which gives us the multi-trace polynomial for our Kondo interaction,
\beq\label{kondo-f}
f_{\rm K}(|\co_{\one}|^2/N^2)=N\l_{\K} \frac{|\co_{\one}|^2}{N^2},
\eeq
where the overall sign differs between eqs.~\eqref{kondodt2} and~\eqref{kondo-f} because $f(\Hat\co)$ was defined as a deformation of the Lagrangian, rather than the Hamiltonian.

To implement this multi-trace deformation holographically, we first need to know the holographic dictionary for $|\co_{\one}|$ in the theory without multi-trace deformation. A scalar field that saturates the BF bound is special because the mass falls in the window for which both asymptotic solutions are normalizable~\cite{Klebanov:1999tb}, and is unique because Dirichlet boundary conditions can be continuously deformed into mixed boundary conditions while keeping the deformation marginally relevant~\cite{Witten:2001ua}. With a Dirichlet boundary condition the renormalized action is the generating functional of $|\co_{\one}|$ with source $\a$, and with $\langle |\co_{\one}|\rangle$ given by the renormalized canonical momentum $\Hat\p_{\f}$ in eq.~\eqref{ren-lim},         
\begin{align}
\label{dirichletO}
\le<\le|\mathcal{O}_{\one}\ri|\ri>=\Hat\p_\f=-4\,N\,\b.
\end{align}

Given the Kondo multi-trace polynomial $f_{\K}(|\co_{\one}|^2/N^2)$ in eq.~\eqref{kondo-f} and the identifications of one-point functions for Dirichlet boundary conditions in subsection~\ref{dirichletbc}, the transformation rules in eq.~\eqref{multi-trace} instruct us to add to $S_{\mathrm{ren}}$ the finite boundary term
\begin{align}
\label{kondoboundary}
S_\mathrm{\K}&=-N\cdot N\l_K\int\dif t\,\le(\frac{\Hat{\pi}_\f}{N}\ri)^2,
\end{align}
where the overall factor of $N$ keeps the bulk action $\co(N)$ in the large-$N$ limit. If we define an $\mathcal{O}(N^0)$ ``holographic Kondo coupling,''
\beq
\label{holokondocoupling}
\k\equiv-8N\l_{\K},
\eeq
then using eq.~\eqref{dirichletO} we can write $S_{\K}$ in eq.~\eqref{kondoboundary} as
\begin{align}
\label{holokondocouplingbetasquared}
S_\mathrm{\K}&=2N\int\dif t\,\kappa\b^2.
\end{align}
Moreover, eq.~\eqref{multi-trace} indicates that although $\langle|\co_{\one}|\rangle=-4N\b$ is unchanged by the multi-trace deformation, the source conjugate to $|\co_{\one}|$ changes as 
\beq
\label{kondosourcechange}
\a\longrightarrow \a-2N\l_K\frac{\Hat\p_\f}{N}=\a-\k\b.
\eeq 
Indeed, we can verify eq.~\eqref{kondosourcechange} explicitly: a variation of the generating functional in the deformed theory gives
\begin{align}
\d\le(S_\mathrm{ren}+S_\mathrm{K}\ri)=\int\dif t\,\le(-4N\b\ri)\,\d\le(\a-\kappa\b\ri),
\end{align}
indicating that $\le<\le|\mathcal{O}_{\one}\ri|\ri>=-4\,N\,\b$ remains unchanged, while $(\a-\k\b)$ is now the source for $|\co_{\one}|$, as advertised. As a result, states in the theory that include the double-trace Kondo coupling will be described by bulk solutions for $\f$ satisfying the boundary condition $\a=\k \b$. In the space of such solutions, the ratio $\b/\a$ that appears in the counterterms of eq.~\eqref{counterterms} is equal to $1/\k$, which is kept fixed, hence the counterterms are well-defined, as mentioned above.

In summary, to implement the double-trace Kondo coupling in our holographic model, we must add to $S_{\mathrm{ren}}$ the finite boundary term $S_{\K}$ in eq.~\eqref{kondoboundary}, where the coefficient $\kappa$ is related to the Kondo coupling constant $\l_{\K}$ via eq.~\eqref{holokondocoupling}, and which requires us to impose the mixed boundary condition on $\f$ that $\kappa = \a/\b$ is fixed. With that boundary condition, $\langle|\mathcal{O}_{\one}|\rangle = - 4 N \beta$, as in eq.~\eqref{dirichletO}\footnote{In the holographic single-impurity Kondo model of ref.~\cite{Erdmenger:2013dpa}, the finite boundary term involving the scalar field was different from our $S_{\K}$ in eq.~\eqref{holokondocouplingbetasquared}: the finite boundary term in ref.~\cite{Erdmenger:2013dpa} had the same form as $S_{\K}$ in eq.~\eqref{holokondocouplingbetasquared}, but with $\b \to \a$ and $\k \to 1/\k$. In that case, the linear combination of $\a$ and $\b$ held fixed in the variational principle would not be $\a - \k \b$. Nevetheless, $\a - \k \b$ was held fixed in ref.~\cite{Erdmenger:2013dpa}. Those two wrongs made a right, in the following sense: the finite boundary term in ref.~\cite{Erdmenger:2013dpa}, when evaluated on $\a=\k\b$, actually agrees with our $S_{\K}$, so the solutions for the scalar and the value of the on-shell action of ref.~\cite{Erdmenger:2013dpa} actually agree with those obtained using our $S_{\K}$. Similarly, the one-point function identified as $\langle \mathcal{O} \rangle \propto N \a$ in ref.~\cite{Erdmenger:2013dpa} agrees with our eq.~\eqref{dirichletO} when evaluated on $\a = \k\b$. In other words, our results ultimately agree with those of ref.~\cite{Erdmenger:2013dpa} where they overlap, despite the difference in finite boundary terms.}.

As discussed in subsection~\ref{tikm}, Abrikosov pseudo-fermions also allow us to write the RKKY interaction as a double-trace deformation of the Hamiltonian, of the form in eq.~\eqref{RKKYdt}, which after dropping some unimportant constants becomes
\begin{align}
\label{RKKYdt1}
	\ch_\mathrm{RKKY}=-\frac{1}{2}N\l_\mathrm{RKKY}\le(\le(\frac{R^1}{N}\ri)^2+\le(\frac{R^2}{N}\ri)^2\ri)\;,
\end{align}
where we chose to normalize the operators $R^1$ and $R^2$ to be $\co(N)$ as $N\to\infty$. We also kept $N\l_\mathrm{RKKY}$ fixed as $N \to \infty$, so that $\ch_\mathrm{RKKY}$ is $\co(N^0)$ in the large-$N$ limit, in contrast to the RKKY interaction in eq.~\eqref{RKKYdt}, which is $\co(N)$. Our ansatz in eq.~\eqref{ansatz} only allows us to describe the subspace of the Hilbert space where $R^1=R^2$. Recalling from the definition of $\ca_t$ in eq.~\eqref{caladef} that
\beq
\car = \sqrt{2} \, \mathrm{sgn}(Q^0Q^1)R^1=\sqrt{2} \,\mathrm{sgn}(Q^0Q^2)R^2,
\eeq
in our case the RKKY interaction in eq.~\eqref{RKKYdt1} becomes
\begin{align}
	\ch_\mathrm{RKKY}=-\frac12 N\l_\mathrm{RKKY}\le(\frac{\car}{N}\ri)^2.
\end{align}
We thus identify the multi-trace polynomial for our RKKY interaction as
\beq\label{f-RKKY}
f_{\rm RKKY}(\car/N)=\frac12 N\l_\mathrm{RKKY}\le(\frac{\car}{N}\ri)^2.
\eeq 

To implement this multi-trace deformation holographically, we first need to know the holographic dictionary for $\car$ in the theory without multi-trace deformation. The charge $\car$ has conformal dimension zero, and hence $\langle \car\rangle$ must be determined by $\cq$, the coefficient of the leading term in $\ca_t$'s asymptotics. The undeformed theory thus corresponds to a Neumann boundary condition for $\ca_t$, which for a gauge field in $AdS_2$ is normalizable, as we show in the appendix. The generating functional is thus given by the Legendre transform of $S_{\mathrm{ren}}+S_{\K}$,
\bal
\label{LTaction}
\Hat{S}=N\int\dif t\,\cq\le(\frac{\Hat{\pi}^t_{\ca_t}}{N}\ri)-(S_\mathrm{ren}+S_\mathrm{K}),
\eal
with variation
\be\label{current-source}
	\d \Hat{S}=\int\dif t\,N\,\cq\,\frac{\d\Hat{\pi}^t_{\ca_t}}{N}+\int\dif t\,4N\b\,\d\le(\a-\kappa\b\ri).
\ee
Using eq.~\eqref{calamomdef} for $\Hat{\pi}^t_{\ca_t}$, we thus identify the source $\cj$ and one-point function of $\car$ as, respectively,
\begin{subequations}
\beq\label{calRcurrent}
\cj=\frac{\Hat{\pi}^t_{\ca_t}}{N}=-\m-2\cq^+\le(2\a^2-2\a\b+\b^2-\frac{1}{3}\frac{\b^3}{\a}\ri),
\eeq
\beq
\label{calRvev}
\langle \car\rangle=N\,\cq.
\eeq
\end{subequations}

Given the RKKY multi-trace polynomial $f_{\rm RKKY}(\car/N)$ in eq.~\eqref{f-RKKY} and the identifications of one-point functions for Neumann boundary conditions, the transformation rules in eq.~\eqref{multi-trace} instruct us to add to $\Hat{S}$ the finite boundary term
\beq
\label{RKKYboundary}
S_\mathrm{RKKY}=-N\int\dif t\,\frac12 N\l_\mathrm{RKKY}\cq^2.
\eeq
If we define an $\mathcal{O}(N^0)$ ``holographic RKKY coupling,''
\beq
\label{holoRKKY}
\l \equiv N \l_{\RKKY},
\eeq
then eq.~\eqref{multi-trace} indicates that $\langle \car\rangle=N\,\cq$ is unchanged by the multi-trace deformation, while the source conjugate to $\car$ changes as
\beq
\label{RKKYsourcechange}
\cj\longrightarrow \cj_\l=\frac{\Hat{\pi}^t_{\ca_t}}{N}-\l\,\cq.
\eeq
We can verify~eq.~\eqref{RKKYsourcechange} explicitly: a variation of the generating functional with respect to $\ca_t$ in the deformed theory gives
\beq
\d\le(\Hat{S}+S_\mathrm{RKKY}\ri)=\int\dif t\,N\,\cq\,\d\le(\frac{\Hat{\pi}^t_{\ca_t}}{N}-\l \cq\ri)+\int\dif t\,4N\b\,\d\le(\a-\kappa\b\ri)\;,
\eeq
indicating that $\langle \car\rangle=N\,\cq$ remains unchanged, while $\cj_{\l}$ is now the source for $\car$, as advertised. As a result, states in the theory that include the double-trace RKKY coupling will be described by bulk solutions for $\car$ satisfying the boundary condition $\cj_{\l}=\frac{\Hat{\pi}^t_{\ca_t}}{N}-\l\,\cq=0$, that is, solutions with a fixed value of
\beq
\label{lambdafixed}
\lambda = -\frac{1}{\mathcal{Q}} \left[\mu+ 2\mathcal{Q}^+\left(2 \alpha^2 - 2 \alpha \beta + \beta^2 - \frac{1}{3}\frac{\beta^3}{\alpha}\right)\right].
\eeq

In summary, to implement the double-trace RKKY coupling in our holographic model, we must add to the Legendre-transformed, renormalized action $\Hat{S}$ of eq.~\eqref{LTaction} the finite boundary term $S_{\RKKY}$ in eq.~\eqref{RKKYboundary}, which requires us to impose the mixed boundary condition on $\ca_t$ that $\frac{\Hat{\pi}^t_{\ca_t}}{N}-\l\,\cq=0$, with $\frac{\Hat{\pi}^t_{\ca_t}}{N}$ in eq.~\eqref{calRcurrent}, which means holding fixed our holographic RKKY coupling $\l$ in eq.~\eqref{lambdafixed}. With that boundary condition, $\langle\mathcal{R}\rangle = N \mathcal{Q}$, as in eq.~\eqref{calRvev}. These identifications are the main results of this section. To our knowledge, they provide the first explicit identification of an RKKY coupling in any holographic system.

We have thus determined the boundary conditions on $\phi$ and $\mathcal{A}_t$ that implement the double-trace Kondo and RKKY couplings in our holographic model. The third field in our ansatz is $a_t^0$, for which we fix the coefficient of the leading term in $a_t^0$'s asymptotics, $Q^0$, which in the dual field theory fixes the representation of the impurities, as mentioned below eq.~\eqref{asympflux}. However, in the equations of motion, eq.~\eqref{EOMS}, the gauge fields defined in eq.~\eqref{gaugedefs}, $\mathcal{A}_t^{\pm} \equiv \frac{1}{2}(a_t^0\pm \mathcal{A}_t)$, are more convenient, because $\mathcal{A}_t^-$ decouples from $\mathcal{A}_t^+$ and $\phi$, which then have equations of motion identical in form to those of the holographic single-impurity Kondo model of ref.~\cite{Erdmenger:2013dpa}. Crucially, however, the equations of motion ref.~\cite{Erdmenger:2013dpa} had $a_t^0$ in place of $\mathcal{A}_t^+$. Our model thus has very different boundary conditions from those of ref.~\cite{Erdmenger:2013dpa}: where ref.~\cite{Erdmenger:2013dpa} had only $a_t^0$ and $\phi$, and held fixed $Q^0$ and $\k=\a/\b$, we have $\mathcal{A}_t^-$, $\mathcal{A}_t^+$, and $\phi$, and hold fixed $\l$ in eq.~\eqref{lambdafixed}, $Q^0$, and $\k=\a/\b$. In particular, the boundary condition on $\mathcal{A}_t$,~\textit{i.e.}\ fixing $\lambda$ as in eq.~\eqref{lambdafixed}, involves not only $\a$, $\b$, and $\mathcal{Q}^+$, but also $\mathcal{Q}^-$, through $\mathcal{Q} = \mathcal{Q}^+-\mathcal{Q}^-$, and so effectively couples all of $\mathcal{A}_t^-$, $\mathcal{A}_t^+$ and $\phi$, as mentioned below eq.~\eqref{EOMS}. We will thus find different physics from that of ref.~\cite{Erdmenger:2013dpa} in section~\ref{phase}.

\subsection{RG Transformations}
\label{RGtransfos}

In this subsection we will check that the Kondo and RKKY couplings identified in the previous subsection behave correctly under RG transformations. Specifically, we will check that under an RG transformation the holographic Kondo coupling $\k$ in eq.~\eqref{holokondocoupling} is either marginally relevant (AFM, $\k<0$) or marginally irrelevant (FM, $\k>0$), while the holographic RKKY coupling $\lambda$ in eq.~\eqref{holoRKKY} has scaling dimension one. We will also precisely define the Kondo temperature, $T_{\K}$, from the RG running of a marginally relevant holographic Kondo coupling.

In holography, RG transformations can be thought of as Weyl transformations on the boundary, which can be implemented in the bulk by so-called Penrose-Brown-Henneaux (PBH) diffeomorphisms~\cite{Imbimbo:1999bj,Schwimmer:2000cu}. For asymptotic solutions that are independent of coordinates in field theory directions, in our case meaning the time coordinate, $t$, PBH diffeomorphisms reduce to translations of the radial coordinate:
\beq
\label{PBH}
r\longrightarrow r+\ln(L),
\eeq
where $L$ is an arbitrary renormalization length scale. The argument of the logarithm in eq.~\eqref{PBH} is made dimensionless by a factor of the $AdS_3$ radius, which we have set to unity. Such a PBH diffeomorphism changes the induced asymptotic metric: $-e^{2r} dt^2 \to -e^{2r}L^{-2} dt^2$. Sending $L\to0$ thus amounts to ``zooming in'' on the UV of the field theory, while sending $L \to \infty$ amounts to ``zooming out'' to the IR.

By performing the PBH diffeomorphism in eq.~\eqref{PBH}, we can derive the RG transformations of the asymptotic coefficients of the scalar and gauge field, and hence derive the RG transformations of our holographic Kondo and RKKY couplings. A scalar field such as our $\phi(r)$ is by definition invariant under any diffeomorphism. Invariance of $\phi(r)$ under the PBH diffeomorphism in eq.~\eqref{PBH} implies
\beq
\label{scalarPBH}
\a\rightarrow L^{1/2}\a, \qquad \b \to L^{1/2}\left(\b+\a\ln(L)\right).
\eeq
In radial gauge, $\ca_t^{\pm}$ are also invariant under the PBH diffeomorphism in eq.~\eqref{PBH}, hence
\begin{subequations}
\label{calaPBH}
\begin{alignat}{2}
&\cq^+ \rightarrow L^{-1}\cq^+,\\
&\m^+ \rightarrow \m^++\frac23\cq^+\left[\left((\ln(L))^2-3\ln(L)+6\right)\a^2-3(2-\ln(L))\a\b+3\b^2\right]\ln(L),\\
&\cq^- \rightarrow  L^{-1}\cq^-,\\
&\m^- \rightarrow \m^-.
\end{alignat}
\end{subequations}
Using the identifications of our holographic Kondo and RKKY couplings from the previous subsection, $\kappa = \a/\b$ and $\l$ in eq.~\eqref{lambdafixed}, the transformations in eqs.~\eqref{scalarPBH} and~\eqref{calaPBH} immediately give us the RG transformations
\begin{subequations}
\label{RG}
\beq
\label{kRG}
\k=\frac{\a}{\b} \quad \longrightarrow \quad \kappa(L) \equiv \frac{\k}{1+\k\ln(L)},
\eeq
\beq
\label{lRG}
\l=\frac{-\m-2\cq^+\le(2\a^2-2\a\b+\b^2-\frac{1}{3}\frac{\b^3}{\a}\ri)}{\cq} \quad \longrightarrow \quad L\l.
\eeq
\end{subequations}
Although $\m$ and $\b$ have inhomogeneous RG transformations, our holographic RKKY coupling $\l$ has a homogeneous RG transformation, thanks to the counterterms in eq.~\eqref{counterterms} that renormalize $\ca_t$'s canonical momentum $\p^t_{\ca_t}$ to the expression in eq.~\eqref{calamomdef}. Eq.~\eqref{kRG} thus shows that $\k$ is marginally relevant or irrelevant if in the UV $\k<0$ (AFM Kondo coupling) or $\k>0$ (FM Kondo coupling), respectively,\footnote{In contrast, if we had started with a Neumann boundary condition for $\phi$ instead of Dirichlet, then $\k$'s RG transformation would be
\be
\k \longrightarrow \frac{1+\k\ln(L)}{\k},
\ee
which is always marginally relevant, since $\k$ grows in the IR, $L\to\infty$, for both $\k<0$ and $\k>0$ in the UV.} while eq.~\eqref{lRG} shows that $\l$ is a relevant coupling of dimension one, as expected. Indeed, $\k$'s holographic beta function is $\b_{\k}=-L\pa_L\k(L)=\k^2(L)$, which is exact, \textit{i.e.} valid to all orders in $\k$, thanks to the large-$N$ limit~\cite{Witten:2001ua}. The beta function $\b_{\k}$ also shows that an AFM Kondo coupling is asymptotically free, and diverges in the IR at a dynamically-generated scale, which allows us to define the Kondo temperature, $T_{\K}$, as follows. Fixing the value of $\kappa$ at a fixed but arbitrary length scale $L'$, from eq.~\eqref{kRG} we find
\beq
\label{kappaLprime}
\kappa(L) = \frac{\kappa(L')}{1+\kappa(L')\ln (L/L')}.
\eeq
Clearly $\kappa(L)$ diverges (the denominator in eq.~\eqref{kappaLprime} vanishes) at the length scale $L' e^{-1/\kappa(L')}$, which is a physical quantity, being invariant under re-scalings of $L'$ by virtue of $\kappa$'s transformation in eq.~\eqref{kRG}. We then define the Kondo temperature as
\beq
\label{kondoTdef}
T_K\equiv \frac{1}{2 \pi} \, \frac{1}{L'}\, e^{1/\kappa(L')},
\eeq
 where the factor of $1/(2\pi)$ will be convenient in section~\ref{phase}.

\section{The Phase Diagram}
\label{phase}

In this section we determine the phase diagram of our holographic two-impurity Kondo model in the plane of $\l/(2\pi T_{\K})$ versus $T/T_{\K}$, with our holographic RKKY coupling $\lambda$ defined in eq.~\eqref{holoRKKY} and the Kondo temperature $T_{\K}$ defined in eq.~\eqref{kondoTdef}. To do so, we will solve the equations of motion for $\mathcal{A}_t^\pm$ and $\phi$, eqs.~\eqref{EOMa} and~\eqref{EOMphi} respectively (using the holographic radial coordinate $z$ of eq.~\eqref{ads3metric}, not the coordinate $r$ of eq.~\eqref{FG-gauge}), subject to the boundary conditions discussed in subsection~\ref{dtbc}.

We know one exact solution, which exists for all values of $Q^0$, $\l/(2\pi T_{\K})$, and $T/T_{\K}$, and obeys our boundary conditions, namely the trivial solution, $a_t^0=\mathcal{A}_t^++\mathcal{A}_t^-=Q^0/z+\mu^0$, with $\f(z)=0$ and $\mathcal{A}_t(z)=\mathcal{A}_t^+(z)-\mathcal{A}_t^-(z)=0$. The trivial solution is dual to a trivial state, with $\langle |\mathcal{O}_{\one}|\rangle = \langle |\mathcal{O}_{\two}|\rangle=0$ and $\langle R^1 \rangle= \langle R^2 \rangle = 0$, where via eq.~\eqref{RKKYdt} the latter implies $\langle S_{\one}^A S^A_{\two}\rangle=0$, so the two impurity spins are neither Kondo screened nor correlated with one another. To describe non-trivial states, with non-zero $\langle |\mathcal{O}_{\one}|\rangle = \langle |\mathcal{O}_{\two}|\rangle$ and/or non-zero $\langle R^1 \rangle =  \langle R^2 \rangle$, we must construct non-trivial solutions for $a_t^0$, $\mathcal{A}_t$, and $\phi$. Moreover, to determine whether a non-trivial state has lower free energy than the trivial state, and hence is thermodynamically preferred, we must determine whether the non-trivial solutions have smaller on-shell Euclidean action than the trivial solution. 

To solve the equations of motion, we need boundary conditions. The equations of motion are second order, so we need two boundary conditions for each field, $a_t^0$, $\mathcal{A}_t$ and $\phi$. As summarized at the end of subsection~\ref{dtbc}, fixing the size of the impurities' representation and the Kondo and RKKY couplings in the UV gives us three conditions at the $AdS_2$ boundary. We impose the remaining three conditions at the horizon, $z=z_H$. First, we demand regularity of $a_t^0$ as a one-form at the horizon, which requires $a_t^0(z_H)=0$. Second, we require regularity of $\phi$ at the horizon. In a near-horizon expansion of $\phi$, the leading modes are $\ln (z-z_H)$ and a constant. Regularity requires the coefficient of the $\ln (z-z_H)$ term to vanish, which in turn implies $\mathcal{A}_t^+(z_H)=0$. That implies, via the definitions in eq.~\eqref{gaugedefs}, $\mathcal{A}_t^-(z_H)=0$ and hence $\mathcal{A}_t(z_H)=0$. We will impose these boundary conditions in all that follows.

In our model, an analysis of linearized fluctuations about the trivial solution reveals an instability, indicating that a phase transition must occur. In fact, a straightforward exercise shows that the equation of motion for a fluctuation of $\phi$ about the trivial solution is identical, to linear order in the fluctuation, to that in the holographic single-impurity Kondo model of ref.~\cite{Erdmenger:2013dpa}. In ref.~\cite{Erdmenger:2013dpa}, an exact solution for that fluctuation of $\phi$, obeying the boundary conditions described above, was found, and was shown to become unstable at sufficiently low $T$: its amplitude grows rather than decays as a function of time, for any $\mathcal{Q}^+$, including $\mathcal{Q}^+=0$, and for any non-zero value of our holographic Kondo coupling, $\k$. As a result, in both the holographic single- and two-impurity Kondo models, a phase transition must occur. However, the linearized analysis tells us neither the order of the transition nor the transition temperature. Moreover, at linearized order the fluctuations of the scalar and the gauge field do not couple, so the linearized analysis tells us nothing about how the phase transition depends on $\l/(2\pi T_{\K})$. To determine these, we will construct non-trivial solutions and compare their on-shell Euclidean action to that of the trivial solution, as described above.

As mentioned below eq.~\eqref{EOMS}, we have only been able to construct non-trivial solutions numerically. However, in subsection~\ref{properties} we will constrain the properties of non-trivial solutions as much as we can without numerics. We then resort to numerics in subsection~\ref{numerics}, where we discuss our main numerical results, including the phase diagram of our model.

\subsection{Properties of Non-trivial Solutions}
\label{properties}

In the trivial solution, $\phi=0$ and hence $J_t^0=0$ in $a_t^0$'s equation of motion, eq.~\eqref{YM3} with $b=0$. In that case, $a_t^0$'s equation of motion requires the flux of $a_t^0$ to be constant in $z$. In other words, in the absence of charged sources, Gauss's law requires the flux to be a constant, fixed by the flux $Q^0$ at the $AdS_2$ boundary. Translating to the dual field theory, in the trivial state the impurities are unscreened, and in particular their representation in the IR is the same as in the UV, $\rho_{\textrm{IR}} = \rho_{\textrm{UV}}$. On the other hand, a non-trivial solution has $\phi \neq 0$ and $J_t^0\neq0$, so $a_t^0$'s flux will change as $z$ increases, as the charged matter field $\phi$ removes flux. A decreasing flux corresponds to screening of the impurities, \textit{i.e.}\ the dimension of their representation is smaller in the IR than in the UV, $\textrm{dim}(\rho_{\textrm{IR}}) < \textrm{dim}(\rho_{\textrm{UV}})$. In short, the flux of $a_t^0$ can only change if $\phi$ becomes non-trivial.

However, for our model we can prove that $\phi=0$ if and only if $a_t^1=a_t^2=0$, and conversely that $\phi \neq 0$ if and only if $a_t^1=a_t^2\neq0$, as follows. First suppose $\phi=0$. In that case, the equations of motion of all the gauge fields can be solved exactly, and in particular $\mathcal{A}_t = \mathcal{Q}/z + \mu$. The condition $\mathcal{A}_t(z_H)=0$ implies $\mu = - \mathcal{Q}/z_H$. However, when $\phi=0$ the boundary condition for the RKKY coupling reduces to $\mu = - \l \mathcal{Q}$. Clearly, in the generic case $\l \neq 1/z_H$, the only consistent solution has $\mathcal{Q} =0$ and $\mu=0$, and hence $\mathcal{A}_t=0$. Eqs.~\eqref{ansatz} and~\eqref{caladef} then imply $a_t^1 = a_t^2=0$. Conversely, suppose $\phi \neq 0$. In that case, the solution for $\mathcal{A}^+_t$ is not $\mathcal{Q}^+/z + \mu^+$, and so $\mu^+ \neq -\mathcal{Q}^+/z_H$. Using $\mathcal{A}_t^- = \mathcal{A}_t^+ - \mathcal{A}_t$ and $\mathcal{A}_t^-=\mathcal{Q}^-/z + \mu^-$, with $\mu^- = - \mathcal{Q}^-/z_H=(\mathcal{Q}-\mathcal{Q}^+)/z_H$, we find $(\mathcal{Q}-\mathcal{Q}^+)/z_H=\m^+-\m$. As a result, $\mu^+ \neq - \mathcal{Q}^+/z_H$ implies $\mu\neq -\mathcal{Q}/z_H$. That condition forbids $\mathcal{A}_t=0$, which has $\mu =0$ and $\mathcal{Q}=0$, and hence trivially has $\mu= -\mathcal{Q}/z_H$. We thus learn that $\phi \neq 0$ implies $\mathcal{A}_t\neq0$, and vice versa. We can also reverse the logic to infer that $\mathcal{A}_t=0$ implies $\phi=0$, completing our proof.

Translating to the field theory, we have learned that, for our model and within our ansatz, if either of $\mathcal{O}_{\one}=\mathcal{O}_{\two}$ or $R^1=R^2$ has vanishing expectation value, then so does the other, while if either acquires a non-zero expectation value, then so must the other. In other words, the absence of Kondo screening is always accompanied by the absence of correlations between the two impurity spins, and vice versa, while Kondo screening is always accompanied by non-zero correlations between the two impurity spins, and vice versa.

Our model thus admits two possible phases, distinguished by their symmetries. The RKKY interaction explicitly breaks the auxiliary $U(2)$ down to $U(1)^0 \times U(1)^3$, where $U(1)^0$ and $U(1)^3$ are generated by $\tau^0$ and $\tau^3$, respectively. The trivial solution preserves $U(1)^0\times U(1)^3$ and the Chern-Simons $U(1)$, dual to the electromagnetic $U(1)$, while a non-trivial solution breaks $U(1)^0 \times U(1)^3 \times U(1)$ to a subgroup. Specifically, both $\phi \neq 0$ and $\mathcal{A}_t\neq0$ break $U(1)^3$ completely, and $\phi \neq 0$ breaks $U(1)^0 \times U(1)$ to the diagonal. In field theory terms, the two possible phases in our system are the trivial phase, where $\langle |\mathcal{O}_{\one}|\rangle = \langle |\mathcal{O}_{\two}|\rangle=0$ and $\langle R^1 \rangle = \langle R^2 \rangle =0$, and so $U(1)^0 \times U(1)^3 \times U(1)$ is preserved, and the non-trivial phase, where $\langle |\mathcal{O}_{\one}|\rangle = \langle |\mathcal{O}_{\two}|\rangle\neq0$ and $\langle R^1 \rangle = \langle R^2 \rangle \neq0$ spontaneously break $U(1)^3$ completely and break $U(1)^0 \times U(1)$ to the diagonal.

These two phases are also distinguished by their phase shifts. As in the holographic single-impurity Kondo model of ref.~\cite{Erdmenger:2013dpa}, in our model the phase shift that accompanies Kondo screening appears holographically as a non-zero Wilson line of the Chern-Simons gauge field in the $x$ direction. Specifically, if we compactify the $x$ direction, then the phase shift is $\propto \oint_x A$. The equation for $A_x$, eq.~\eqref{CS2}, clearly shows that if $\mathcal{A}_t^+=0$ and $\phi=0$ then $A_x=0$, while if $\mathcal{A}_t^+\neq0$ and $\phi\neq0$, then $A_x\neq0$. Translating to the field theory, we find that in the trivial state no phase shift occurs, while in the non-trivial state a phase shift occurs. In fact, in both the holographic single-impurity Kondo model of ref.~\cite{Erdmenger:2013dpa} and in our model, the Chern-Simons gauge field's only role is to implement the phase shift. The Chern-Simons gauge field will play no further role in the remainder of this section.

We can constrain the possible phases in our model even further just using simple group theory arguments, as follows. We have two impurity spins, each in a totally anti-symmetric representation of $SU(N)$, whose Young tableau consists of a single column with $q$ boxes. We take $N \to \infty$ with $q/N$ of order one. Furthermore, we compute only the leading contributions to $\langle |\mathcal{O}_{\one}|\rangle = \langle |\mathcal{O}_{\two}|\rangle$ and $\langle R^1 \rangle = \langle R^2 \rangle$, which in the probe limit are order $N$, as shown in eqs.~\eqref{dirichletO} and~\eqref{calRvev}. In these limits the spin-spin correlator is thus, via eqs.~\eqref{RKKYdt} and~\eqref{calRvev},
\beq
\langle S^A_{\one} S^A_{\two} \rangle = -\frac{1}{2} N^2 \mathcal{Q}^2 + \mathcal{O}(N), \label{spinlimits}
\eeq
where in the large-$N$ counting, $\mathcal{Q}$ is order one. Eq.~\eqref{spinlimits} shows that we will have access only to the order $N^2$ contribution of AFM spin-spin correlations, $\langle S^A_{\one} S^A_{\two} \rangle<0$.

In fact we can show, using group theory alone, that at large $N$ the leading contribution to $\langle S^A_{\one} S^A_{\two} \rangle$ in FM eigenstates is always order $N$, and in AFM eigenstates is order $N^2$, and that the vast majority of eigenstates are in fact AFM. To our knowledge, the following results have never before appeared in the literature about the Kondo effect.

The tensor product of two identical anti-symmetric representations $\r_{\textrm{UV}}$, each with a Young tableau with $q$ boxes, is
\beq
\r_{\textrm{UV}} \otimes \r_{\textrm{UV}} = \sum_{p=0}^{p_{\textrm{max}}} \r_p, \label{tensor}
\eeq
where the irreducible representation $\r_p$ has a Young tableau with two columns, the first with $(q+p)$ boxes and the second with $(q-p)$ boxes, and where
\beq
p_{\textrm{max}} = \begin{cases} q, \qquad \qquad \,q\leq N/2 \\ N-q, \qquad q>N/2.\end{cases} \label{pmax}
\eeq
For a given representation $\r_p$, we can express $\langle S^A_{\one} S^A_{\two} \rangle$ in terms of the quadratic Casimir of that representation, $C(\r_p)$:
\bea
\left . \langle S^A_{\one} S^A_{\two} \rangle \right |_{\r_p} & = & \frac{1}{2} \left . \left( (S_{\one}+S_{\two})^A(S_{\one}+S_{\two})^A - S_{\one}^A S_{\one}^A - S_{\two}^A S_{\two}^A\right) \right |_{\r_p} \nn \\
& = & \frac{1}{2} C(\r_p)  - \frac{1}{2} \left . \left(S_{\one}^A S_{\one}^A + S_{\two}^A S_{\two}^A\right) \right |_{\r_p}.
\eea
Using~\cite{Barut:1986}
\beq
C(\r_p)= N(N+2) \frac{q}{N} \left(1-\frac{q}{N}\right) - p(p+1),
\eeq
as well as, for any of the $\rho_p$,
\beq
\langle S^A_{\one}S^A_{\one} \rangle = \frac{1}{2} (N+1)\, \chi_{\one \a}^{\dagger} \chi_{\one \alpha} \left(1 - \frac{\chi_{\one \a}^{\dagger} \chi_{\one \alpha}}{N}\right) = \frac{1}{2} N (N+1)\, \frac{q}{N} \left(1-\frac{q}{N}\right),
\eeq
and, for our identical impurity spins, $S^A_{\one}S^A_{\one}=S^A_{\two}S^A_{\two}$, we find
\beq
\left . \langle S^A_{\one} S^A_{\two} \rangle \right |_{\r_p} = \frac{1}{2} N \, \frac{q}{N} \left(1-\frac{q}{N}\right) - \frac{1}{2} p (p+1). \label{spincorrelator}
\eeq
Clearly, $\left . \langle S^A_{\one} S^A_{\two} \rangle \right |_{\r_p}$ decreases monotonically as $p$ increases. As a result, the FM ground state, which maximizes $\langle S^A_{\one} S^A_{\two} \rangle$, has $p=0$, while the AFM ground state, which minimizes $\langle S^A_{\one} S^A_{\two} \rangle$, has $p=p_{\textrm{max}}$. In fact, in the large-$N$ limit with $q/N$ of order one, eq.~\eqref{spincorrelator} shows that for any FM representation the leading contribution to $\langle S^A_{\one} S^A_{\two} \rangle$ is order $N$. Moreover, $\langle S^A_{\one} S^A_{\two} \rangle>0$ only for $p$ up to a critical value,
\beq
p_{\textrm{crit}} = \frac{1}{2} \left(\sqrt{1+4N \frac{q}{N}\left(1-\frac{q}{N}\right)}-1\right),
\eeq
which scales as $\sqrt{N}$ when $N\to \infty$ with $q/N$ of order one. The total number of representations in eq.~\eqref{tensor} scales as $N$ as $N\to\infty$ ($p_{\textrm{max}}$ scales as $N$), so only a small fraction of representations, of order $\sqrt{N}/N = 1/\sqrt{N}$, are FM. For an AFM ground state, using $p_{\textrm{max}}$ from eq.~\eqref{pmax}, we find
\beq
\left . \langle S^A_{\one} S^A_{\two} \rangle \right |_{\r_{p_{\textrm{max}}}} = \begin{cases} -\frac{1}{2}N(N+1)\left(\frac{q}{N}\right)^2, \qquad \quad \,\,\,\, q \leq N/2, \\ -\frac{1}{2}N(N+1)\left(1-\frac{q}{N}\right)^2, \qquad q>N/2, \end{cases}
\eeq
which clearly scales as $N^2$ as $N\to\infty$ with $q/N$ of order one. Moreover, inserting $p= P \, p_{\textrm{max}}$ with $0\leq P\leq 1$ into eq.~\eqref{spincorrelator} reveals that $\langle S^A_{\one} S^A_{\two}\rangle<0$ and $\langle S^A_{\one} S^A_{\two} \rangle$ scales linearly in $N$ when $N \to \infty$ with $q/N$ of order one only for the small fraction of eigenstates for which $P$ is order $1/\sqrt{N}$. In other words, for the vast majority of AFM eigenstates, $\langle S^A_{\one} S^A_{\two} \rangle$ scales as $N^2$ in the large-$N$ limit with $q/N$ of order one.

We have thus shown that, in the large-$N$ limit with $q/N$ order one, the leading contribution to $\langle S^A_{\one} S^A_{\two} \rangle$ is order $N$ in FM eigenstates and order $N^2$ in the vast majority of AFM eigenstates, and that the latter vastly outnumber the former.

In our holographic model, we thus expect to find non-trivial solutions only for AFM RKKY coupling, $\l/T_K>0$, and only the trivial solution for FM RKKY coupling, $\l/T_K<0$. In field theory terms, we expect to find Kondo screening, non-zero AFM spin-spin correlations of order $N^2$, and non-zero phase shift only for AFM RKKY coupling, and no Kondo screening, no spin-spin correlations, and no phase shift for FM RKKY coupling. More precisely, we will only be able to distinguish between superpositions of AFM eigenstates with $\langle S^A_{\one} S^A_{\two} \rangle \neq 0$ of order $N^2$, and uncorrelated spins, $\langle S^A_{\one} S^A_{\two} \rangle=0$.

The appearance of non-trivial solutions for only one sign of a double-trace coupling is in fact generic in large-$N$ field theory and in holography~\cite{Papadimitriou:2007sj,Klebanov:1999tb,Witten:2001ua}. In the field theory, a Legendre transform of the generating functional produces the quantum effective action, which is minimized by the ground state. Adding a double-trace coupling shifts the quantum effective potential, and generically will change the ground state only for one sign of the double-trace coupling constant, much the way a mass term added to a scalar field theory with quartic interaction will trigger scalar condensation only for negative mass-squared. To translate to the dual gravity theory, recall that the field theory generating functional is proportional to the gravity theory's on-shell action. The ground state is a regular solution of the gravity theory's equations of motion, and the double-trace coupling appears as a boundary term. Our field theory intuition thus suggests that non-trivial solutions will exist only for one sign of the double-trace coupling constant. Indeed, in the holographic single-impurity Kondo model of ref.~\cite{Erdmenger:2013dpa}, non-trivial solutions appeared only for AFM Kondo coupling, $\l_{\K}= -\k/(8N)>0$, as expected. We expect the same in our model: non-trivial solutions will exist only when both the Kondo and RKKY couplings are AFM.

To summarize, in our model and with our ansatz, we expect to find only two classes of solutions, dual to two distinct phases. The first class are the trivial solutions, with $\phi$, $a_t^1 = a_t^2$, and $A_x$ simultaneously vanishing, dual to a phase with no Kondo screening, no spin-spin correlations, and no phase shift. The second class are non-trivial solutions, which should exist only with AFM Kondo and RKKY couplings, $\l_{\K}=-\k/(8N)>0$ and $\l_{\RKKY}/T_{\K}=\l/(N T_{\K})>0$, with $\phi$, $a_t^1=a_t^2$, and $A_x$ simultaneously non-vanishing, dual to a phase with Kondo screening, AFM spin-spin correlations of order $N^2$, and a non-zero phase shift. Our numerical results of the following subsection will confirm these expectations.

Given all of the above, only one quantum phase transition is possible in our model, namely a transition from the trivial phase to the non-trivial phase. Although our numerical results will not extend down to exactly $T=0$, we will find highly suggestive evidence for such a quantum phase transition in our model.

In contrast, as reviewed in subsection~\ref{tikm} the original large-$N$ two-impurity Kondo model of refs.~\cite{PhysRevB.39.3415,Millis1990} exhibits a quantum phase transition as $\l_{\RKKY}/T_{\K}$ increases from $-\infty$ to $+\infty$, from a FM ground state, with Kondo screening and non-zero phase shift, to an AFM ground state, with neither Kondo screening nor a phase shift. The quantum phase transition was thus characterized by the phase shift: $\pi/2$ in the FM ground state, zero in the AFM ground state. However, in the original two-impurity Kondo model the large-$N$ limit is vector-like, which allows access to the order $N$ FM spin-spin correlations. Furthermore refs.~\cite{PhysRevB.39.3415,Millis1990} focused exclusively on the very special case that the two impurity spins were in totally anti-symmetric representations, each with a Young tableau of exactly $q=N/2$ boxes. In that special case, in the AFM RKKY limit, $\l_{\RKKY}/T_{\K}\to+\infty$, the two impurity spins lock into the anti-symmetric singlet of $SU(N)$, and hence neither Kondo screening nor a phase shift occurs. We instead employ a matrix-like large-$N$ limit, and a probe limit, so that we have access only to the order $N^2$ contribution to $\langle S^A_{\one} S^A_{\two} \rangle$, and hence only to AFM spin-spin correlations, as explained above. Moreover, our bottom-up model is too crude to allow fine-tuning the impurity spin representation to the special case that allows a singlet AFM ground state: generically, even with strong AFM RKKY coupling, our ground state will not be a singlet, hence some Kondo screening and non-zero phase shift will occur. In short, although our model and the original large-$N$ two-impurity Kondo model of refs.~\cite{PhysRevB.39.3415,Millis1990} are different, no contradiction exists: each model captures the Kondo and RKKY phenomena expected for its choice of parameters and in its respective limit.

\subsection{Numerical Results}
\label{numerics}

We now turn to the numerical solution of the equations of motion for $\mathcal{A}_t^{\pm}$ and $\phi$ in eq.~\eqref{EOMS}, and to the numerical evaluation of the on-shell Euclidean action for these solutions.

We first re-scale all dimensionful quantities by powers of $z_H = 1/(2\pi T)$, to obtain dimensionless coordinates and dimensionless fields,
\beq
(z/z_H,t/z_H)\rightarrow (z,t), \qquad z_H \mathcal{A}_t^{\pm}\rightarrow \mathcal{A}_t^{\pm}, \qquad \f\rightarrow \f. \label{rescalings}
\eeq
After these re-scalings, the $AdS_3$ boundary is at $z=0$ while the horizon is at $z=1$.

The asymptotic expansions of $\mathcal{A}_t^+$ and $\phi$ in eq.~\eqref{UV-asymptotics-nl} involve powers of $r$, which asymptotically approaches $- \ln(z)$, as shown in eq.~\eqref{rasymptotics}, so re-scaling $z$ as in eq.~\eqref{rescalings} will shift the values of the constant coefficients appearing in those asymptotic expansions. Specifically, after the re-scalings in eq.~\eqref{rescalings}, the dimensionless $\phi$ has an asymptotic expansion
\beq
\f=\sqrt{z}\le(\a_T\ln(z)+\b_T\ri)+\ldots
\eeq
where $\ldots$ represents terms sub-leading in $z$ as $z \to 0$, compared to the terms shown, and where $\a_T$ and $\b_T$ are related to the original $\a$ and $\b$ in eq.~\eqref{UV-asymptotics-nl} as
\beq
\label{aTbT}
\a_T\equiv\sqrt{z_H}\,\a, \qquad \b_T\equiv\sqrt{z_H}\le(\b+\a\ln(z_H)\ri).
\eeq
We then define $\kappa_T \equiv \a_T/\b_T$, which is related to $\kappa = \a/\b$ as
\beq
\kappa_T \equiv \frac{\a_T}{\b_T}=\frac{\kappa}{1+\kappa\ln(z_H)}. \label{readOffCouplings}
\eeq
Comparing eq.~\eqref{readOffCouplings} to eq.~\eqref{kRG} or~\eqref{kappaLprime} reveals that $\k_T$ is $\k(L)$ evaluated at the length scale $L=z_H = 1/(2\pi T)$. Using our definition of $T_{\K}$ in eq.~\eqref{kondoTdef}, we can write $\k_T$ as a function of $T/T_{\K}$,
\beq
\label{kappaTkondo}
\k_T=\frac{1}{\ln(T_{\K}/T)},
\eeq
whose simple form justifies the choice of the $1/(2\pi)$ factor in eq.~\eqref{kondoTdef}. The dimensionless $\mathcal{A}_t^+$ has an asymptotic expansion
\begin{subequations}
\beq
\label{calaplusTdef}
\mathcal{A}^+_t=\frac{\mathcal{Q}^+}{z}+\mathcal{Q}^+\le[c_T^{(3)}(\ln (z))^3+c_T^{(2)}(\ln (z))^2+c_T^{(1)}\ln (z)\ri]+\m_T^++\ldots,
\eeq
\beq
\m_T^+\equiv z_H\m^++\mathcal{Q}^+\le[c_T^{(3)}(\ln(z_H))^3-c_T^{(2)}(\ln(z_H))^2+c_T^{(1)}\ln(z_H)\ri],
\eeq
\beq
c_T^{(3)}\equiv\frac{2}{3}\a_T^2, \qquad c_T^{(2)}\equiv-2\a_T^2+2\a_T\b_T, \qquad c_T^{(1)}\equiv4\a_T^2-4\a_T\b_T+2\b_T^2,
\eeq
\end{subequations}
where $\ldots$ represents terms sub-leading in $z$ as $z \to 0$, compared to the terms shown. After the re-scalings in eq.~\eqref{rescalings}, the boundary condition $\l = \hat{\pi}_{\mathcal{A}_t}^t/(N\mathcal{Q})$ becomes
\begin{subequations}
\beq
\label{rescaledRKKY}
\frac{\l}{2\pi T}=\frac{1}{2\pi T}\frac{\hat{\pi}_{\mathcal{A}_t}^t}{N\mathcal{Q}}=-\frac{1}{\mathcal{Q}}\le[\mu_T+2\mathcal{Q}^+\le(2\a_T^2-2\a_T\b_T+\b_T^2-\frac{1}{3}\frac{\b_T^3}{\a_T}\ri)\ri],
\eeq
\beq
\mu_T \equiv \mu_T^+ - z_H \, \mu^+.
\eeq
\end{subequations}

As mentioned below eq.~\eqref{EOMS}, $\aminus$ decouples from $\aplus$ and $\f$, and indeed $\aminus$'s equation of motion, eq.~\eqref{EOMatilde}, is trivial, so we can solve for $\aminus$ exactly: the solution obeying the boundary condition $\mathcal{A}_t^-(z=1)=0$ is
\beq
\label{calaminusTdef}
\mathcal{A}_t^- = \mathcal{Q}^- \left(\frac{1}{z}-1\right).
\eeq
The solution for $\mathcal{A}^-$ is thus completely determined by $\mathcal{Q}^-$.

As also mentioned below eq.~\eqref{EOMS}, the equations of motion for $\aplus$ and $\phi$, eqs.~\eqref{EOMa} and~\eqref{EOMphi} are in fact identical in form to those in the holographic single-impurity Kondo model of ref.~\cite{Erdmenger:2013dpa}. However, as discussed at the end of subsection~\ref{dtbc}, the boundary conditions here are very different from those of ref.~\cite{Erdmenger:2013dpa}: where ref.~\cite{Erdmenger:2013dpa} had only $a_t^0$ and $\phi$, and held fixed $Q^0$ and $\k=\a/\b$, we have $\mathcal{A}_t^-$, $\mathcal{A}_t^+$, and $\phi$, and hold fixed $\l$ in eq.~\eqref{lambdafixed}, $Q^0$, and $\k=\a/\b$, which effectively couples all three fields. In practice, we obtain $\mathcal{Q}^{\pm}$, $\mathcal{Q}$, and $Q^1=Q^2$ by fixing $\l/(2\pi T)$, $Q^0$, and $\k_T$, solving for $\mathcal{A}_t^+$ and $\phi$, and then extracting $\mathcal{Q}^+$ from the solution, which gives us $\mathcal{Q}^-=Q^0-\mathcal{Q}^+$ and $\mathcal{Q}=2\mathcal{Q}^+-Q^0$, and from eq.~\eqref{caladef}, $Q^1=Q^2$.

We have been able to obtain only numerical solutions for $\aplus$ and $\f$. In principle, to obtain numerical solutions we could ``shoot from the boundary,'' that is, we could dial through values of $\l/(2\pi T)$, $Q^0$, and $\kappa_T$, numerically integrating the equations of motion, and retaining only those solutions satisfying the regularity conditions at the horizon, namely that $\aplus(z=1)=0$ and $\phi(z=1)$ is finite. In practice, however, such shooting is numerically costly: a generic numerical solution for $\phi$ will not be regular, and in particular will grow near the horizon due to the $\ln (z-1)$ term in $\phi$'s near-horizon expansion, mentioned above. Such behavior is in fact common for scalar fields in holography, and indeed occurs in the holographic single-impurity Kondo model of ref.~\cite{Erdmenger:2013dpa}. We will instead shoot from the horizon, since then we can demand that the coefficient of the $\ln(z-1)$ term in $\phi$'s near-horizon expansion vanishes from the start.

However, as mentioned in subsection~\ref{asymptotic}, our model has a major difference from the holographic single-impurity Kondo model of ref.~\cite{Erdmenger:2013dpa}. In the model of ref.~\cite{Erdmenger:2013dpa}, $M^2 = -1/4+(Q^0)^2$, where $Q^0$ was fixed from the start, that is, $Q^0$ was an \textit{input} for the calculation. In contrast, in our model $M^2 = -1/4 + (\mathcal{Q}^+)^2$, where $\mathcal{Q}^+$ will be an \textit{output} of the calculation. That poses a practical challenge in solving the equations of motion for $\aplus$ and $\phi$. The differential operator in $\phi$'s equation of motion depends on $\mathcal{Q}^+$, via $M^2 = -1/4 + (\mathcal{Q}^+)^2$, so we cannot solve $\phi$'s equation of motion until we know $\mathcal{Q}^+$, but we will not know $\mathcal{Q}^+$ until we have solved the equations of motion.

We address that challenge using the following numerical procedure. First, we fix $Q^0$, and then choose a target value for $\mathcal{Q}^+$. As explained above, we then know $\mathcal{Q}^-$ and $\mathcal{Q}$, and hence via eqs.~\eqref{calRdef},~\eqref{calRvev}, and~\eqref{spinlimits} we also know $\langle R^1 \rangle = \langle R^2 \rangle$ and $\langle S_{\one}^A S_{\two}^A\rangle$. Choosing $\mathcal{Q}^+$ also fixes $M^2 = -1/4+(\mathcal{Q}^+)^2$, so that $\phi$'s equation of motion is fixed. We then demand that the coefficient of the $\ln(z-1)$ term in $\phi$'s near-horizon expansion vanish and that $\aplus(z=1)=0$. Two free parameters then remain at the horizon, $\phi(z=1)\equiv \phi_H$ and $\partial_z \aplus(z=1)$. We fix $\phi_H$ and dial through $\partial_z \aplus(z=1)$ values, for each value obtaining numerical solutions for $\aplus$ and $\phi$, but retaining only those solutions with the target value of $\mathcal{Q}^+$. We then extract $\a_T$ and $\b_T$ from the asymptotics of $\phi$'s numerical solution, which gives us $\kappa_T = \a_T/\b_T$ and hence, after translating from $\beta_T$ to $\beta$ via eq.~\eqref{aTbT}, $\langle |\mathcal{O}_{\one}|\rangle = \langle |\mathcal{O}_{\two}|\rangle = - 4 N \beta$. We also extract $\mu_T$ from the solution for $\mathcal{A}_t = \aplus - \aminus$, using eqs.~\eqref{gaugedefs},~\eqref{calaplusTdef}, and~\eqref{calaminusTdef}, which gives us  $\l/(2\pi T)$ via eq.~\eqref{rescaledRKKY}. Using $T/T_{\K} = e^{-1/\kappa_T}$ from eq.~\eqref{kappaTkondo}, we then also find $\frac{T}{T_{\K}}\frac{\l}{2 \pi T}=\frac{\l}{2 \pi T_{\K}}$. We thus obtain all the one-point functions for a given point $(\l/(2 \pi T_{\K}),T/T_{\K})$ in the phase diagram. We then change $\phi_H$ and repeat the process of dialing through $\partial_z \aplus(z=1)$ values to obtain the same target value of $\mathcal{Q}^+$, but now obtaining new values of $T/T_{\K}$ and $\l/(2\pi T_{\K})$. In this way, we generate the phase diagram in the $(\l/(2 \pi T_{\K}),T/T_{\K})$ plane by moving along curves of constant $\mathcal{Q}^+$, or equivalently of constant $\mathcal{Q} = 2 \mathcal{Q}^+ - Q^0$, which via eqs.~\eqref{calRdef},~\eqref{calRvev}, and~\eqref{spinlimits} mean curves of constant $\langle R^1 \rangle = \langle R^2 \rangle$ and $\langle S_{\one}^A S_{\two}^A\rangle$.

We also numerically compute the renormalized free energy, $\mathcal{F}$, of each non-trivial solution. If we Wick-rotate to Euclidean signature and then compactify our dimensionless Euclidean time direction into a circle of circumference $2\pi$, then $\mathcal{F}$ is $T$ times the renormalized Euclidean on-shell action. All of our solutions are static, so we can always trivially perform the integration over the compact Euclidean time direction, producing an overall factor of $2 \pi$. As a result, $\mathcal{F}$ is simply $2 \pi T$ times an integral over $z$ in the Euclidean on-shell action, which we performed numerically, plus the boundary terms described in subsection~\ref{dtbc}. As in the holographic single-impurity Kondo model of ref.~\cite{Erdmenger:2013dpa}, in our model the Chern-Simons gauge field's contribution to $\mathcal{F}$ vanishes.

The trivial solution, $a_t^0=Q^0 (\frac{1}{z}-1)$ with all other fields vanishing, obeys all of our boundary conditions and exists everywhere in the $(\l/(2 \pi T_{\K}),T/T_{\K})$ plane. The trivial solution has $\mathcal{F}=-(Q^0)^2/2$. If we define $\Delta \mathcal{F}$ as $-(Q^0)^2/2$ minus the value of $\mathcal{F}$ for a non-trivial solution, then $\Delta \mathcal{F} > 0$ means the non-trivial solution is thermodynamically preferred over the trivial solution, and vice-versa for $\Delta \mathcal{F} < 0$.

However, our numerical results were not always sufficiently accurate to determine the sign of $\Delta \mathcal{F}$. After a large number of iterations of our numerical shooting, the change in our numerical results for $\a_T$, $\b_T$, and $\mu_T$ between iterations stabilized to roughly $10^{-7}$. Assuming the iterations were converging to the actual values, we thus took $10^{-7}$ as the uncertainty in our numerical results for $\a_T$, $\b_T$, and $\mu_T$. Numerically, we found that obtaining $\l/(2\pi T_{\K})$ of order one required $\mathcal{Q}^+ \geq 10^{-3}$. These two bounds together imply an uncertainty in our numerical results for $\l/(2\pi T_{\K})$ of roughly $10^{-4}$. The on-shell action includes a boundary term $\propto \l/(2\pi T_{\K})$, eq.~\eqref{RKKYboundary}, so our numerical results for $\Delta \mathcal{F}$ were also accurate up to a threshold of only $10^{-4}$. In some cases, our numerical result for $|\Delta \mathcal{F}|$ was less than $10^{-4}$, so that we could not determine the sign of $\Delta \mathcal{F}$, and hence not conclude whether the non-trivial solution was preferred over the trivial solution.

The equations of motion for $\aminus$, $\aplus$, and $\phi$ in eq.~\eqref{EOMS}, and the bulk integral over $z$ in $\Delta \mathcal{F}$, are invariant under three distinct $\mathbb{Z}_2$ symmetries, each of which reverses the overall sign of one field while leaving the other two fields invariant. The boundary terms in $\Delta \mathcal{F}$ are invariant under two of these $\mathbb{Z}_2$ symmetries, which we can thus use to restrict the ranges of the free parameters in our numerical analysis, without loss of generality. Taking $\phi \to -\phi$, leaving $\aminus$ and $\aplus$ unchanged, will take $\phi_H \to -\phi_H$, $\a_T \to -\a_T$, and $\b_T \to -\b_T$. However, $\kappa_T \equiv \a_T/\b_T$ will be invariant, as will $\l/(2\pi T)$ in eq.~\eqref{rescaledRKKY}. As a result, the boundary terms in $\Delta \mathcal{F}$ will also be invariant. We thus restricted to $\phi_H >0$ in our numerical solutions, without loss of generality. Similarly, taking $\aminus \to -\aminus$ and simultaneously $\aplus \to -\aplus$, while leaving $\phi$ unchanged, sends $\mu_T \to -\mu_T$, $\mathcal{Q}^- \to -\mathcal{Q}^-$, and $\mathcal{Q}^+ \to -\mathcal{Q}^+$, and hence $Q^0 \to -Q^0$ and $\mathcal{Q} \to - \mathcal{Q}$, with $\a_T$ and $\b_T$ unchanged. Again, $\kappa_T$ and $\l/(2\pi T)$ will both be unchanged, and hence the boundary terms in $\Delta \mathcal{F}$ will be unchanged. We thus also restricted to $Q^0<0$ in our numerical solutions, without loss of generality.

\begin{figure}
\begin{center}
\begin{tabular}{cc}
\includegraphics[width=.48\textwidth]{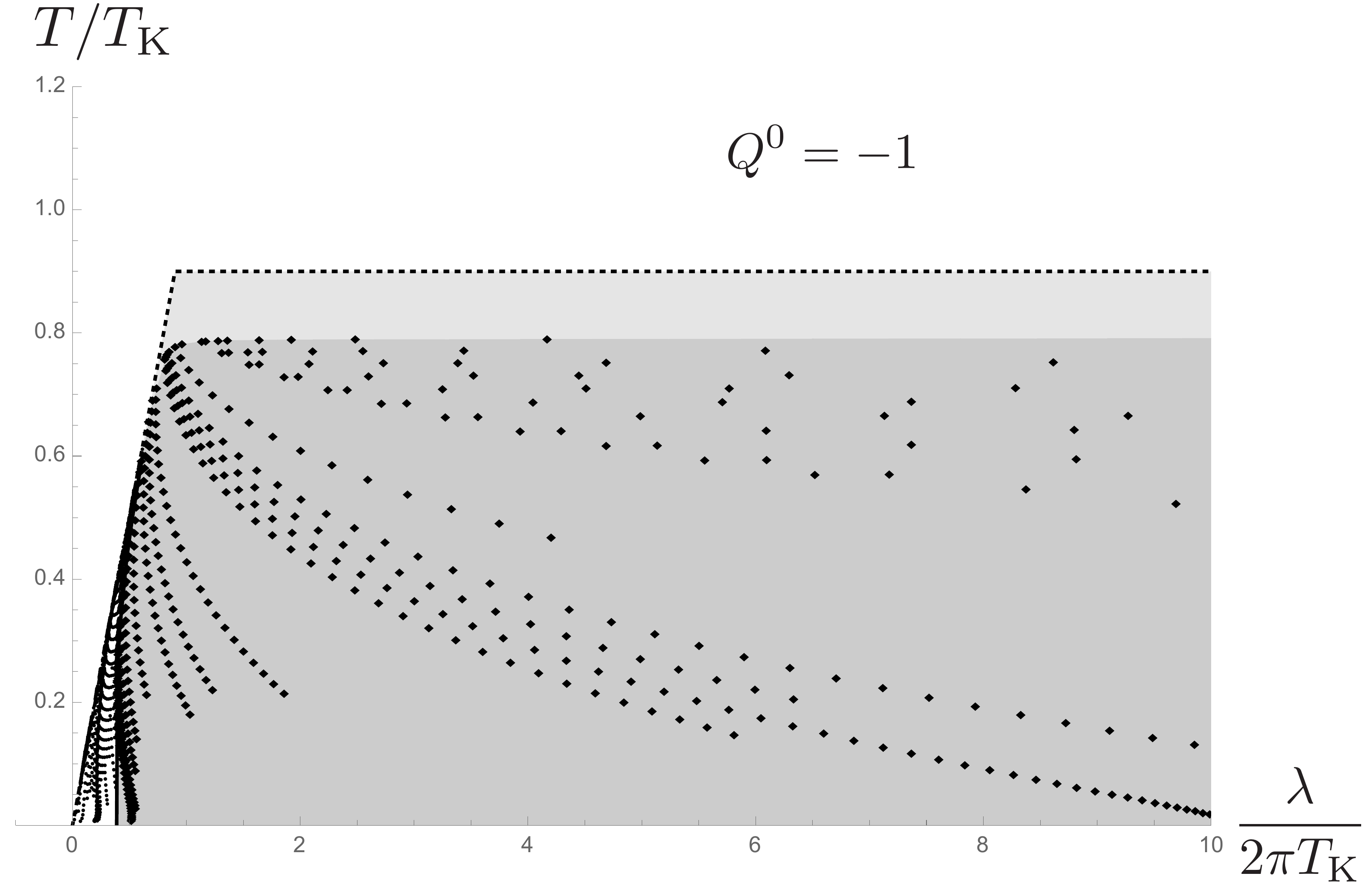} & \includegraphics[width=0.48\textwidth]{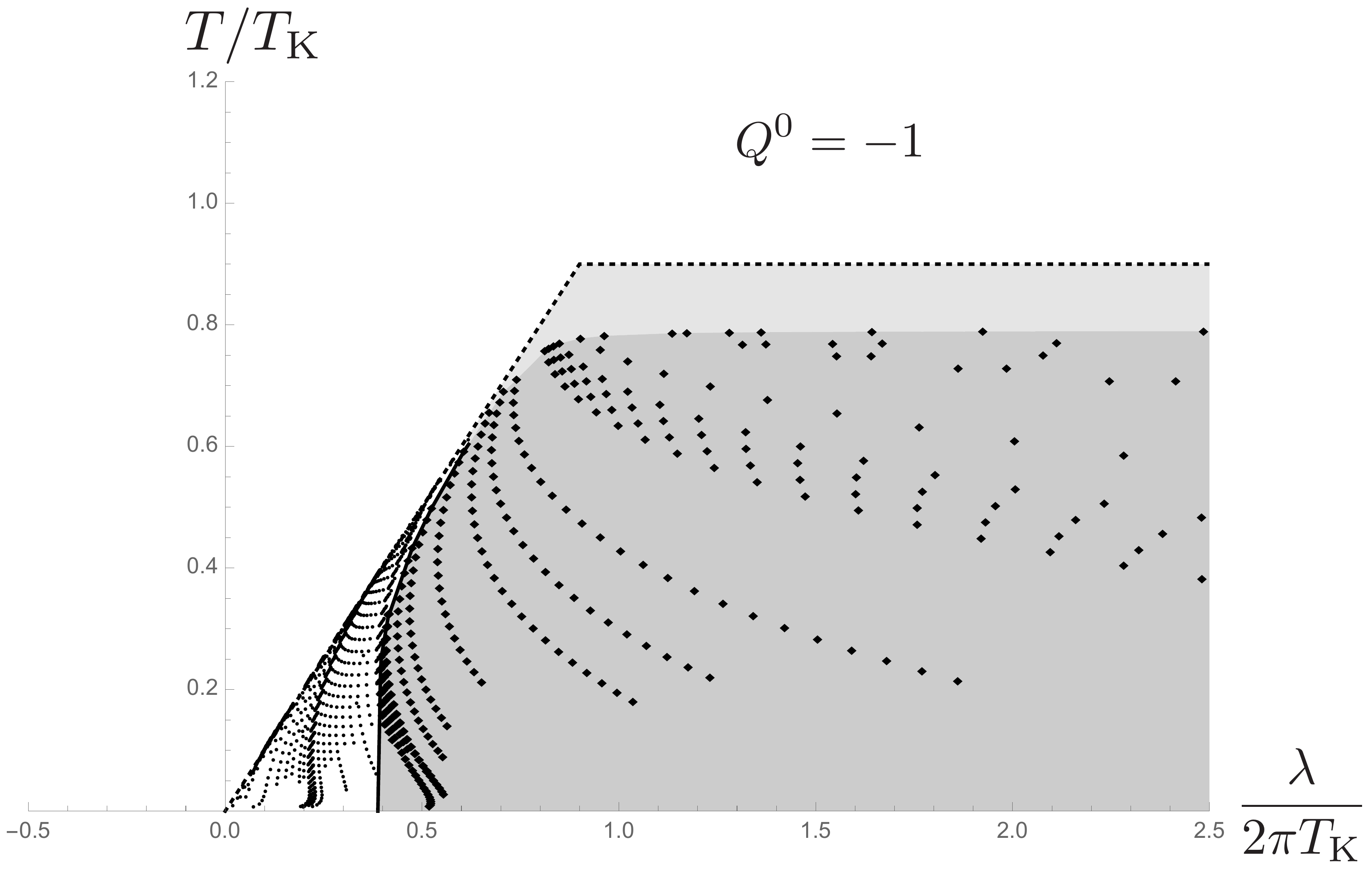} \\ (a.) & (b.) \\
\includegraphics[width=.48\textwidth]{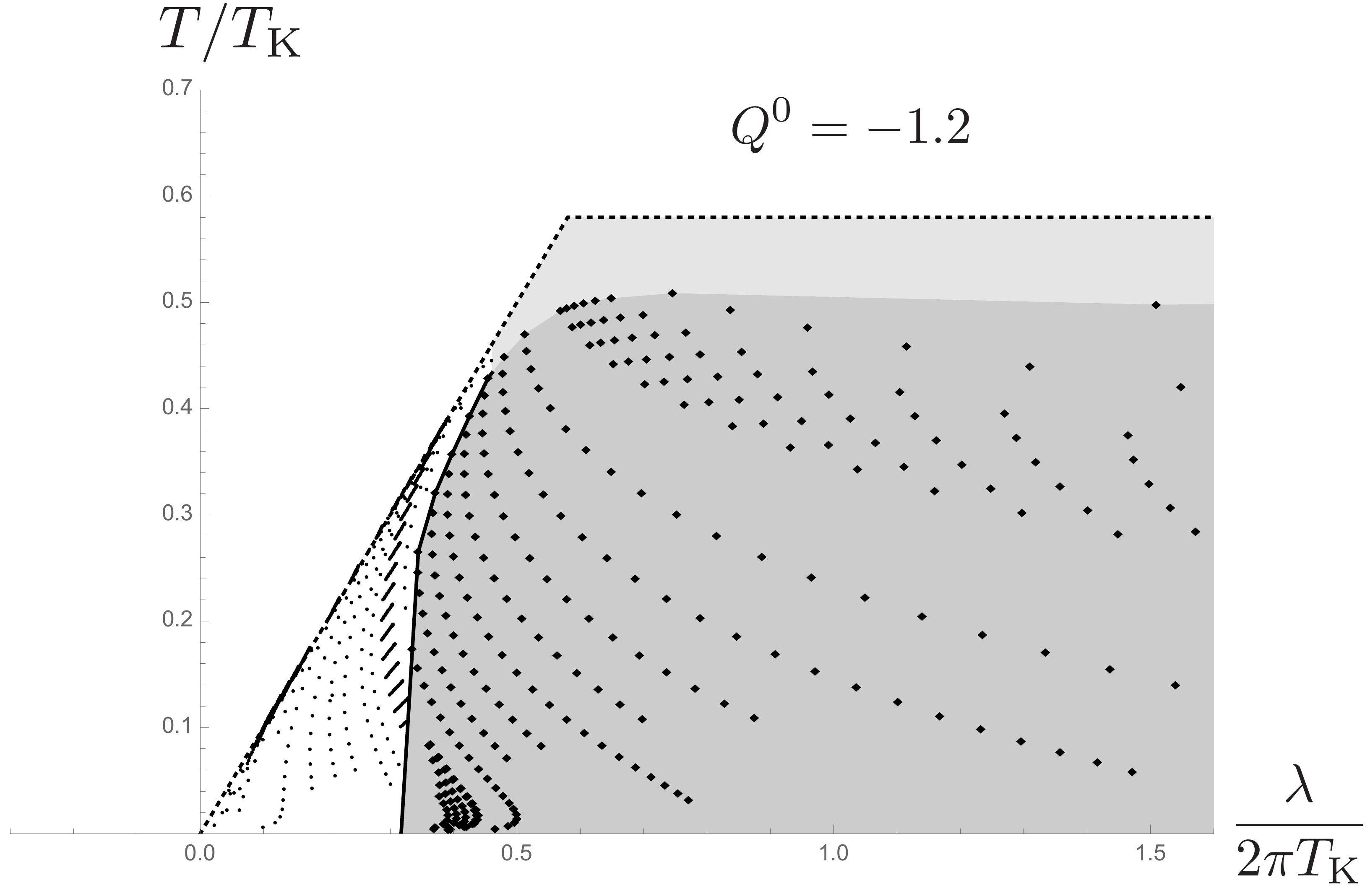} & \includegraphics[width=.48\textwidth]{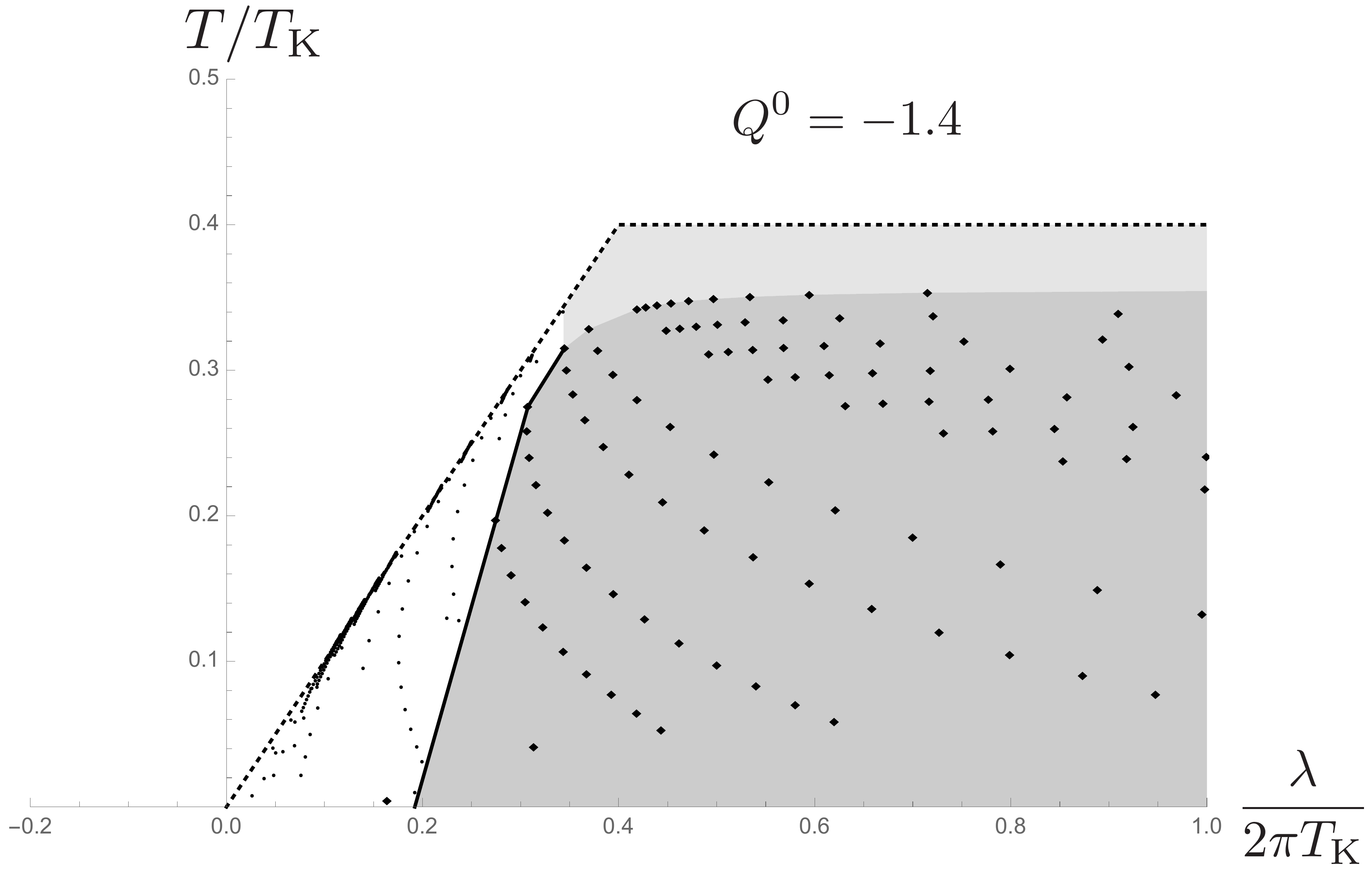} \\ (c.) & (d.)
\end{tabular}
\caption{\label{fig:phase} Phase diagrams of our model, in the plane of $\l/(2\pi T_{\K})$ versus $T/T_{\K}$, for (a.) $Q^0=-1$ with $\l/(2\pi T_{\K}) \in (-0.5,10)$, (b.) $Q^0=-1$ again, but zooming in on $\l/(2\pi T_{\K}) \in (-0.5,2.5)$ to reveal greater detail, (c.) $Q^0=-1.2$ with $\l/(2\pi T_{\K}) \in (-0.3,1.6)$, and (d.) $Q^0=-1.4$ with $\l/(2\pi T_{\K}) \in (-0.2,1.0)$. In all four figures, each black dot represents a non-trivial numerical solution, which we obtained only inside the regions bounded by the dotted lines. In the light gray region, within our numerical accuracy threshold we could not determine whether the non-trivial solutions were thermodynamically preferred over the trivial solution. In the dark gray region, the non-trivial solutions were thermodynamically preferred over the trivial solution, while in the white region, they were not. In each phase diagram, a solid black line denotes the boundary between the dark gray and white regions. For $Q^0=-1.2$ and $Q^0=-1.4$ we also determined, but have not shown, the phase diagram for values of $\l/(2\pi T_{\K})$ up to four times larger than those shown above. In each case we found that, the light and dark gray regions simply extended all the way to the highest $\l/(2\pi T_{\K})$ values that we considered, similar to the $Q^0=-1$ case in (a.).}
\end{center}
\end{figure}

Our main result is fig.~\ref{fig:phase}, the phase diagrams of our model in the $(\l/(2 \pi T_{\K}),T/T_{\K})$ plane, for $Q^0=-1$, $-1.2$, and $-1.4$. In fig.~\ref{fig:phase}, each black dot represents a non-trivial numerical solution. As anticipated in subsection~\ref{properties}, every non-trivial solution we found had both $\phi \neq 0$ and $a_t^1 = a_t^2 \neq 0$. As also anticipated in subsection~\ref{properties} we found non-trivial solutions only for AFM RKKY coupling, $\l/(2\pi T_{\K})>0$. In fact, we found non-trivial solutions only inside the region bounded by the dotted lines in each of figs~\ref{fig:phase} (a.), (b.), (c.), and (d.). In each case, the dotted diagonal line, emanating from the origin, is $T/T_{\K} = \l/(2\pi T_{\K})$, while the horizontal dotted line was determined by a linearized stability analysis, as follows.

As mentioned above, in our model a fluctuation of $\phi$ about the trivial solution obeys the same equation of motion and boundary conditions as in the holographic single-impurity Kondo model of ref.~\cite{Erdmenger:2013dpa}, but with $\aplus$ replacing $a_t^0$. A central result of ref.~\cite{Erdmenger:2013dpa} was that the fluctuation of $\phi$ became unstable, with amplitude growing rather than decaying in time, below a critical $T$ that depended on $Q^0$ (implicitly, through a critical value of $\kappa_T$ in eq.~(4.32) of ref.~\cite{Erdmenger:2013dpa}). Moreover, that critical $T$ decreased as $|Q^0|$ increased, intuitively because an impurity spin in a larger-dimensional representation is ``more difficult'' to screen. The linearized analysis thus guaranteed that a phase transition must occur. The same result applies in our model, but with $\mathcal{Q}^+$ replacing $Q^0$. However, as mentioned above, the instability of the fluctuation of $\phi$ is independent of $\l/(2 \pi T_{\K})$, so the linearized analysis tells us nothing about the values of $\l/(2 \pi T_{\K})$ where the phase transition will occur. Given $\mathcal{Q}^+ \equiv \frac{1}{2} \left(Q^0 + \mathcal{Q}\right)$, the minimal value of $|\mathcal{Q}^+|$ is $|Q^0/2|$, which determines the maximal critical $T/T_{\K}$ for the instability, that is, the highest $T/T_{\K}$ at which the linearized fluctuation becomes unstable. In each of figs.~\ref{fig:phase} (a.), (b.), (c.), and (d.), the horizontal dotted lines denote that maximal critical $T/T_{\K}$. As we increase $|Q^0|$, going from fig.~\ref{fig:phase} (c.) to fig.~\ref{fig:phase} (d.), the horizontal dotted line moves to smaller $T/T_{\K}$, as expected. Fig.~\ref{fig:phase} shows that for sufficiently large $\l/(2 \pi T_{\K})$, to the right of the diagonal dotted line, non-trivial solutions appear at the maximal critical $T/T_{\K}$, that is, as soon as the instability occurs. In contrast, for smaller $\l/(2 \pi T_{\K})$, and specifically along the diagonal dotted line, non-trivial solutions only appear at $T/T_{\K}$ below the maximal critical value.

In each phase diagram in fig.~\ref{fig:phase}, we have divided the region bounded by the dotted lines into three sub-regions, coded by shading: light gray, dark gray, and white, with a solid black line separating the dark gray and white regions. In the light gray regions, we found non-trivial solutions, but $|\Delta \mathcal{F}|$ was smaller than our numerical accuracy threshold of $10^{-4}$, hence we could not conclude whether the non-trivial solution was thermodynamically preferred over the trivial solution. In the dark gray region, the non-trivial solutions had $|\Delta \mathcal{F}|>10^{-4}$ and $\Delta \mathcal{F}>0$, so the non-trivial solution was thermodynamically preferred over the trivial solution. In the white region, $|\Delta \mathcal{F}|>10^{-4}$ but $\Delta \mathcal{F}<0$, so the non-trivial solution was not thermodynamically preferred over the trivial solution.

The solid black line separating the dark gray and white regions is an interpolation between thermodynamically preferred solutions: to the left of that line, towards smaller $\l/(2 \pi T_{\K})$, the next nearest numerical solution that we obtained was not thermodynamically preferred. The actual boundary between thermodynamically favored and dis-favored solutions is thus either at the solid black line, or somewhere between the solid black line and the first black dots to its left.

In any of the phase diagrams in fig.~\ref{fig:phase}, imagine fixing $\l/(2 \pi T_{\K})$ and reducing $T/T_{\K}$, that is, imagine moving down along a vertical line in the phase diagram. Our results demonstrate that, for sufficiently large AFM $\l/(2 \pi T_{\K})$, a phase transition will occur, from the trivial state, with no Kondo screening, zero spin-spin correlations, and zero phase shift, to the non-trivial state, with Kondo screening, non-zero AFM spin-spin correlations of order $N^2$, and non-zero phase shift. As discussed in subsection~\ref{properties}, in our model and with our ansatz, these are the only two possible states.

The order of these phase transitions depends on $\l/(2 \pi T_{\K})$. For example, in one of the phase diagrams in fig.~\ref{fig:phase}, suppose we fix $\l/(2 \pi T_{\K})$ and reduce $T/T_{\K}$, moving down along a vertical line, such that we hit the diagonal dotted line. In that case, as we reduce $T/T_{\K}$ non-trivial solutions appear at the diagonal dotted line, but are not thermodynamically preferred. The non-trivial solutions become thermodynamically preferred only at the critical $T/T_{\K}$ where our vertical line hits the dark gray region. As a result, as $T/T_{\K}$ drops below the critical $T/T_{\K}$, all one-point functions and the phase shift will jump from zero to non-zero values, indicating a first-order transition. Suppose we then increase $\l/(2 \pi T_{\K})$ and repeat the process, such that now as we reduce $T/T_{\K}$ we hit the horizontal dotted line. In that case, although we cannot say for certain due to the limitations of our numerical accuracy, our numerical results are consistent with the non-trivial solutions being thermodynamically preferred as soon as they appear. In those cases, we expect the one-point functions and phase shift to increase smoothly from zero, indicating a continuous transition.

\begin{figure}
\begin{center}
\begin{tabular}{c @{\hspace{0.3in}} c}
\includegraphics[width=.48\textwidth]{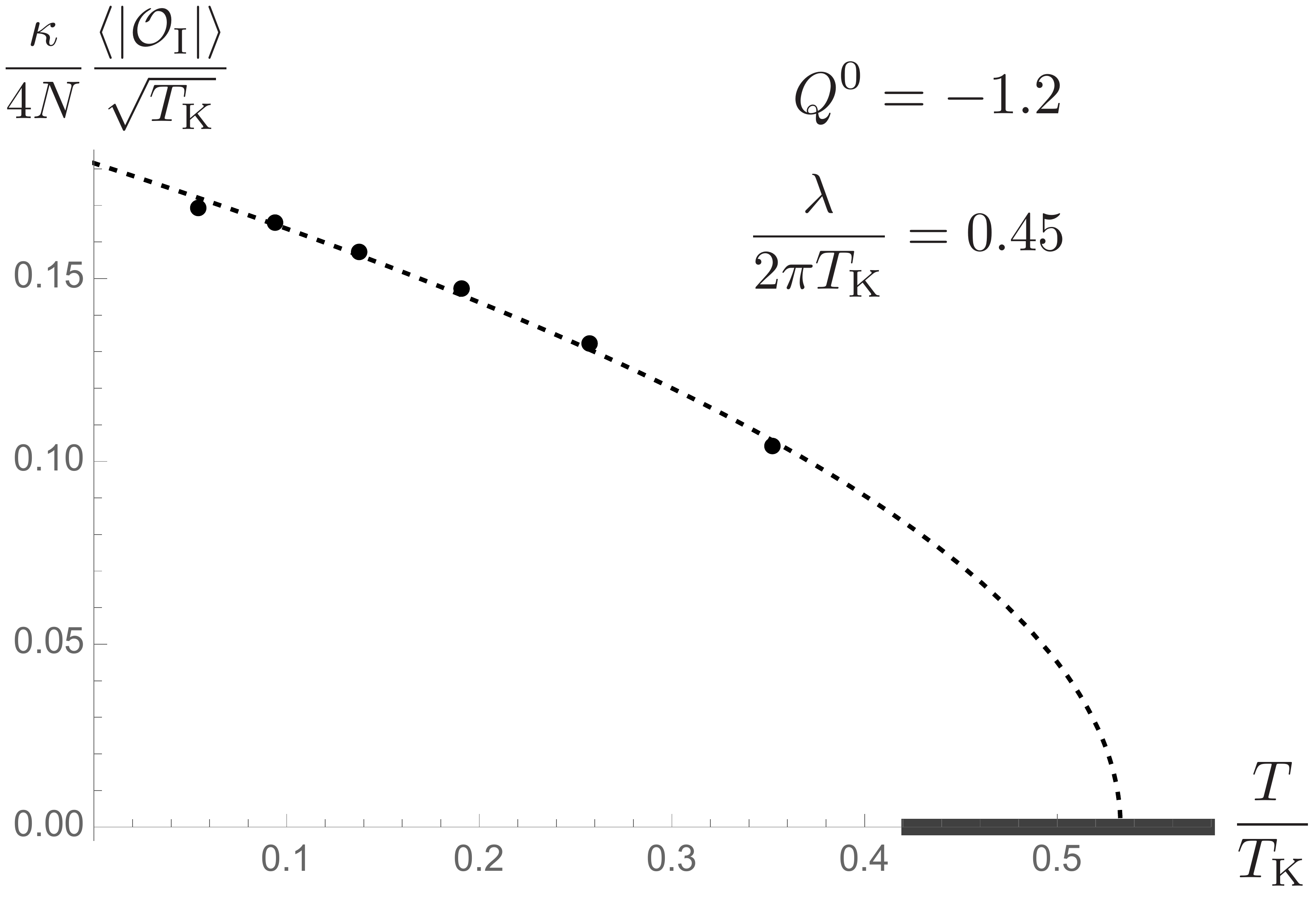}
&
\includegraphics[width=.48\textwidth]{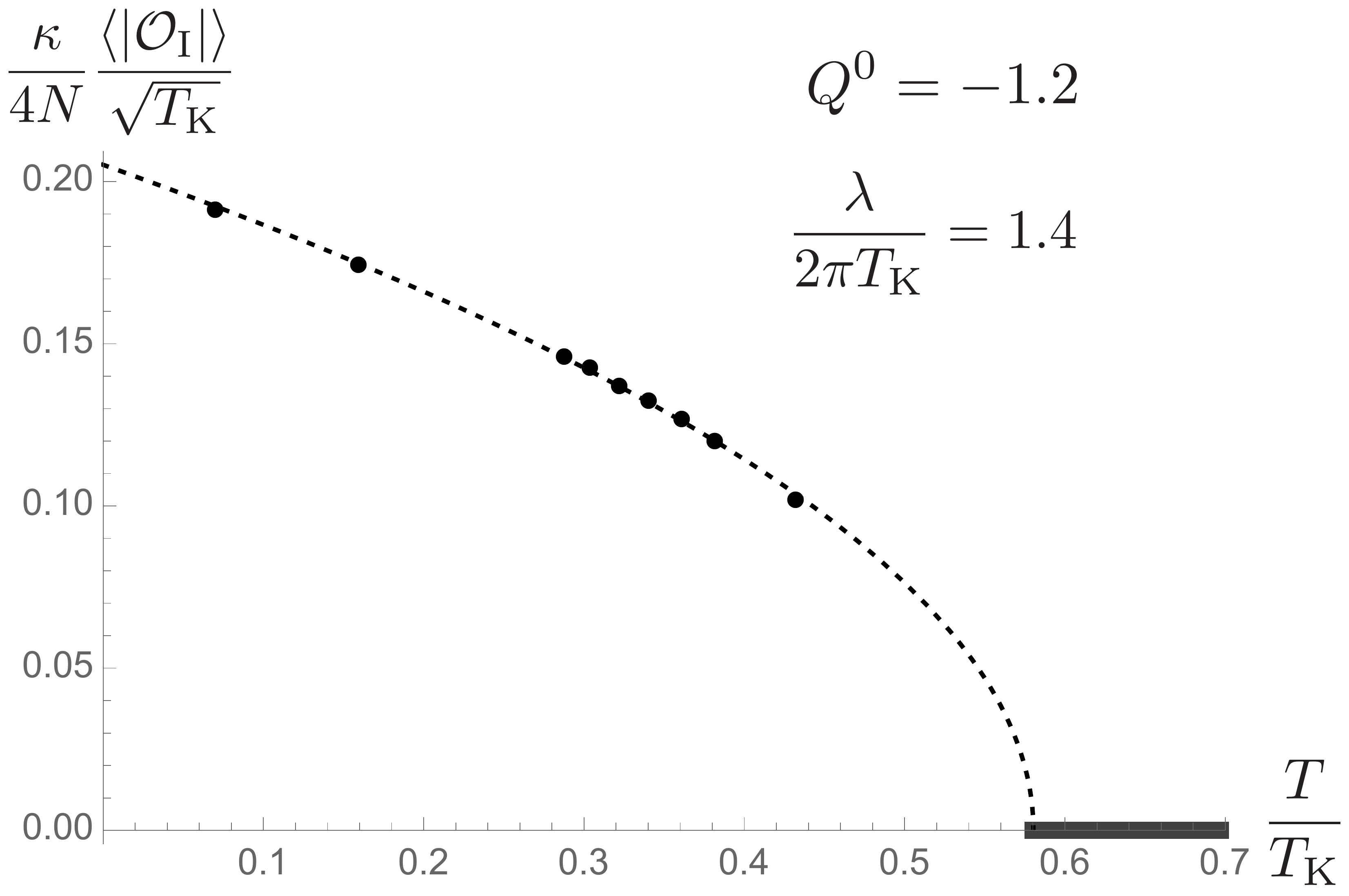} \\ (a.) & (b.)
\end{tabular}
\caption{\label{fig:ovev}  Our numerical results for $\frac{\kappa}{4N}\frac{\langle |\mathcal{O}_{\one}| \rangle}{\sqrt{T_{\K}}} = - \a_T \sqrt{2 \pi T/T_{\K}}$ as a function of $T/T_{\K}$ for $Q^0=-1.2$, with (a.) $\l/(2\pi T_{\K})=0.45$, and (b.) $\l/(2\pi T_{\K})=1.4$. In each figure, the black dots are our numerical data, the dotted line is a numerical fit to our data of the form in a mean-field second-order transition, with critical exponent $1/2$, and the heavy gray line segment at $\langle |\mathcal{O}_{\one}| \rangle=0$ represents the trivial solution, for $T/T_{\K}$ values where the trivial solution is thermodynamically preferred, as far as we can determine within our numerical accuracy. As we decrease $T/T_{\K}$, a phase transition occurs approximately where the heavy gray line segment ends. Our results suggest a first-order transition in (a.) and a second-order mean-field transition in (b.), consistent with our expectations from fig.~\ref{fig:phase} (c.).}
\end{center}
\end{figure}

Our results for the one-point functions are consistent with such an interpretation. Fig.~\ref{fig:ovev} shows our numerical results for $\frac{\kappa}{4N}\frac{\langle |\mathcal{O}_{\one}| \rangle}{\sqrt{T_{\K}}} = - \a_T \sqrt{2 \pi T/T_{\K}}$ as a function of $T/T_{\K}$ for $Q^0=-1.2$, with $\l/(2\pi T_{\K})=0.45$ in fig.~\ref{fig:ovev} (a.) and $\l/(2\pi T_{\K})=1.4$ in fig.~\ref{fig:ovev} (b.). In each figure, each black dot represents a non-trivial numerical solution, the dotted curve is a numerical fit to our data of the form in a mean-field second-order transition, with critical exponent $1/2$, and the heavy gray line segment at $\langle |\mathcal{O}_{\one}| \rangle=0$ represents the trivial solution, for $T/T_{\K}$ values where the non-trivial solution is preferred, as far as we can determine within our numerical accuracy. In other words, in fig.~\ref{fig:ovev} (a.) the heavy gray line segment extends to the value of $T/T_{\K}$ of the solid black line at $\l/(2\pi T_{\K})=0.45$ in fig.~\ref{fig:phase} (c.), while in fig.~\ref{fig:ovev} (b.) the heavy gray line segment extends down to the $T/T_{\K}$ value of the horizontal dotted line in fig.~\ref{fig:phase} (c.). For $\l/(2\pi T_{\K})=0.45$, the phase diagram in fig.~\ref{fig:phase} (c.) suggests a first-order transition, and indeed fig.~\ref{fig:ovev} (a.) suggests that the transition cannot be continuous: the fit to our data suggests that $\langle | \mathcal{O}_{\one} | \rangle$ jumps $\langle | \mathcal{O}_{\one} | \rangle=0$ to $\langle | \mathcal{O}_{\one} | \rangle \neq 0$ when the transition occurs. On the other hand, for $\l/(2\pi T_{\K})=1.4$, the phase diagram in fig.~\ref{fig:phase} (c.) suggests a continuous transition, as indeed implied by fig.~\ref{fig:ovev} (b.): the fit to our data suggests that $\langle | \mathcal{O}_{\one} | \rangle$ may in fact rise smoothly from zero starting at the transition, with second-order mean-field exponent.

\begin{figure}
\begin{center}
\includegraphics[width=1\textwidth]{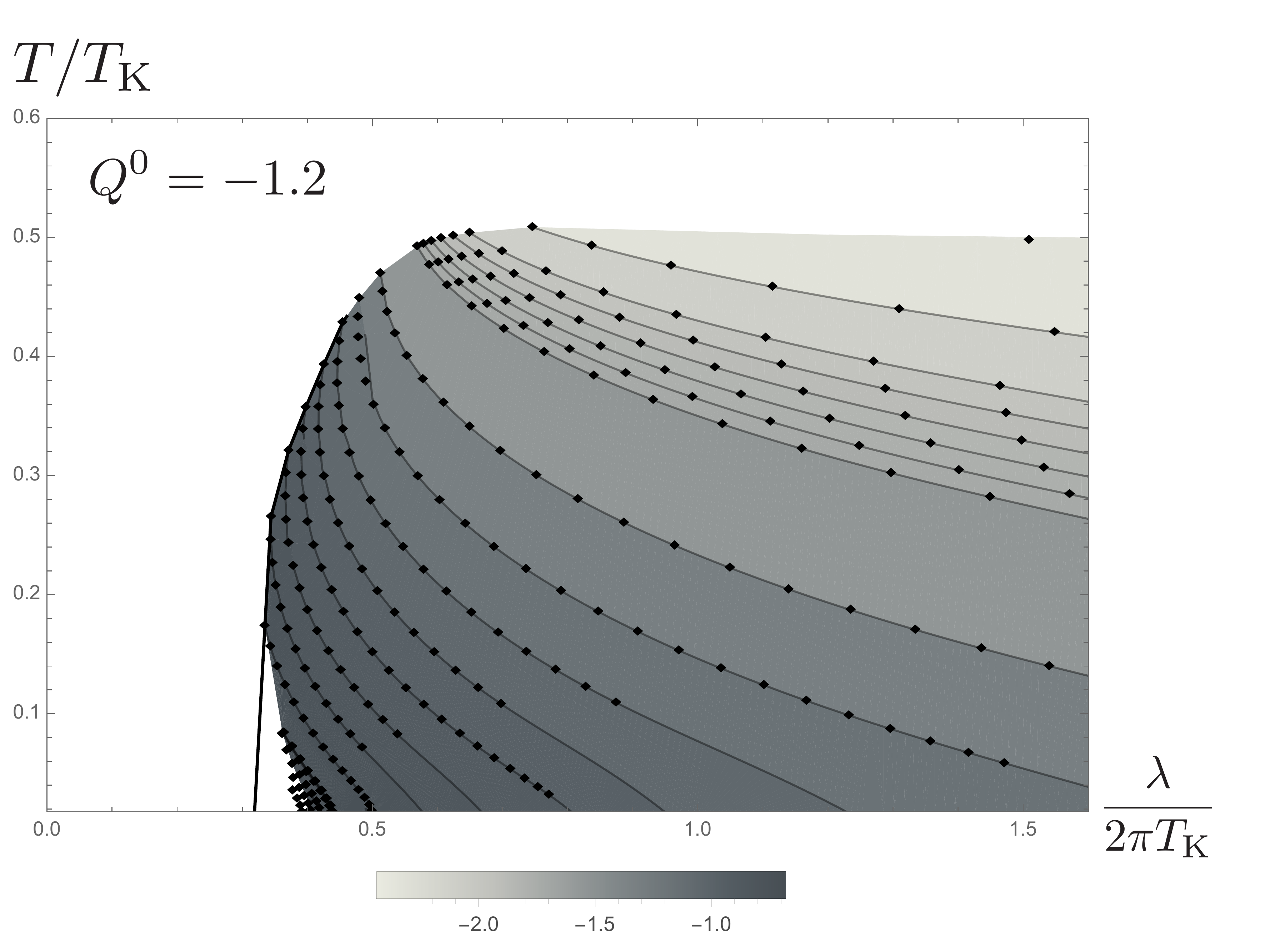}
\caption{\label{fig:contour} Contour plot of $\log_{10}(|\langle\mathcal{R}\rangle|/N)=\log_{10} (|\mathcal{Q}|)$ in the plane of $\l/(2 \pi T_{\K})$ versus $T/T_{\K}$, for $Q^0=-1.2$ and for thermodynamically preferred solutions only, \textit{i.e.}\ for solutions in the dark gray region of the corresponding phase diagram, fig.~\ref{fig:phase} (c.). Clearly spin-spin correlations, as measured by $|\langle\mathcal{R}\rangle|$, grow as $T/T_{\K}$ decreases. Moreover, for decreasing $T/T_{\K}$, our numerical results are consistent with a first order phase transition for $0.3 \lesssim \l/(2 \pi T_{\K}) \lesssim 0.6$, in which $|\langle\mathcal{R}\rangle|$ jumps from zero to non-zero values, and a continuous transition for $\l/(2 \pi T_{\K})\gtrsim 0.6$, in which $|\langle\mathcal{R}\rangle|$ rises smoothly from zero. These results are consistent with our expectations from the corresponding phase diagram, fig.~\ref{fig:phase} (c.).}
\end{center}
\end{figure}

Fig.~\ref{fig:contour} is a contour plot of our numerical results for $\log_{10}(|\langle\mathcal{R}\rangle|/N)=\log_{10} (|\mathcal{Q}|)$ in the plane of $\l/(2 \pi T_{\K})$ versus $T/T_{\K}$, for $Q^0=-1.2$ and for thermodynamically preferred solutions only, \textit{i.e.}\ for solutions in the dark gray region of the corresponding phase diagram, fig.~\ref{fig:phase} (c.). In fig.~\ref{fig:contour}, the black dots again represent non-trivial numerical solutions, which we generated along curves of constant $\mathcal{Q}$, as explained above. Fig.~\ref{fig:contour} makes clear that spin-spin correlations, as measured by $|\langle\mathcal{R}\rangle|$, grow as $T/T_{\K}$ decreases, as expected. Moreover, fig.~\ref{fig:contour} suggests that as $T/T_{\K}$ decreases through the critical value, the transition will most likely be first-order for $0.3 \lesssim \l/(2 \pi T_{\K}) \lesssim 0.6$, with $|\langle\mathcal{R}\rangle|$ jumping from zero to non-zero values, and continuous for $\l/(2 \pi T_{\K})\gtrsim 0.6$, with $|\langle\mathcal{R}\rangle|$ rising smoothly from zero, all of which is consistent with our expectations from the corresponding phase diagram, fig.~\ref{fig:phase} (c.).

\begin{figure}
\begin{center}
\begin{tabular}{c @{\hspace{0.3in}} c}
\includegraphics[width=.48\textwidth]{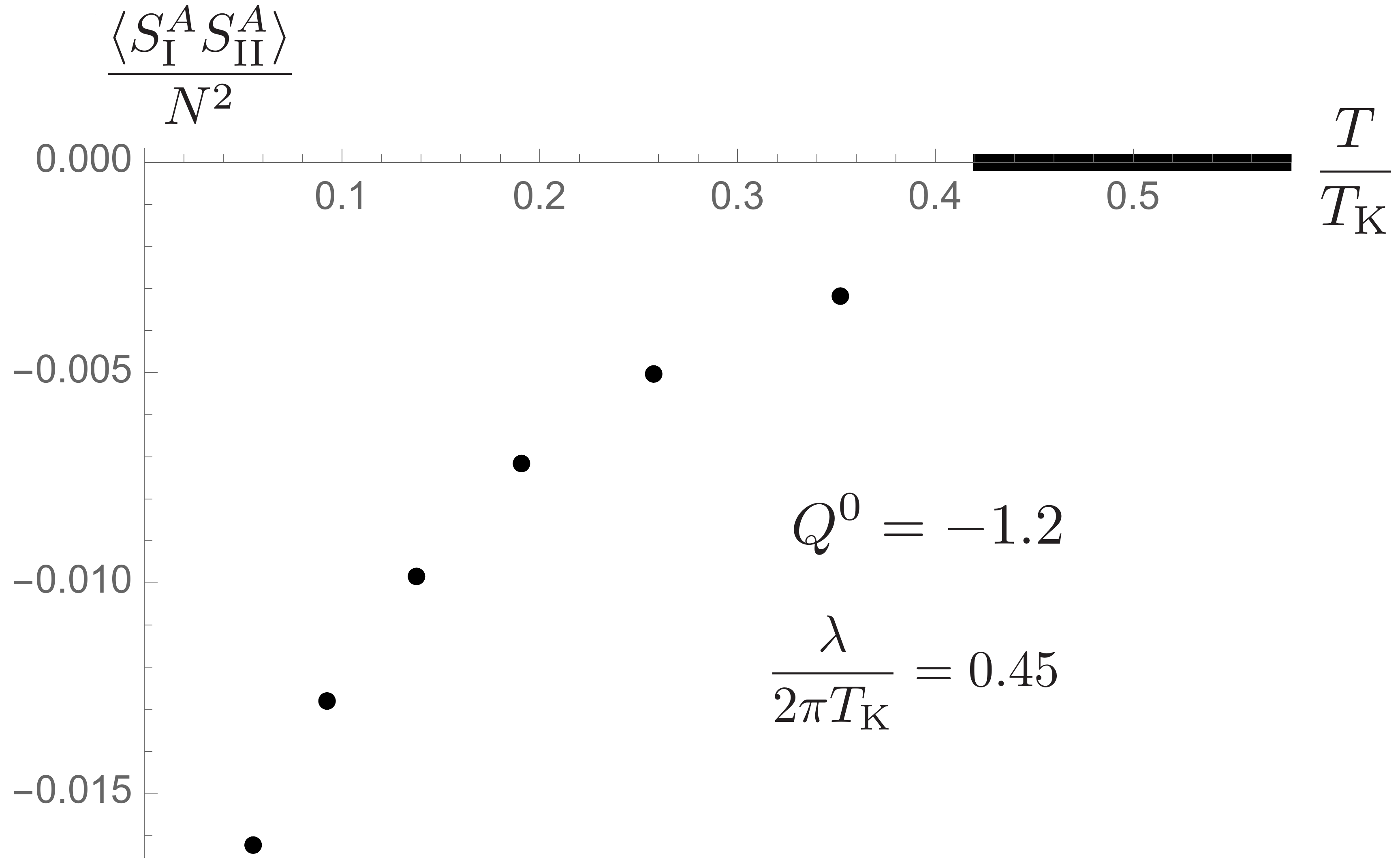}
&
\includegraphics[width=.48\textwidth]{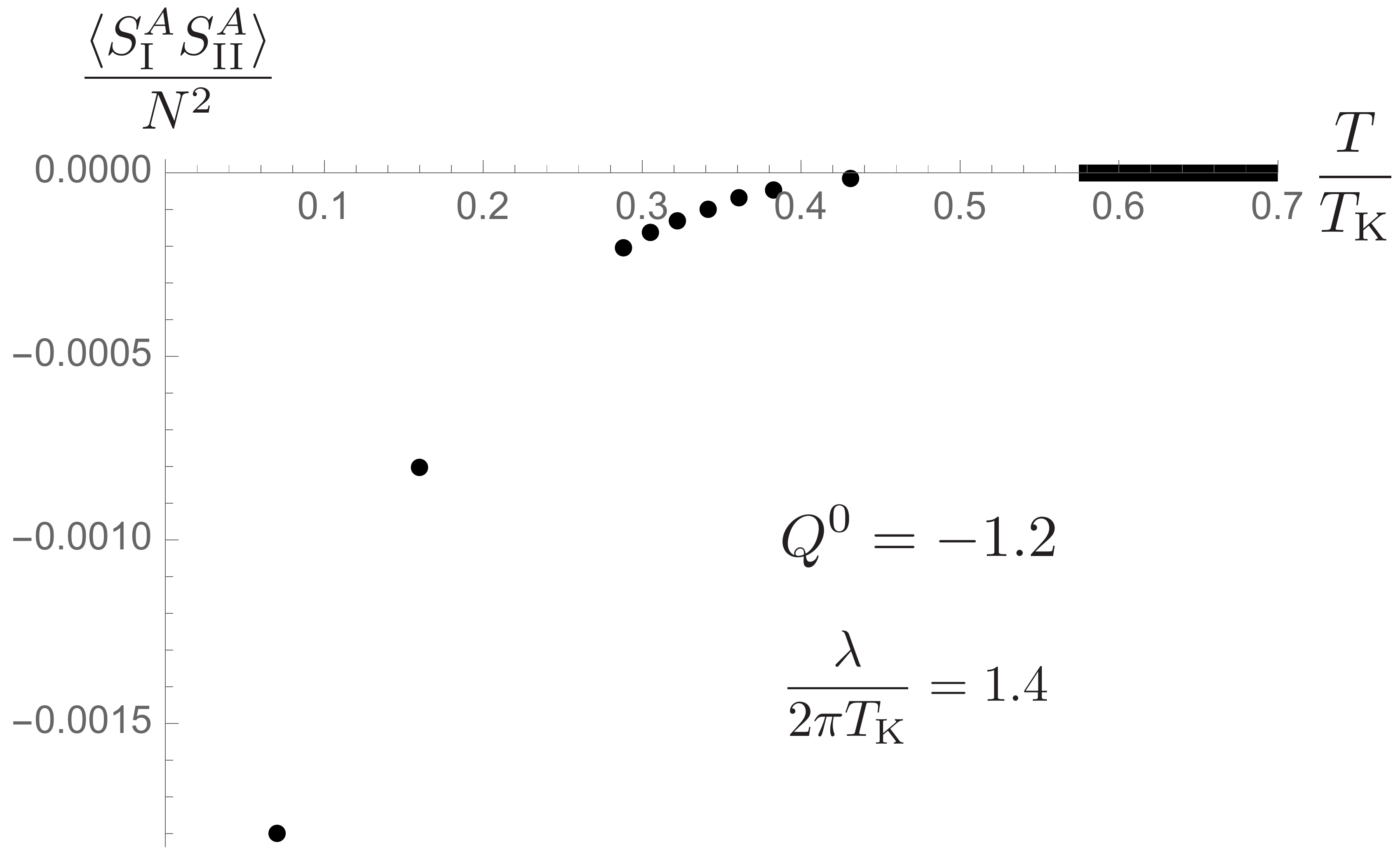} \\ (a.) & (b.)
\end{tabular}
\caption{\label{fig:spins} Our numerical results for $\langle S^A_{\one} S^A_{\two}\rangle/N^2 = - \mathcal{Q}^2/2$, as a function of $T/T_{\K}$ for $Q^0=-1.2$, with (a.) $\l/(2\pi T_{\K})=0.45$ and (b.) $\l/(2\pi T_{\K})=1.4$. In each figure, the black dots are our numerical data and the heavy gray line segment at $\langle S^A_{\one} S^A_{\two}\rangle/N^2 =0$ represents the trivial solution, for $T/T_{\K}$ values where the trivial solution is thermodynamically preferred, as far as we can determine within our numerical accuracy. As we decrease $T/T_{\K}$, a phase transition occurs approximately where the heavy gray line segment ends. Our results suggest a first-order transition in (a.) and a continuous transition in (b.), consistent with our expectations from the corresponding phase diagram, fig.~\ref{fig:phase} (c.).}
\end{center}
\end{figure}

 Fig.~\ref{fig:spins} shows our numerical results for the order $N^2$ contribution to the spin-spin correlator, given by eq.~\eqref{spinlimits}, $\langle S^A_{\one} S^A_{\two}\rangle/N^2 = - \mathcal{Q}^2/2$, as a function of $T/T_{\K}$ for $Q^0=-1.2$, with $\l/(2\pi T_{\K})=0.45$ in fig.~\ref{fig:spins} (a.) and $\l/(2\pi T_{\K})=1.4$ in fig.~\ref{fig:spins} (b.). As in fig.~\ref{fig:ovev}, in fig.~\ref{fig:spins} the black dots represent non-trivial numerical solutions, and the heavy gray line represents the trivial solution, which has $\langle S^A_{\one} S^A_{\two}\rangle/N^2 =0$, for $T/T_{\K}$ values where the trivial solution is thermodynamically preferred, as far as we can determine within our numerical accuracy. Our numerical results in fig.~\ref{fig:spins} (a.) for $\l/(2\pi T_{\K})=0.45$ suggest that most likely $\langle S^A_{\one} S^A_{\two}\rangle/N^2$ jumps from $\langle S^A_{\one} S^A_{\two}\rangle/N^2=0$ to $\langle S^A_{\one} S^A_{\two}\rangle/N^2\neq0$ when the transition occurs, consistent with a first-order transition, while fig.~\ref{fig:spins} (b.) for $\l/(2\pi T_{\K})=1.4$ suggests that $\langle S^A_{\one} S^A_{\two}\rangle/N^2$ will rise smoothly from zero through the transition, consistent with a continuous transition. These results conform to our expectations from the corresponding phase diagram, fig.~\ref{fig:phase} (c.).

Most importantly, the phase diagrams in fig.~\ref{fig:phase} strongly suggest that in our model a quantum phase transition occurs as a function of increasing $\l/(2 \pi T_{\K})$, from the trivial state to the non-trivial state. Moreover, fig.~\ref{fig:phase} suggests that such putative quantum phase transitions occur at non-zero AFM values of $\l/(2 \pi T_{\K})$. For example, when $Q^0=-1$, fig.~\ref{fig:phase} (b.) suggests that a transition may occur approximately where the solid black line hits the horizontal axis, $\l/(2 \pi T_{\K}) \approx 0.4$. The putative quantum phase transitions in our model also appear to be first order: as we increase $\l/(2 \pi T_{\K})$ through the critical value, all one-point functions and the phase shift will jump from zero (white region) to non-zero values (dark gray region). Similarly, the quantum phase transition in the large-$N$ two-impurity Kondo model of refs.~\cite{PhysRevB.39.3415,Millis1990} occurred at a non-zero AFM value of $\l_{\textrm{RKKY}}/T_{\K}$, and was first order. 

However, we were unable to obtain non-trivial numerical solutions at exactly $T=0$. In general, as $T/T_{\K}$ decreases our numerical solutions for $\phi$ tend to grow, apparently without bound. Indeed, such growth is typical for scalar fields in the probe limit, when the scalar potential includes only a mass term: see for example ref.~\cite{Horowitz:2009ij}. Most likely, obtaining reliable non-trivial numerical solutions at $T=0$ will require leaving the probe limit, \textit{i.e.}\ including the back-reaction of the matter fields on the metric. For the holographic single-impurity Kondo model of ref.~\cite{Erdmenger:2013dpa}, such back-reaction was studied in ref.~\cite{Erdmenger:2014xya}. We leave the analogous study for our model, and more generally the fate of our model at $T=0$, for future research.

\section{Summary and Outlook}
\label{outlook}

We proposed a holographic two-impurity Kondo model, building on the holographic single-impurity Kondo model of ref.~\cite{Erdmenger:2013dpa}, based on the CFT and large-$N$ approaches to the Kondo problem. In field theory terms, our model begins with the CFT description of the original two-impurity Kondo model, as $(1+1)$-dimensional free left-moving chiral fermions with two identical Kondo couplings to two identical impurity spins at the same location, plus an RKKY coupling between the impurity spins. We then gauged the spin $SU(N)$ symmetry, employed a probe limit to avoid a gauge anomaly, and added adjoint degrees of freedom such that the 't Hooft large-$N$ limit and large 't Hooft coupling produced a holographic dual with Einstein-Hilbert action. Our model thus had three couplings: the 't Hooft coupling, which was single-trace with respect to $SU(N)$ and was always large, and the Kondo and RKKY couplings, which were double-trace and had the RG flows expected from field theory, as we showed in subsection~\ref{RGtransfos}.

The gravity dual, described in section~\ref{model}, consisted of a complex scalar field and a $U(2)$ YM gauge field in $AdS_2$ coupled as a defect to a Chern-Simons gauge field in $AdS_3$. The complex scalar was bi-fundamental under the YM and Chern-Simons gauge groups. We had two main results. First, in section~\ref{holorg} we performed the holographic renormalization of our model, which allowed us to identify the Kondo and RKKY couplings as boundary conditions on the scalar field and YM gauge field, respectively. This was the first identification of an RKKY coupling in holography. Second, in section~\ref{phase} we solved the bulk equations of motion and evaluated the on-shell action numerically, which allowed us to determine the phase diagram of our model in the plane of RKKY coupling, $\l/(2 \pi T_{\K})$, versus $T/T_{\K}$. For sufficiently large AFM RKKY coupling we identified phase transitions as $T/T_{\K}$ decreases, from a trivial state, with no Kondo screening, no spin-spin correlations, and no phase shift, to a non-trivial state, with Kondo screening, AFM spin-spin correlations of order $N^2$, and a non-zero phase shift. Indeed, we argued, just using $SU(N)$ representation theory, that at leading order in the large-$N$ limit and for totally anti-symmetric impurity spins, FM spin-spin correlations will be absent, and only AFM spin-spin correlations of order $N^2$ will be visible. We also found numerical evidence for a first-order quantum phase transition in our model from the trivial state to the non-trivial state as $\l/(2 \pi T_{\K})$ increases through a non-zero, AFM critical value. These results are consistent with field theory expectations. For example, the quantum phase transition in the large-$N$ two impurity Kondo model of refs.~\cite{PhysRevB.39.3415,Millis1990} occurred at a non-zero AFM value of the RKKY coupling, and was first order. More generally, the coexistence of Kondo and inter-impurity screening is generic in two-impurity Kondo models, and the coexistence of Kondo and inter-impurity screening is believed to occur in the Kondo lattice~\cite{2009arXiv0912.0040S,2010uqpt.book..193S,Jones:2007}.

Our results show that holographic Kondo models can capture essential two-impurity Kondo phenomena, including most importantly a quantum phase transition characterized by a jump in the phase shift. Our results thus demonstrate that holographic Kondo models can provide a foundation for future model building, with the ultimate goal of building and solving a holographic Kondo lattice. Of course, much work remains to reach that goal.

In particular, to describe a Kondo lattice we must separate the impurities in space. Doing so in holographic Kondo models produces a number of problems, which we now discuss, roughly in order of increasing severity.

What would separating the impurities look like in our holographic model? For guidance, we turn to the top-down construction of the holographic single-impurity Kondo model of ref.~\cite{Erdmenger:2013dpa}. In that case the $AdS_2$ fields were the worldvolume fields on a D-brane dual to the impurity. A single D-brane has a $U(1)$ worldvolume gauge field. To describe two coincident impurities, we need two coincident D-branes, so that the gauge group is enhanced to $U(2)$. In fact, the worldvolume gauge multiplet will also include adjoint scalars, dual to scalar fields valued in the adjoint of the auxiliary $U(2)$ symmetry, and whose eigenvalues describe the positions of the two D-branes in transverse directions. To separate the D-branes, we can thus give non-zero expectation values to those adjoint scalar eigenvalues, which will break $U(2)$ down to $U(1) \times U(1)$ and will give masses to $a_t^1$ and $a_t^2$ via a Higgs mechanism. Among other things, those masses alter $a_t^1$ and $a_t^2$'s asymptotics.

Of course, in a regime where the supergravity approximation to string theory is reliable, as soon as the D-branes are more than a string length apart the open strings between them, whose lightest excitations include $a_t^1$ and $a_t^2$, will have enormous masses, of string scale, and hence could be integrated out. The ultimate low-energy effective description would then include two D-branes separated in space, each with its own $U(1)$ worldvolume gauge field, but now with no couplings between those worldvolume fields. What are the corresponding statements in the dual field theory? The non-zero expectation values for the adjoint scalar eigenvalues are non-normalizable modes dual to non-zero sources for the $U(2)$ adjoint scalar operators. Those non-zero sources explicitly break the auxiliary $U(2)$ symmetry down to $U(1) \times U(1)$, in which case no symmetry protects the dimensions of $R^1$ and $R^2$. Holography is apparently telling us that indeed those dimensions receive large corrections in the 't Hooft coupling, so that $R^1$ and $R^2$ effectively decouple from the dual field theory.

However, in a bottom-up model we are free to re-scale parameters however we like in order to keep $a_t^1$ and $a_t^2$'s masses at the AdS curvature scale, rather than the string scale. Using the D-brane description as a guide, but assuming such a re-scaling, we attempted to separate our impurities by adding to our holographic model a complex scalar field in $AdS_2$, valued in the adjoint of $U(2)$, and giving non-zero expectation values to the eigenvalues of that adjoint scalar. We then attempted holographic renormalization, but encountered two formidable obstacles, both arising from $a_t^1$ and $a_t^2$'s altered asymptotics. First, $Q^1$ and $Q^2$ acquire negative mass dimensions. Second, in $\Phi$'s equation of motion, near the $AdS_2$ boundary the terms involving $a_t^1$ and $a_t^2$ actually grow larger than the $M^2$ term, posing a major challenge to formulating a well-posed variational problem.

Separating the impurities in our holographic model has a more fundamental problem, however: separating the impurities in the fashion above radically changes the field theory interpretation of our model. In particular, we could no longer interpret the dual field theory as a $(1+1)$-dimensional CFT description of a two-impurity Kondo system, since that description is based on an $s$-wave reduction in the limit that the separation between the impurities is negligible. Instead, our model would be more similar to, though distinct from, a chiral Luttinger liquid coupled to two separated impurities. More generally, for the Kondo lattice a partial wave decomposition and the resulting $(1+1)$-dimensional description provides no obvious advantage, and is thus best abandoned. In other words, to build a holographic Kondo lattice we should commit to holographic models in which the field theory has two or more spatial dimensions, and introduce an infinite number of impurities. As mentioned in ref.~\cite{Jensen:2011su}, with an infinite number of impurities, the probe limit is by definition invalid.

Actually, perhaps the most fundamental problem with all holographic quantum impurity models to date is that the spin group is the gauge group, $SU(N)$. In that case, a genuine AFM phase is impossible, simply because if $N>2$ then two impurity spins, both in the fundamental representation of $SU(N)$, cannot lock into a singlet. An alternative is the symplectic large-$N$ limit: instead of replacing $SU(2)$ with $SU(N)$ and then taking $N \to \infty$, the symplectic large-$N$ limit is based on identifying $SU(2) \simeq Sp(1)$, replacing $Sp(1)$ with $Sp(N)$, and then taking $N \to \infty$~\cite{2008NatPh...4..643F,2007arXiv0710.1128F}. The symplectic large-$N$ limit allows for two impurity spins in the fundamental representation to lock into a singlet, and hence allows for a genuine AFM phase. Top-down holographic duals of strongly-coupled gauge theories with symplectic gauge groups can be realized, for example by introducing orientifolds in the bulk~\cite{Aharony:1999ti}.

However, as mentioned in ref.~\cite{Erdmenger:2013dpa}, in any model where the spin group is the gauge group, the impurity spin operator is the gauge current of the auxiliary fermions or bosons, and hence is not gauge-invariant. Holography only provides access to correlators of gauge-invariant operators, so in any holographic model where spin is the gauge group, access to observables involving spin will be indirect at best, and completely absent at worst. Moreover, if spin is the gauge group and we separate two impurities, then to make the RKKY interaction gauge-invariant we must connect the two spin operators with an open Wilson line. In other words, when spin is the gauge group, the RKKY coupling between separated impurities is non-local. In top-down holographic models, an open Wilson line is dual to an open string stretched between D-branes~\cite{Aharony:2008an}.

As already proposed in ref.~\cite{Erdmenger:2013dpa}, a holographic model in which the spin symmetry group is a global symmetry would have many advantages. For example, we would no longer be restricted to the large-$N$ limit of the spin group: we could demand that the spin group be $SU(2)$, while the large-$N$ strongly-coupled gauge theory sector would merely provide a classical gravity dual. Moreover, many observables involving the spin operator, such as the magnetic susceptibility, could then be computed using holography.

We believe that the long-term goal of solving a (holographic) Kondo lattice provides sufficient motivation to pursue solutions to all of the problems above, using our holographic model as a starting point. Indeed, we plan to study these, and many other issues in holographic quantum impurity models, in the near future.


\section*{Acknowledgements}

We thank J.~Erdmenger, M.~Flory, C.~Hoyos, N.~Iqbal, M.~Newrzella, A.~Starinets, and J.M.S.~Wu for useful discussions. We also thank the Galileo Galilei Institute for Theoretical Physics for hospitality and the INFN for partial support during the completion of this work. The work of A.~O'B. was supported by a University Research Fellowship from the Royal Society and a Junior Research Fellowship from Balliol College. I.~P. acknowledges partial support through COST Action MP 1210 Short Term Scientific Mission COST-STSM-ECOST-STSM-MP1210-301114-051520 and thanks the University of Oxford for hospitality during the completion of this work. The work of J.~P. was supported by a Clarendon scholarship from the Clarendon Fund and St John's College, Oxford, and in part by the European Research Council under the European Union's Seventh Framework Programme (ERC Grant agreement 307955).

\appendix
\section*{Appendix: Normalizability of AdS$_2$ Gauge Fields}
\addcontentsline{toc}{section}{Appendix: Normalizability of AdS$_2$ Gauge Fields}
\renewcommand{\theequation}{\arabic{equation}}

In this appendix we discuss the normalizability of a massless or massive Abelian gauge field in asymptotically locally $AdS_{d+1}$ spacetime. Our goal is to determine what boundary conditions are permitted as a function of $d$ and the mass, particularly in the case discussed in the main text, $d=1$,~\textit{i.e.}\ asymptotically locally $AdS_2$ spacetime.

Consider an Abelian gauge field $a_m$ of mass $m_a$, with field strength $f_{mn}$, in an asymptotically locally $AdS_{d+1}$ spacetime with $d \geq 1$, with metric $g_{mn}$ and $g \equiv \textrm{det}(g_{mn})$. The action and equation of motion of such a gauge field are
\begin{subequations}
\beq
\label{massvecaction}
S=-\int d^{d+1} x\sqrt{-g}\left(\frac14 f_{mn}f^{mn}+\frac12 m_a^2 \, g^{mn}a_m a_n\right),
\eeq
\beq
\label{massvecEOM}
\nabla^m f_{mn}-m_a^2a_n=0.
\eeq
\end{subequations}

To define normalizability, we need a norm on field space. Following the definition of the norm for scalar fields in ref.~\cite{Klebanov:1999tb}, we define a norm from the action in eq.~\eqref{massvecaction} by integrating by parts and dropping all boundary terms. (For massless Abelian gauge fields, the definition of normalizability in ref.~\cite{Marolf:2006nd}, namely finiteness of symplectic flux at the asymptotically $AdS_{d+1}$ boundary, produces the same results.) Our norm is thus
\beq
\label{norm}
S'=\frac12\int d ^{d+1}x \sqrt{-g} \, a_p \, g^{pn}\left(\nabla^m f_{mn}-m_a^2a_n\right).
\eeq
In other words, a field configuration $a_m$ is normalizable if and only if $S'$ evaluated on that configuration is finite. Crucially, such a definition is non-trivial only for off-shell modes, since $S'$ vanishes when evaluated on a solution to the equation of motion, eq.~\eqref{massvecEOM}.

If we choose FG coordinates, like those in eq.~\eqref{FG-gauge}, and the radial gauge $a_r=0$, then the two linearly independent asymptotic solutions for $a_m$ are of the form $e^{\delta_{\pm} r}$ with
\be
\label{massvecpowers}
\d_\pm=-\frac{d-2}{2}\pm\sqrt{\left(\frac{d-2}{2}\right)^2+m_a^2}.
\ee
Demanding that $\d_{\pm}$ are real produces the BF bound for massive vector fields, $m_a^2 \geq -(d-2)^2/4$, which we will assume is obeyed in what follows. The most general asymptotic field configuration is a linear combination of the $e^{\d_{\pm}r}$ solutions, with coefficients that can depend on all of the other coordinates besides $r$. Plugging the most general asymptotic field configuration into the norm $S'$ in eq.~\eqref{norm} and demanding that terms involving derivatives in directions besides $r$ remain finite, we find that both asymptotic behaviors are normalizable only when $\d_{\pm}<\frac{4-d}{2}$, which immediately translates into an upper bound on the mass, $m_a^2<\frac{d(4-d)}{4}$. Crucially, both asymptotic behaviors are normalizable only for $d \leq 3$. Combined with the BF bound, we thus find that if $d \leq 3$ and
\beq
\label{massveclimits}
-\frac{(d-2)^2}{4}\leq m_a^2<\frac{d(4-d)}{4},
\eeq
then both asymptotic behaviors $e^{\delta^{\pm} r}$ are normalizable, and hence any of Dirichlet, Neumann, or mixed boundary conditions is permitted.

Of course, $(1+1)$-dimensional gauge fields on their own, whether massless or massive, Abelian or non-Abelian, have no propagating modes. In particular, in $AdS_2$, choosing the FG coordinates in eq.~\eqref{FG-gauge} and $a_r=0$ gauge, the equation of motion eq.~\eqref{massvecEOM} for $a_t$ becomes
\beq
\label{massvecEOMsimple}
\partial_r^2a_t- \partial_r a_t-m_a^2 \, a_t = 0,
\eeq
which has no time derivatives, indicating the absence of propagating modes. As a result, in asymptotically locally $AdS_2$ spacetime, and in contrast to $d>1$, the normalizability condition is satisfied trivially: when $d=1$, as $r \to \infty$ the integrand of $S'$ in eq.~\eqref{norm} approaches the equation of motion in eq.~\eqref{massvecEOMsimple}, and in particular no terms involving time derivatives appear, so that $S'$ remains finite for both solutions, $e^{\delta^{\pm} r}$.

However, gauge fields in asymptotically locally $AdS_2$ spacetimes coupled to other fields can have propagating modes. Indeed, our holographic two-impurity Kondo model is an example. In such cases, as long as the interactions are sub-leading as $r \to \infty$ relative to the terms in the action $S$ of eq.~\eqref{massvecaction}, so that the asymptotic solutions remain $e^{\delta^{\pm}r}$ with $\delta^{\pm}$ in eq.~\eqref{massvecpowers}, then our results for normalizability remain valid. In our holographic two-impurity Kondo model, the interactions are indeed sub-leading relative to the terms in eq.~\eqref{massvecaction}, hence the results above apply.

\bibliographystyle{JHEP}
\bibliography{kondotwo}

\providecommand{\href}[2]{#2}\begingroup\raggedright\begin{thebibliography}{100}

\bibitem{2006cond.mat.12006C}
P.~Coleman, {\it {Heavy Fermions: Electrons at the Edge of Magnetism}},  in
  {\em Handbook of Magnetism and Advanced Magnetic Materials: Fundamentals and
  Theory} (Kronmuller and Parkin, eds.), vol.~1, pp.~95--148.
\newblock John Wiley and Sons, 2007.
\newblock \href{http://arxiv.org/abs/[arxiv:cond-mat/0612006]}{{\tt
  [arxiv:cond-mat/0612006]}}.

\bibitem{2008NatPh...4..186G}
P.~Gegenwart, Q.~Si, and F.~Steglich, {\it {Quantum Criticality in
  Heavy-fermion Metals}},  {\em Nature Physics} {\bf 4} (Mar., 2008) 186--197,
  [\href{http://arxiv.org/abs/0712.2045}{{\tt arXiv:0712.2045}}].

\bibitem{2009arXiv0912.0040S}
Q.~{Si}, {\it {Quantum Criticality and Global Phase Diagram of Magnetic Heavy
  Fermions}},  {\em ArXiv e-prints} (Dec., 2009)
  [\href{http://arxiv.org/abs/0912.0040}{{\tt arXiv:0912.0040}}].

\bibitem{2010uqpt.book..193S}
Q.~Si, {\it {Quantum Criticality and the Kondo Lattice}},  {\em Understanding
  Quantum Phase Transitions.~Series: Condensed Matter Physics, CRC Press}
  (Nov., 2010) 193--216, [\href{http://arxiv.org/abs/1012.5440}{{\tt
  arXiv:1012.5440}}].

\bibitem{2010Sci...329.1161S}
Q.~{Si} and F.~{Steglich}, {\it {Heavy Fermions and Quantum Phase
  Transitions}},  {\em Science} {\bf 329} (Sept., 2010) 1161--,
  [\href{http://arxiv.org/abs/1102.4896}{{\tt arXiv:1102.4896}}].

\bibitem{2015arXiv150905769C}
P.~{Coleman}, {\it {Heavy Fermions and the Kondo Lattice: a 21st Century
  Perspective}},  {\em ArXiv e-prints} (Sept., 2015)
  [\href{http://arxiv.org/abs/1509.05769}{{\tt arXiv:1509.05769}}].

\bibitem{2014arXiv1409.4673K}
B.~{Keimer}, S.~A. {Kivelson}, M.~R. {Norman}, S.~{Uchida}, and J.~{Zaanen},
  {\it {High Temperature Superconductivity in the Cuprates}},  {\em ArXiv
  e-prints} (Sept., 2014) [\href{http://arxiv.org/abs/1409.4673}{{\tt
  arXiv:1409.4673}}].

\bibitem{1977PhyBC..91..231D}
S.~{Doniach}, {\it {The Kondo lattice and weak antiferromagnetism}},  {\em
  Physica B+C} {\bf 91} (July, 1977) 231--234.

\bibitem{Goldhaber1998}
D.~Goldhaber-Gordon, H.~Shtrikman, D.~Mahalu, D.~Abusch-Magder, U.~Meirav, and
  M.~A. Kastner, {\it {Kondo Effect in a Single-electron Transistor}},  {\em
  Nature} {\bf 391} (1998) 156--159.

\bibitem{Cronenwett24071998}
S.~Cronenwett, T.~Oosterkamp, and L.~Kouwenhoven, {\it {A Tunable Kondo Effect
  in Quantum Dots}},  {\em Science} {\bf 281} (1998), no.~5376 540--544.

\bibitem{vanderWiel22092000}
W.~G. van~der Wiel, S.~D. Franceschi, T.~Fujisawa, J.~M. Elzerman, S.~Tarucha,
  and L.~P. Kouwenhoven, {\it {The Kondo Effect in the Unitary Limit}},  {\em
  Science} {\bf 289} (2000), no.~5487 2105--2108.

\bibitem{PTP.32.37}
J.~Kondo, {\it {Resistance Minimum in Dilute Magnetic Alloys}},  {\em Prog.
  Theo. Phys.} {\bf 32} (1964), no.~1 37--49.

\bibitem{0034-4885-37-2-001}
C.~Rizzuto, {\it {Formation of Localized Moments in Metals: Experimental Bulk
  Properties}},  {\em Rep. Prog. Phys.} {\bf 37} (1974), no.~2 147.

\bibitem{GrŸner1978591}
G.~Gr{\"u}ner and A.~Zawadowski, {\it {Low Temperature Properties of Kondo
  Alloys}},  in {\em Progress in Low Temperature Physics} (D.~Brewer, ed.),
  vol.~7, Part B, pp.~591 -- 647.
\newblock Elsevier, 1978.

\bibitem{Wilson:1974mb}
K.~G. Wilson, {\it {The Renormalization Group: Critical Phenomena and the Kondo
  Problem}},  {\em Rev.Mod.Phys.} {\bf 47} (1975) 773.

\bibitem{PhysRevB.21.1003}
H.~R. Krishna-murthy, J.~W. Wilkins, and K.~G. Wilson, {\it
  Renormalization-group approach to the anderson model of dilute magnetic
  alloys. i. static properties for the symmetric case},  {\em Phys. Rev. B}
  {\bf 21} (Feb, 1980) 1003--1043.

\bibitem{PhysRevB.21.1044}
H.~R. Krishna-murthy, J.~W. Wilkins, and K.~G. Wilson, {\it
  Renormalization-group approach to the anderson model of dilute magnetic
  alloys. ii. static properties for the asymmetric case},  {\em Phys. Rev. B}
  {\bf 21} (Feb, 1980) 1044--1083.

\bibitem{PhysRevLett.45.379}
N.~Andrei, {\it {Diagonalization of the Kondo Hamiltonian}},  {\em Phys. Rev.
  Lett.} {\bf 45} (Aug, 1980) 379--382.

\bibitem{Wiegmann:1980}
P.~Wiegmann, {\it {Exact Solution of s-d Exchange Model at T=0}},  {\em Sov.
  Phys. JETP Lett.} {\bf 31} (1980) 364.

\bibitem{RevModPhys.55.331}
N.~Andrei, K.~Furuya, and J.~H. Lowenstein, {\it {Solution of the Kondo
  problem}},  {\em Rev. Mod. Phys.} {\bf 55} (Apr, 1983) 331--402.

\bibitem{doi:10.1080/00018738300101581}
A.~Tsvelick and P.~Wiegmann, {\it {Exact Results in the Theory of Magnetic
  Alloys}},  {\em Advances in Physics} {\bf 32} (1983), no.~4 453--713.

\bibitem{0022-3719-19-17-017}
P.~Coleman and N.~Andrei, {\it {Diagonalisation of the Generalised Anderson
  Model}},  {\em Jour. Phys.} {\bf C19} (1986) 3211--3233.

\bibitem{1994cond.mat..8101A}
N.~{Andrei}, {\it {Integrble Models in Condensed Matter Physics}},  {\em eprint
  arXiv:cond-mat/9408101} (Aug., 1994)
  [\href{http://arxiv.org/abs/cond-mat/9408101}{{\tt cond-mat/9408101}}].

\bibitem{ZinnJustin1998}
P.~Zinn-Justin and N.~Andrei, {\it {The Generalized Multi-channel Kondo Model:
  Thermodynamics and Fusion Equations}},  {\em Nucl. Phys.} {\bf B528} (1998)
  648--682, [\href{http://arxiv.org/abs/cond-mat/9801158}{{\tt
  cond-mat/9801158}}].

\bibitem{PhysRevB.58.3814}
A.~Jerez, N.~Andrei, and G.~Zar\'and, {\it {Solution of the Multichannel
  Coqblin-Schrieffer Impurity Model and Application to Multilevel Systems}},
  {\em Phys. Rev.} {\bf B58} (1998) 3814--3841,
  [\href{http://arxiv.org/abs/cond-mat/9803137}{{\tt cond-mat/9803137}}].

\bibitem{PhysRevB.35.5072}
P.~Coleman, {\it {Mixed Valence as an Almost Broken Symmetry}},  {\em Phys.
  Rev.} {\bf B35} (Apr, 1987) 5072--5116.

\bibitem{RevModPhys.59.845}
{Bickers, N.}, {\it {Review of Techniques in the Large-N Expansion for Dilute
  Magnetic Alloys}},  {\em Rev. Mod. Phys.} {\bf 59} (Oct, 1987) 845--939.

\bibitem{1997PhRvL..79.4665P}
O.~{Parcollet} and A.~{Georges}, {\it {Transition from Overscreening to
  Underscreening in the Multichannel Kondo Model: Exact Solution at Large N}},
  {\em Phys. Rev. Lett.} {\bf 79} (1997) 4665--4668,
  [\href{http://arxiv.org/abs/cond-mat/9707337}{{\tt cond-mat/9707337}}].

\bibitem{1998PhRvB..58.3794P}
O.~Parcollet, A.~Georges, G.~Kotliar, and A.~Sengupta, {\it {Overscreened
  Multi-channel SU(N) Kondo Model: Large-N Solution and Conformal Field
  Theory}},  {\em Phys. Rev.} {\bf B58} (Aug., 1998) 3794--3813,
  [\href{http://arxiv.org/abs/{arXiv:cond-mat/9711192}}{{\tt
  {arXiv:cond-mat/9711192}}}].

\bibitem{Affleck:1990zd}
I.~Affleck, {\it {A Current Algebra Approach To The Kondo Effect}},  {\em Nucl.
  Phys.} {\bf B336} (1990) 517.

\bibitem{Affleck:1990by}
I.~Affleck and A.~Ludwig, {\it {The Kondo Effect, Conformal Field Theory and
  Fusion Rules}},  {\em Nucl.Phys.} {\bf B352} (1991) 849--862.

\bibitem{Affleck:1990iv}
I.~Affleck and A.~Ludwig, {\it {Critical Theory of Overscreened Kondo Fixed
  Points}},  {\em Nucl.Phys.} {\bf B360} (1991) 641--696.

\bibitem{Affleck:1991tk}
I.~Affleck and A.~Ludwig, {\it {Universal Non-integer 'Ground State Degeneracy'
  in Critical Quantum Systems}},  {\em Phys.Rev.Lett.} {\bf 67} (1991)
  161--164.

\bibitem{PhysRevB.48.7297}
I.~Affleck and A.~Ludwig, {\it {Exact Conformal-field-theory Results on the
  Multichannel Kondo Effect: Single-fermion Green's function, Self-energy, and
  Resistivity}},  {\em Phys.Rev.} {\bf B48} (1993) 7297--7321.

\bibitem{Affleck:1995ge}
I.~Affleck, {\it {Conformal Field Theory Approach to the Kondo Effect}},  {\em
  Acta Phys. Polon.} {\bf B26} (1995) 1869--1932,
  [\href{http://arxiv.org/abs/cond-mat/9512099}{{\tt cond-mat/9512099}}].

\bibitem{Hewson:1993}
A.~Hewson, {\it {The Kondo Model to Heavy Fermions}},  {\em {Cambridge
  University Press}} (1993).

\bibitem{doi:10.1080/000187398243500}
D.~L. Cox and A.~Zawadowski, {\it {Exotic Kondo Effects in Metals: Magnetic
  Ions in a Crystalline Electric Field and Tunnelling Centres}},  {\em Advances
  in Physics} {\bf 47} (1998), no.~5 599--942,
  [\href{http://arxiv.org/abs/{arxiv:cond-mat/9704103}}{{\tt
  {arxiv:cond-mat/9704103}}}].

\bibitem{2009arXiv0911.2209A}
I.~{Affleck}, {\it {The Kondo screening cloud: what it is and how to observe
  it}},  {\em ArXiv e-prints} (Nov., 2009)
  [\href{http://arxiv.org/abs/0911.2209}{{\tt arXiv:0911.2209}}].

\bibitem{RevModPhys.68.13}
A.~Georges, G.~Kotliar, W.~Krauth, and M.~J. Rozenberg, {\it Dynamical
  mean-field theory of strongly correlated fermion systems and the limit of
  infinite dimensions},  {\em Rev. Mod. Phys.} {\bf 68} (Jan, 1996) 13--125.

\bibitem{PhysRevLett.47.737}
C.~Jayaprakash, H.~Krishna-murthy, and J.~Wilkins, {\it {Two-Impurity Kondo
  Problem}},  {\em Phys. Rev. Lett.} {\bf 47} (Sep, 1981) 737--740.

\bibitem{PhysRevLett.58.843}
B.~Jones and C.~Varma, {\it {Study of Two Magnetic Impurities in a Fermi Gas}},
   {\em Phys. Rev. Lett.} {\bf 58} (Mar, 1987) 843--846.

\bibitem{PhysRevB.35.4901}
R.~M. Fye, J.~E. Hirsch, and D.~J. Scalapino, {\it Kondo effect versus indirect
  exchange in the two-impurity anderson model: A monte carlo study},  {\em
  Phys. Rev. B} {\bf 35} (Apr, 1987) 4901--4908.

\bibitem{PhysRevLett.61.125}
B.~Jones, C.~Varma, and J.~Wilkins, {\it {Low-Temperature Properties of the
  Two-Impurity Kondo Hamiltonian}},  {\em Phys. Rev. Lett.} {\bf 61} (Jul,
  1988) 125--128.

\bibitem{PhysRevB.39.3415}
B.~A. Jones, B.~G. Kotliar, and A.~J. Millis, {\it Mean-field analysis of two
  antiferromagnetically coupled anderson impurities},  {\em Phys. Rev. B} {\bf
  39} (Feb, 1989) 3415--3418.

\bibitem{PhysRevB.40.324}
B.~Jones and C.~Varma, {\it {Critical Point in the Solution of the Two Magnetic
  Impurity Problem}},  {\em Phys. Rev.} {\bf B40} (Jul, 1989) 324--329.

\bibitem{PhysRevB.40.4780}
R.~M. Fye and J.~E. Hirsch, {\it Quantum monte carlo study of the two-impurity
  kondo hamiltonian},  {\em Phys. Rev. B} {\bf 40} (Sep, 1989) 4780--4796.

\bibitem{Jones1990}
B.~Jones, {\it {Antiferromagnetic Phase Instability in the Two-Impurity Kondo
  Problem}},  in {\em Field Theories in Condensed Matter Physics: A Workshop}
  (Z.~Tesanovic, ed.), pp.~87--103.
\newblock Addison-Wesley, 1990.

\bibitem{Millis1990}
A.~Millis, B.~Kotliar, and B.~Jones, {\it {The Two Kondo Impurity Problem: A
  Large N Biased Review}},  in {\em Field Theories in Condensed Matter Physics:
  A Workshop} (Z.~Tesanovic, ed.), pp.~159--166.
\newblock Addison-Wesley, 1990.

\bibitem{Affleck:1991yq}
I.~Affleck and A.~W.~W. Ludwig, {\it {Exact critical theory of the two impurity
  Kondo model}},  {\em Phys. Rev. Lett.} {\bf 68} (1992) 1046--1049.

\bibitem{PhysRevLett.72.916}
R.~M. Fye, {\it ``anomalous fixed point behavior'' of two kondo impurities: A
  reexamination},  {\em Phys. Rev. Lett.} {\bf 72} (Feb, 1994) 916--919.

\bibitem{1995PhRvB..52.9528A}
I.~Affleck, A.~Ludwig, and B.~Jones, {\it {Conformal-field-theory Approach to
  the Two-impurity Kondo Problem: Comparison with Numerical
  Renormalization-group Results}},  {\em Phys. Rev.} {\bf B52} (Oct., 1995)
  9528--9546, [\href{http://arxiv.org/abs/{arXiv:cond-mat/9409100}}{{\tt
  {arXiv:cond-mat/9409100}}}].

\bibitem{INGERSENT1994402}
K.~Ingersent and B.~A. Jones, {\it Low-temperature physics of the two-impurity,
  two-channel kondo model},  {\em Physica B: Condensed Matter} {\bf 199} (1994)
  402 -- 405.

\bibitem{PhysRevLett.74.2583}
J.~Gan, {\it Mapping the critical point of the two-impurity kondo model to a
  two-channel problem},  {\em Phys. Rev. Lett.} {\bf 74} (Mar, 1995)
  2583--2586.

\bibitem{PhysRevB.51.8287}
J.~Gan, {\it Solution of the two-impurity kondo model: Critical point,
  fermi-liquid phase, and crossover},  {\em Phys. Rev. B} {\bf 51} (Apr, 1995)
  8287--8309.

\bibitem{PhysRevLett.74.2808}
A.~Georges and A.~M. Sengupta, {\it Solution of the two-impurity, two-channel
  kondo model},  {\em Phys. Rev. Lett.} {\bf 74} (Apr, 1995) 2808--2811.

\bibitem{PhysRevLett.76.275}
J.~B. Silva, W.~L.~C. Lima, W.~C. Oliveira, J.~L.~N. Mello, L.~N. Oliveira, and
  J.~W. Wilkins, {\it Particle-hole asymmetry in the two-impurity kondo model},
   {\em Phys. Rev. Lett.} {\bf 76} (Jan, 1996) 275--278.

\bibitem{Jones:2007}
B.~Jones, {\it {The Kondo Effect}},  in {\em Handbook of Magnetism and Advanced
  Magnetic Materials: Fundamentals and Theory} (Kronmuller and Parkin, eds.),
  vol.~1, pp.~149--163.
\newblock John Wiley and Sons, 2007.

\bibitem{Maldacena:1997re}
J.~M. Maldacena, {\it {The Large N Limit of Superconformal Field Theories and
  Supergravity}},  {\em Adv. Theor. Math. Phys.} {\bf 2} (1998) 231--252,
  [\href{http://arxiv.org/abs/hep-th/9711200}{{\tt hep-th/9711200}}].

\bibitem{Gubser:1998bc}
S.~S. Gubser, I.~R. Klebanov, and A.~M. Polyakov, {\it {Gauge Theory
  Correlators from Non-critical String Theory}},  {\em Phys. Lett.} {\bf B428}
  (1998) 105--114, [\href{http://arxiv.org/abs/hep-th/9802109}{{\tt
  hep-th/9802109}}].

\bibitem{Witten:1998qj}
E.~Witten, {\it {Anti-de Sitter Space and Holography}},  {\em Adv. Theor. Math.
  Phys.} {\bf 2} (1998) 253--291,
  [\href{http://arxiv.org/abs/hep-th/9802150}{{\tt hep-th/9802150}}].

\bibitem{Aharony:1999ti}
O.~Aharony, S.~S. Gubser, J.~M. Maldacena, H.~Ooguri, and Y.~Oz, {\it {Large N
  field theories, string theory and gravity}},  {\em Phys. Rept.} {\bf 323}
  (2000) 183--386, [\href{http://arxiv.org/abs/hep-th/9905111}{{\tt
  hep-th/9905111}}].

\bibitem{Aharony:2008ug}
O.~Aharony, O.~Bergman, D.~L. Jafferis, and J.~Maldacena, {\it {N=6
  superconformal Chern-Simons-matter theories, M2-branes and their gravity
  duals}},  {\em JHEP} {\bf 10} (2008) 091,
  [\href{http://arxiv.org/abs/0806.1218}{{\tt arXiv:0806.1218}}].

\bibitem{Kachru:2009xf}
S.~Kachru, A.~Karch, and S.~Yaida, {\it {Holographic Lattices, Dimers, and
  Glasses}},  {\em Phys.Rev.} {\bf D81} (2010) 026007,
  [\href{http://arxiv.org/abs/0909.2639}{{\tt arXiv:0909.2639}}].

\bibitem{Sachdev:2010um}
S.~Sachdev, {\it {Holographic Metals and the Fractionalized Fermi Liquid}},
  {\em Phys.Rev.Lett.} {\bf 105} (2010) 151602,
  [\href{http://arxiv.org/abs/1006.3794}{{\tt arXiv:1006.3794}}].

\bibitem{Kachru:2010dk}
S.~Kachru, A.~Karch, and S.~Yaida, {\it {Adventures in Holographic Dimer
  Models}},  {\em New J.Phys.} {\bf 13} (2011) 035004,
  [\href{http://arxiv.org/abs/1009.3268}{{\tt arXiv:1009.3268}}].

\bibitem{Sachdev:2010uj}
S.~Sachdev, {\it {Strange Metals and the AdS/CFT Correspondence}},  {\em
  J.Stat.Mech.} {\bf 1011} (2010) P11022,
  [\href{http://arxiv.org/abs/1010.0682}{{\tt arXiv:1010.0682}}].

\bibitem{Mueck:2010ja}
W.~M{\"u}ck, {\it {The Polyakov Loop of Anti-symmetric Representations as a
  Quantum Impurity Model}},  {\em Phys.Rev.} {\bf D83} (2011) 066006,
  [\href{http://arxiv.org/abs/1012.1973}{{\tt arXiv:1012.1973}}].

\bibitem{Faraggi:2011bb}
A.~Faraggi and L.~Pando~Zayas, {\it {The Spectrum of Excitations of Holographic
  Wilson Loops}},  {\em JHEP} {\bf 1105} (2011) 018,
  [\href{http://arxiv.org/abs/1101.5145}{{\tt arXiv:1101.5145}}].

\bibitem{Jensen:2011su}
K.~Jensen, S.~Kachru, A.~Karch, J.~Polchinski, and E.~Silverstein, {\it
  {Towards a Holographic Marginal Fermi Liquid}},  {\em Phys.Rev.} {\bf D84}
  (2011) 126002, [\href{http://arxiv.org/abs/1105.1772}{{\tt
  arXiv:1105.1772}}].

\bibitem{Karaiskos:2011kf}
N.~Karaiskos, K.~Sfetsos, and E.~Tsatis, {\it {Brane Embeddings in Sphere
  Submanifolds}},  {\em Class.Quant.Grav.} {\bf 29} (2012) 025011,
  [\href{http://arxiv.org/abs/1106.1200}{{\tt arXiv:1106.1200}}].

\bibitem{Harrison:2011fs}
S.~Harrison, S.~Kachru, and G.~Torroba, {\it {A Maximally Supersymmetric Kondo
  Model}},  {\em Class.Quant.Grav.} {\bf 29} (2012) 194005,
  [\href{http://arxiv.org/abs/1110.5325}{{\tt arXiv:1110.5325}}].

\bibitem{Benincasa:2011zu}
P.~Benincasa and A.~Ramallo, {\it {Fermionic Impurities in Chern-Simons-Matter
  Theories}},  {\em JHEP} {\bf 1202} (2012) 076,
  [\href{http://arxiv.org/abs/1112.4669}{{\tt arXiv:1112.4669}}].

\bibitem{Faraggi:2011ge}
A.~Faraggi, W.~M{\"u}ck, and L.~Pando~Zayas, {\it {One-loop Effective Action of
  the Holographic Antisymmetric Wilson Loop}},  {\em Phys.Rev.} {\bf D85}
  (2012) 106015, [\href{http://arxiv.org/abs/1112.5028}{{\tt
  arXiv:1112.5028}}].

\bibitem{Benincasa:2012wu}
P.~Benincasa and A.~Ramallo, {\it {Holographic Kondo Model in Various
  Dimensions}},  {\em JHEP} {\bf 1206} (2012) 133,
  [\href{http://arxiv.org/abs/1204.6290}{{\tt arXiv:1204.6290}}].

\bibitem{Matsueda:2012gc}
H.~Matsueda, {\it {Multiscale Entanglement Renormalization Ansatz for Kondo
  Problem}},  \href{http://arxiv.org/abs/1208.2872}{{\tt arXiv:1208.2872}}.

\bibitem{Itsios:2012ev}
G.~Itsios, K.~Sfetsos, and D.~Zoakos, {\it {Fermionic Impurities in the
  Unquenched ABJM}},  {\em JHEP} {\bf 1301} (2013) 038,
  [\href{http://arxiv.org/abs/1209.6617}{{\tt arXiv:1209.6617}}].

\bibitem{Erdmenger:2013dpa}
J.~Erdmenger, C.~Hoyos, A.~O'Bannon, and J.~Wu, {\it {A Holographic Model of
  the Kondo Effect}},  {\em JHEP} {\bf 12} (2013) 086,
  [\href{http://arxiv.org/abs/1310.3271}{{\tt arXiv:1310.3271}}].

\bibitem{Maldacena:1998im}
J.~M. Maldacena, {\it {Wilson Loops in Large N Field Theories}},  {\em
  Phys.Rev.Lett.} {\bf 80} (1998) 4859--4862,
  [\href{http://arxiv.org/abs/hep-th/9803002}{{\tt hep-th/9803002}}].

\bibitem{Rey:1998ik}
S.-J. Rey and J.-T. Yee, {\it {Macroscopic Strings as Heavy Quarks in Large N
  Gauge Theory and Anti-de Sitter Supergravity}},  {\em Eur.Phys.J.} {\bf C22}
  (2001) 379--394, [\href{http://arxiv.org/abs/hep-th/9803001}{{\tt
  hep-th/9803001}}].

\bibitem{Camino:2001at}
J.~Camino, A.~Paredes, and A.~Ramallo, {\it {Stable Wrapped Branes}},  {\em
  JHEP} {\bf 05} (2001) 011, [\href{http://arxiv.org/abs/hep-th/0104082}{{\tt
  hep-th/0104082}}].

\bibitem{Yamaguchi:2006tq}
S.~Yamaguchi, {\it {Wilson Loops of Anti-symmetric Representation and
  D5-branes}},  {\em JHEP} {\bf 0605} (2006) 037,
  [\href{http://arxiv.org/abs/hep-th/0603208}{{\tt hep-th/0603208}}].

\bibitem{Gomis:2006sb}
J.~Gomis and F.~Passerini, {\it {Holographic Wilson Loops}},  {\em JHEP} {\bf
  08} (2006) 074, [\href{http://arxiv.org/abs/hep-th/0604007}{{\tt
  hep-th/0604007}}].

\bibitem{Gomis:2006im}
J.~Gomis and F.~Passerini, {\it {Wilson Loops as D3-Branes}},  {\em JHEP} {\bf
  0701} (2007) 097, [\href{http://arxiv.org/abs/hep-th/0612022}{{\tt
  hep-th/0612022}}].

\bibitem{Blake:2014lva}
M.~Blake, A.~Donos, and D.~Tong, {\it {Holographic Charge Oscillations}},  {\em
  JHEP} {\bf 04} (2015) 019, [\href{http://arxiv.org/abs/1412.2003}{{\tt
  arXiv:1412.2003}}].

\bibitem{Nishioka:2009un}
T.~Nishioka, S.~Ryu, and T.~Takayanagi, {\it {Holographic Entanglement Entropy:
  An Overview}},  {\em J. Phys.} {\bf A42} (2009) 504008,
  [\href{http://arxiv.org/abs/0905.0932}{{\tt arXiv:0905.0932}}].

\bibitem{Chesler:2013lia}
P.~M. Chesler and L.~G. Yaffe, {\it {Numerical solution of gravitational
  dynamics in asymptotically anti-de Sitter spacetimes}},  {\em JHEP} {\bf 07}
  (2014) 086, [\href{http://arxiv.org/abs/1309.1439}{{\tt arXiv:1309.1439}}].

\bibitem{Horowitz:2012ky}
G.~T. Horowitz, J.~E. Santos, and D.~Tong, {\it {Optical Conductivity with
  Holographic Lattices}},  {\em JHEP} {\bf 07} (2012) 168,
  [\href{http://arxiv.org/abs/1204.0519}{{\tt arXiv:1204.0519}}].

\bibitem{2003PhRvL..90u6403S}
T.~Senthil, S.~Sachdev, and M.~Vojta, {\it {Fractionalized Fermi Liquids}},
  {\em Phys. Rev. Lett.} {\bf 90} (May, 2003) 216403,
  [\href{http://arxiv.org/abs/{arXiv:cond-mat/0209144}}{{\tt
  {arXiv:cond-mat/0209144}}}].

\bibitem{2004PhRvB..69c5111S}
T.~Senthil, M.~Vojta, and S.~Sachdev, {\it {Weak Magnetism and Non-Fermi
  liquids Near Heavy-fermion Critical Points}},  {\em Phys. Rev.} {\bf B69}
  (Jan., 2004) 035111,
  [\href{http://arxiv.org/abs/{arxiv:cond-mat/0305193}}{{\tt
  {arxiv:cond-mat/0305193}}}].

\bibitem{Kraus:2006wn}
P.~Kraus, {\it {Lectures on Black Holes and the AdS(3)/CFT(2) Correspondence}},
   {\em Lect. Notes Phys.} {\bf 755} (2008) 193--247,
  [\href{http://arxiv.org/abs/hep-th/0609074}{{\tt hep-th/0609074}}].

\bibitem{deHaro:2000xn}
S.~de~Haro, S.~N. Solodukhin, and K.~Skenderis, {\it {Holographic
  reconstruction of space-time and renormalization in the AdS / CFT
  correspondence}},  {\em Commun. Math. Phys.} {\bf 217} (2001) 595--622,
  [\href{http://arxiv.org/abs/hep-th/0002230}{{\tt hep-th/0002230}}].

\bibitem{Bianchi:2001kw}
M.~Bianchi, D.~Z. Freedman, and K.~Skenderis, {\it {Holographic
  renormalization}},  {\em Nucl. Phys.} {\bf B631} (2002) 159--194,
  [\href{http://arxiv.org/abs/hep-th/0112119}{{\tt hep-th/0112119}}].

\bibitem{Papadimitriou:2005ii}
I.~Papadimitriou and K.~Skenderis, {\it {Thermodynamics of asymptotically
  locally AdS spacetimes}},  {\em JHEP} {\bf 08} (2005) 004,
  [\href{http://arxiv.org/abs/hep-th/0505190}{{\tt hep-th/0505190}}].

\bibitem{Papadimitriou:2010as}
I.~Papadimitriou, {\it {Holographic renormalization as a canonical
  transformation}},  {\em JHEP} {\bf 11} (2010) 014,
  [\href{http://arxiv.org/abs/1007.4592}{{\tt arXiv:1007.4592}}].

\bibitem{vanRees:2011fr}
B.~C. van Rees, {\it {Holographic renormalization for irrelevant operators and
  multi-trace counterterms}},  {\em JHEP} {\bf 08} (2011) 093,
  [\href{http://arxiv.org/abs/1102.2239}{{\tt arXiv:1102.2239}}].

\bibitem{vanRees:2011ir}
B.~C. van Rees, {\it {Irrelevant deformations and the holographic
  Callan-Symanzik equation}},  {\em JHEP} {\bf 10} (2011) 067,
  [\href{http://arxiv.org/abs/1105.5396}{{\tt arXiv:1105.5396}}].

\bibitem{PhysRevB.73.224445}
D.~Bensimon, A.~Jerez, and M.~Lavagna, {\it {Intermediate Coupling Fixed Point
  Study in the Overscreened Regime of Generalized Multichannel $\mathrm{SU}(N)$
  Kondo Models}},  {\em Phys. Rev.} {\bf B73} (2006) 224445.

\bibitem{Nozieres:1980}
P.~Nozi{\`e}res and A.~Blandin, {\it {Kondo Effect in Real Metals}},  {\em J.
  Phys. France} {\bf 41} (1980), no.~3 193--211.

\bibitem{2008arXiv0809.4836A}
A.~{Auerbach} and D.~P. {Arovas}, {\it {Schwinger Bosons Approaches to Quantum
  Antiferromagnetism}},  {\em ArXiv e-prints} (Sept., 2008)
  [\href{http://arxiv.org/abs/0809.4836}{{\tt arXiv:0809.4836}}].

\bibitem{2010JMP....51i3504M}
M.~{Mathur}, I.~{Raychowdhury}, and R.~{Anishetty}, {\it {SU(N) irreducible
  Schwinger bosons}},  {\em Journal of Mathematical Physics} {\bf 51} (Sept.,
  2010) 093504, [\href{http://arxiv.org/abs/1003.5487}{{\tt arXiv:1003.5487}}].

\bibitem{Buchbinder:2007ar}
E.~I. Buchbinder, J.~Gomis, and F.~Passerini, {\it {Holographic Gauge Theories
  in Background Fields and Surface Operators}},  {\em JHEP} {\bf 12} (2007)
  101, [\href{http://arxiv.org/abs/0710.5170}{{\tt arXiv:0710.5170}}].

\bibitem{Castro:2008ms}
A.~Castro, D.~Grumiller, F.~Larsen, and R.~McNees, {\it {Holographic
  Description of AdS(2) Black Holes}},  {\em JHEP} {\bf 0811} (2008) 052,
  [\href{http://arxiv.org/abs/0809.4264}{{\tt arXiv:0809.4264}}].

\bibitem{Fujita:2014mqa}
M.~Fujita, S.~Harrison, A.~Karch, R.~Meyer, and N.~M. Paquette, {\it {Towards a
  Holographic Bose-Hubbard Model}},  {\em JHEP} {\bf 04} (2015) 068,
  [\href{http://arxiv.org/abs/1411.7899}{{\tt arXiv:1411.7899}}].

\bibitem{Klebanov:1999tb}
I.~R. Klebanov and E.~Witten, {\it {AdS/CFT correspondence and symmetry
  breaking}},  {\em Nucl. Phys.} {\bf B556} (1999) 89--114,
  [\href{http://arxiv.org/abs/hep-th/9905104}{{\tt hep-th/9905104}}].

\bibitem{Marolf:2006nd}
D.~Marolf and S.~F. Ross, {\it {Boundary Conditions and New Dualities: Vector
  Fields in AdS/CFT}},  {\em JHEP} {\bf 11} (2006) 085,
  [\href{http://arxiv.org/abs/hep-th/0606113}{{\tt hep-th/0606113}}].

\bibitem{Fefferman}
C.~Fefferman and C.~R. Graham, {\it {Conformal Invariants}},  in {\em Elie
  Cartan et les Mathematiques d'aujourd'hui}, p.~95.
\newblock Asterique, 1985.

\bibitem{deBoer:1999xf}
J.~de~Boer, E.~P. Verlinde, and H.~L. Verlinde, {\it {On the holographic
  renormalization group}},  {\em JHEP} {\bf 08} (2000) 003,
  [\href{http://arxiv.org/abs/hep-th/9912012}{{\tt hep-th/9912012}}].

\bibitem{Martelli:2002sp}
D.~Martelli and W.~Mueck, {\it {Holographic renormalization and Ward identities
  with the Hamilton-Jacobi method}},  {\em Nucl. Phys.} {\bf B654} (2003)
  248--276, [\href{http://arxiv.org/abs/hep-th/0205061}{{\tt hep-th/0205061}}].

\bibitem{Papadimitriou:2004ap}
I.~Papadimitriou and K.~Skenderis, {\it {AdS / CFT correspondence and
  geometry}},  in {\em {AdS/CFT correspondence: Einstein metrics and their
  conformal boundaries. Proceedings, 73rd Meeting of Theoretical Physicists and
  Mathematicians, Strasbourg, France, September 11-13, 2003}}, pp.~73--101,
  2004.
\newblock \href{http://arxiv.org/abs/hep-th/0404176}{{\tt hep-th/0404176}}.

\bibitem{Chemissany:2014xsa}
W.~Chemissany and I.~Papadimitriou, {\it {Lifshitz holography: The whole
  shebang}},  {\em JHEP} {\bf 01} (2015) 052,
  [\href{http://arxiv.org/abs/1408.0795}{{\tt arXiv:1408.0795}}].

\bibitem{usfuture}
J.~Erdmenger, C.~Hoyos, A.~O'Bannon, I.~Papadimitriou, J.~Probst, and J.~Wu,
  {\it {In preparation}}, .

\bibitem{Bianchi:2001de}
M.~Bianchi, D.~Z. Freedman, and K.~Skenderis, {\it {How to go with an RG
  flow}},  {\em JHEP} {\bf 08} (2001) 041,
  [\href{http://arxiv.org/abs/hep-th/0105276}{{\tt hep-th/0105276}}].

\bibitem{Papadimitriou:2004rz}
I.~Papadimitriou and K.~Skenderis, {\it {Correlation functions in holographic
  RG flows}},  {\em JHEP} {\bf 10} (2004) 075,
  [\href{http://arxiv.org/abs/hep-th/0407071}{{\tt hep-th/0407071}}].

\bibitem{Henningson:1998gx}
M.~Henningson and K.~Skenderis, {\it {The Holographic Weyl anomaly}},  {\em
  JHEP} {\bf 07} (1998) 023, [\href{http://arxiv.org/abs/hep-th/9806087}{{\tt
  hep-th/9806087}}].

\bibitem{Papadimitriou:2007sj}
I.~Papadimitriou, {\it {Multi-Trace Deformations in AdS/CFT: Exploring the
  Vacuum Structure of the Deformed CFT}},  {\em JHEP} {\bf 0705} (2007) 075,
  [\href{http://arxiv.org/abs/hep-th/0703152}{{\tt hep-th/0703152}}].

\bibitem{Witten:2001ua}
E.~Witten, {\it {Multitrace operators, boundary conditions, and AdS / CFT
  correspondence}},  \href{http://arxiv.org/abs/hep-th/0112258}{{\tt
  hep-th/0112258}}.

\bibitem{Imbimbo:1999bj}
C.~Imbimbo, A.~Schwimmer, S.~Theisen, and S.~Yankielowicz, {\it
  {Diffeomorphisms and holographic anomalies}},  {\em Class. Quant. Grav.} {\bf
  17} (2000) 1129--1138, [\href{http://arxiv.org/abs/hep-th/9910267}{{\tt
  hep-th/9910267}}].

\bibitem{Schwimmer:2000cu}
A.~Schwimmer and S.~Theisen, {\it {Diffeomorphisms, anomalies and the
  Fefferman-Graham ambiguity}},  {\em JHEP} {\bf 08} (2000) 032,
  [\href{http://arxiv.org/abs/hep-th/0008082}{{\tt hep-th/0008082}}].

\bibitem{Barut:1986}
A.~Barut and R.~Raczka, {\it {Theory of Group Representations and
  Applications}},  {\em {World Scientific Publishing Company}} (1986).

\bibitem{Horowitz:2009ij}
G.~T. Horowitz and M.~M. Roberts, {\it {Zero Temperature Limit of Holographic
  Superconductors}},  {\em JHEP} {\bf 11} (2009) 015,
  [\href{http://arxiv.org/abs/0908.3677}{{\tt arXiv:0908.3677}}].

\bibitem{Erdmenger:2014xya}
J.~Erdmenger, M.~Flory, and M.-N. Newrzella, {\it {Bending branes for DCFT in
  two dimensions}},  {\em JHEP} {\bf 01} (2015) 058,
  [\href{http://arxiv.org/abs/1410.7811}{{\tt arXiv:1410.7811}}].

\bibitem{2008NatPh...4..643F}
R.~{Flint}, M.~{Dzero}, and P.~{Coleman}, {\it {Heavy electrons and the
  symplectic symmetry of spin}},  {\em Nature Physics} {\bf 4} (Aug., 2008)
  643, [\href{http://arxiv.org/abs/0710.1126}{{\tt arXiv:0710.1126}}].

\bibitem{2007arXiv0710.1128F}
R.~{Flint}, M.~{Dzero}, and P.~{Coleman}, {\it {Supplementary material to Heavy
  electrons and the symplectic symmetry of spin}},  {\em ArXiv e-prints} (Oct.,
  2007) [\href{http://arxiv.org/abs/0710.1128}{{\tt arXiv:0710.1128}}].

\bibitem{Aharony:2008an}
O.~Aharony and D.~Kutasov, {\it {Holographic Duals of Long Open Strings}},
  {\em Phys. Rev.} {\bf D78} (2008) 026005,
  [\href{http://arxiv.org/abs/0803.3547}{{\tt arXiv:0803.3547}}].

\end{thebibliography}\endgroup

\end{document}